\documentclass[aps,prd,onecolumn,nofootinbib,superscriptaddress]{revtex4}
\usepackage{float}
\usepackage{graphicx}
\usepackage{amsmath}
\usepackage{amsfonts}
\usepackage{amssymb,ulem}
\usepackage{color}%
\usepackage{dcolumn}
\usepackage{subfigure}
\usepackage{multirow}

\usepackage{MnSymbol,wasysym}
\usepackage{braket,diagbox}
\usepackage{eurosym}
\usepackage{calrsfs}
\usepackage[usenames,dvipsnames,svgnames]{xcolor}
\newcommand{\red}{\textcolor[rgb]{1.00,0.00,0.00}}

\newcommand{\del}[1]{\red{\sout{#1}}}

\newcommand{\RNum}[1]{\uppercase\expandafter{\romannumeral #1\relax}}
\usepackage[colorlinks=true,linkcolor=blue,urlcolor=blue,filecolor=black,citecolor=red,
pdfstartview=FitV,pdftitle={},pdfsubject={},pdfkeywords={},pdfpagemode=None,bookmarksopen=true]{hyperref}
\usepackage{float}

\usepackage[title]{appendix}

\begin{document}
\baselineskip=0.5 cm

\title{Images of hairy  Reissner-Nordstr\"{o}m black hole illuminated by static accretions}
\author{Yuan Meng}
\email{mengyuanphy@163.com}
\affiliation{Center for Gravitation and Cosmology, College of Physical Science and Technology, Yangzhou University, Yangzhou, 225009, China}
\author{Xiao-Mei Kuang}
\email{xmeikuang@yzu.edu.cn (corresponding author)}
\affiliation{Center for Gravitation and Cosmology, College of Physical Science and Technology, Yangzhou University, Yangzhou, 225009, China}
\author{Xi-Jing Wang}
\email{xijingwang01@163.com}
\affiliation{Center for Gravitation and Cosmology, College of Physical Science and Technology, Yangzhou University, Yangzhou, 225009, China}
\author{Bin Wang}
\email{wang$_$b@sjtu.edu.cn}
\affiliation{Center for Gravitation and Cosmology, College of Physical Science and Technology, Yangzhou University, Yangzhou, 225009, China}
\affiliation{Shanghai Frontier Science Center for Gravitational Wave Detection, Shanghai Jiao Tong University, Shanghai 200240, China}
\author{Jian-Pin Wu}
\email{jianpinwu@yzu.edu.cn}
\affiliation{Center for Gravitation and Cosmology, College of Physical Science and Technology, Yangzhou University, Yangzhou, 225009, China}

\date{\today}

\begin{abstract}
\baselineskip=0.44 cm
In this paper, we investigate the shadow and optical appearance of the hairy Reissner-Nordstr\"{o}m (RN) black hole illuminated by two toy models of static accretion.  The hairy RN black hole was constructed in the gravitation decoupling approach to describe the deformation of a Schwarzschild black hole due to the inclusion of additional arbitrary source (scalar field, tensor field, fluidlike dark matter, etc.). So it is characterized by the parameters: mass ($M$), deformation factor ($\alpha$), electric charge ($Q$) and the additional hairy charge ($l_o$)., differentiating from the case in RN black hole. Though the specific background theory that
results in this hairy RN black hole is still tricky, here we shall focus on the novel observable features introduced by the hair of  this black hole.
First, we find that for the hairy RN black hole,  the event horizon, radius of photon sphere and critical impact parameter all increase as the increasings of $Q$ and $l_o$, but decrease as $\alpha$ grows. Furthermore,  the three characterized parameters are found to have significant effects on the photon trajectories, and shadows as well as images of the hairy RN black hole surrounded by the static accretion disk and spherical accretion, respectively. In particular, both $Q$ and $l_o$  have mutually reinforcing effects on the optical appearance and shadows of the hairy RN black hole, which implies that we may not distinguish the electric charge and hairy charge from the  shadow and image of black hole in this scenario.
Additionally,  because of the competing effects of the charge parameters ($Q, l_o$) and the deviation parameter $\alpha$ on the observed intensities of brightness, the optical appearance  between the hairy RN black hole and RN black hole could have degeneracies, indicating the indistinguishability. Our current results contribute more to the phenomenal aspects which could be helpful to build the background theory of this hairy RN black hole.

\end{abstract}


\maketitle

\newpage
\tableofcontents

\section{Introduction}
Recently, the Event Horizon Telescope (EHT) has made significant achievements, providing us with a new window to test the essence of gravity in the strong field regime. One of the most impressive achievement among them is the release of images of the supermassive black hole M87* \cite{EventHorizonTelescope:2019dse, EventHorizonTelescope:2019uob, EventHorizonTelescope:2019jan, EventHorizonTelescope:2019ths, EventHorizonTelescope:2019pgp, EventHorizonTelescope:2019ggy} and the central black hole of our Milky Way \cite{EventHorizonTelescope:2022wkp, EventHorizonTelescope:2022apq, EventHorizonTelescope:2022wok, EventHorizonTelescope:2022exc, EventHorizonTelescope:2022urf, EventHorizonTelescope:2022xqj}. The dark regions in the center of the images are the shadow of black hole, which are related to the strong gravitational lensing of photons around the black hole \cite{Virbhadra:1999nm,Virbhadra:2002ju,Virbhadra:2007kw,Synge:1966okc, Bardeen:1972fi,Bozza:2010xqn}. The unstable photon orbit (photon sphere) in the innermost layer around the black hole define the radius of shadow, and the bright rings in the images are related to the outer photon orbits. Moreover, the bright ring-shaped construction is the product of various light rays as the black hole is always illuminated by complex accretion flows.

In fact, early on, the black hole shadow was known as the escaping cone of photon \cite{Synge:1966okc}, which is defined as the impact parameter corresponding to the critical curve.
Furthermore, the expression of the angular radius of the photon capture region for the Schwarzschild black hole  was proposed \cite{Luminet:1979nyg}, and later Bardeen investigated the D-type shadow of the Kerr black hole \cite{Bardeen:1972fi}, which is caused by the dragging effect of the rotating black hole on the light rays. This has inspired a wave of research on the shadow of black hole, and numerous simulations of various black hole shadows have been widely discussed, see for examples \cite{Shen:2005cw,Yumoto:2012kz,Atamurotov:2013sca,Papnoi:2014aaa,Abdujabbarov:2015xqa,Kumar:2018ple} and references therein. Those studies indicate that the properties of  black hole shadows closely depend on the spacetime background. The shadows of black holes in different spacetime dimensions and modified theories of gravity  have also been  attracted extensive discussion, such as conformal gravity \cite{Meng:2022kjs}, Gauss-Bonnet theory \cite{Ma:2019ybz,Guo:2020zmf}, the Chern-Simons type theory \cite{Meng:2023wgi,Ayzenberg:2018jip,Amarilla:2010zq}, f(R) gravity \cite{Addazi:2021pty,Dastan:2016vhb}, and so on \cite{Amarilla:2011fx,Amarilla:2013sj,Amir:2017slq,Mizuno:2018lxz,Eiroa:2017uuq,Vagnozzi:2019apd,Banerjee:2019nnj,Chowdhuri:2020ipb,Konoplya:2019goy,Younsi:2016azx,Olmo:2023lil}. In addition, the shadows of naked singularities have been studied in \cite{Shaikh:2018lcc,Joshi:2020tlq,Dey:2020bgo}, and the wormholes {shadows}  were also explored in \cite{Rahaman:2021web,Kasuya:2021cpk,Shaikh:2018oul,Wielgus:2020uqz,Peng:2021osd,Neto:2022pmu,Tsukamoto:2012xs} and references therein. On the other hand, images from EHT collaboration are simulated via Kerr black holes in general relativity (GR). However, due to the finite resolution of EHT, this also provides some space for alternative theories beyond GR. Therefore, plenty of works have been carried out to use the shadow observations of the EHT to test or constrain modified theories of gravity \cite{Vagnozzi:2022moj,Afrin:2021imp,Kuang:2022ojj,Tang:2022hsu,Tang:2022bcm,Kuang:2022xjp,Kumar:2019pjp,Shaikh:2021yux,Wu:2023yhp,Capozziello:2023tbo,Sui:2023rfh,Pantig:2022qak,Ghosh:2023kge,Tsukamoto:2014tja}.

The bright ring-shaped {constructions} in the EHT images are closely related to the accretion flow radiation around the black hole, and the geometric features of the accretion flow determine the optical appearance of the black hole. However, in the realistic astrophysical environments, the accretion flows around black holes are always very complex such that one need numerical simulation using general {relativistic} magnetohydrodynamic (GRMHD) to extract the black hole images \cite{EventHorizonTelescope:2019pcy}. It is worth noting that usually
in order to investigate the main characteristics of black hole images and the gravity in strong field {regime}, giving some toy models of accretion {structures} is enough to put forward. Thus, Wald {et al.}  first proposed an optically and geometrically thin toy accretion disk around Schwarzschild black holes, and defined direct $(n=1)$, lensed ring $(n=2)$ and  photon ring $(n=3)$ emissions by the number of times ($n$) light intersecting with the accretion disk \cite{Gralla:2019xty}. They found that the total observed brightness of black hole images mainly comes from the  direct emissions, while the contribution of lensed ring emissions is small and photon ring emissions can even be ignored. Subsequently, toy accretion disk models around various black holes have been extensively studied \cite{Dokuchaev:2019pcx,Peng:2020wun,He:2021htq,Eichhorn:2021iwq,Li:2021riw,Wang:2023vcv}.  The spherical accretions, as another toy accretion model,  have also been attracted great interest \cite{Zeng:2020dco,Saurabh:2020zqg,Zeng:2020vsj,Qin:2020xzu,Narayan:2019imo}, in which the size and shape of shadows are mainly determined by the spacetime geometry of the black hole, rather than the details of the accretions. Moreover, the accretion constructions around black holes with two unstable photon spheres were also analyzed in \cite{Gan:2021xdl,Gan:2021pwu,Meng:2023htc}. Compared to black hole with single photon sphere, the black hole with double photon spheres could have additional photon rings in the image, and the photon ring and lensed ring emissions will contribute significantly to the total observed brightness. In general, the optical appearance is a potential probe to distinguish GR black holes from other alternative compact objects \cite{Boshkayev:2022vlv,Xavier:2023exm,Sakai:2014pga,Bacchini:2021fig,Destounis:2023khj} or black holes in alternative {theories} beyond GR \cite{Guo:2022iiy,Guo:2022ghl,Archer-Smith:2020hqq,Okyay:2021nnh,Uniyal:2022vdu,Hou:2022eev,
Uniyal:2023inx,Akbarieh:2023kjv,Gao:2023mjb,Theodosopoulos:2023ice,Zeng:2023fqy,Meng:2023uws}, but this does not always work, as even in GR, the images of black holes may exhibit degeneracies. For example, sufficiently dense {boson} stars \cite{Liebling:2012fv,Rosa:2023qcv} may have shadows similar to those of classical black holes \cite{Rosa:2022tfv} and images of very relativistic rotating boson stars may resemble Kerr black holes \cite{Vincent:2015xta}.

On the other hand, when there are other additional sources around the black hole, the interaction between spacetime and additional sources can cause the black hole to carry global charge. This charge is also known as the hair of black hole, and in this case the spacetime usually will deviate from that in GR. In this scenario, Ovalle and Casadio first proposed the gravitational decoupling (GD) approach \cite{Ovalle:2020kpd,Contreras:2021yxe} to introduce additional sources, and obtain hairy black holes which describes deformations of the black holes in GR. Due to the fact that in the GD approach, the additional sources  can be dark matter, scalar fields, tensor fields etc., so the GD approach can be used to study the general existence problem of solutions for hairy black holes. In particular, they take the Schwarzschild  black hole as the seed black hole, and use the GD approach to construct the hairy Schwarzschild  black hole which obtain extensive attention, such as the thermodynamics of rotating hairy black holes \cite{Mahapatra:2022xea}, scalar perturbations and quasinormal modes of hairy black holes \cite{Cavalcanti:2022cga,Yang:2022ifo,Li:2022hkq}, strong gravitational lensing, black hole shadow and image \cite{Afrin:2021imp,Islam:2021dyk,Meng:2023htc}, precession and Lense-Thirring effect \cite{Wu:2023wld} and gravitational waves from extreme mass ratio inspirals \cite{Zi:2023omh}.

In this work, we will focus on the optical features of the hairy RN black hole, which was also constructed via the GD approach by taking the Schwarzschild black hole in GR as the seed black hole. Since the hairy RN black hole includes electric charge  and additional arbitrary hairy charge, thus, the aim of this work to investigate if the black hole shadow and image can probe the different charges in this scenario. In details, we will firstly explore the effects of characterized parameters of the hairy RN black hole on the light rays, and on the black hole rings and images when it is surrounded by the optically and geometrically thin accretion disks and spherical accretion. Then we shall discuss the possible distinguishable effects of the charges on the hairy RN black hole shadows and images, and potentially differentiate the hairy RN black hole from RN black hole.

This paper is organized as follows. In Sec. \ref{sec:tragectories and photons}, we will study the radius of photon sphere and impact parameters by analyzing the effective potential of photons around the hairy RN black hole. In Sec. \ref{sec:ring and images}, we use the ray-tracing method to discuss the distribution of the light rays around the hairy RN black hole, and then investigate the optical appearance of hairy RN black hole illuminated by the optically and geometrically thin accretion disk. In Sec. \ref{sec:spherical accretions}, we present the black hole shadows and images illuminated by static  spherical accretions. Sec. \ref{sec:conclusion} gives our conclusions.

\section{Photons sphere of the hairy RN black hole}\label{sec:tragectories and photons}
Black holes in classical GR are only described by mass, electric charge and spin because of the no-hair theorem \cite{Ruffini:1971bza}. But the interaction between black hole spacetime and matters may introduce other charges, so the black hole could carry additional hairy charges and the physical effects of  these hairy charges can modify the original background spacetime, namely, hairy black holes may form. Recently, Ovalle et.al used the GD approach to obtain a spherically symmetric black hole with additional hairy charges  \cite{Ovalle:2020kpd,Contreras:2021yxe}, which has the metric
\begin{eqnarray}
ds^2=-f(r)dt^2+\frac{1}{f(r)}dr^2+r^2(d\theta^2+\sin^2\theta d\phi^2),
\label{eq:metric}
\end{eqnarray}
with
\begin{eqnarray}
f(r)=1-\frac{2M}{r}+\frac{Q^2}{r^2}-\frac{\alpha}{r}\left(M-\frac{l_o}{2}\right)e^{-r/(M-l_o/2)}.
\end{eqnarray}
This metric describes the deformation of the Schwarzschild black hole solution due to the introduction of additional sources, which can be dark matter, scalar fields, tensor fields, etc.  Here  $Q$ and $l_0$ represent potential charges, and both of them are proportional to $\alpha$, such that the above solution reduces to Schwarzschild case as $\alpha\to 0$. In appendix \ref{appendix}, we review how the hairy RN black hole \eqref{eq:metric} is solved out from the seed  metric via the GD approach. The technics  of this approach to obtain the deformed solutions sound good, but the specific background theory, which can give  the additional required tensor field in the approach, remains open. The authors of \cite{Ovalle:2020kpd} briefly discussed the potential background theories which still call for further investigations. Nevertheless, the uncertainty of the background theories did not hinder physicists' attention on this type of deformed black hole, and lots of theoretical and observable properties of the deformed black hole have been investigated \cite{Mahapatra:2022xea,Cavalcanti:2022cga,Yang:2022ifo,Li:2022hkq,Afrin:2021imp,Islam:2021dyk,
Meng:2023htc,Wu:2023wld,Zi:2023omh}. The reasons may stem from two aspects: (i) Due to the uncertainty of the ingredients in our Universe and the dispute on no-hair theorem of black hole, people can temporarily suspend the unclear background theory and go ahead to study its possible outputs.  (ii) More complete understanding of the hairy black hole could give more information to build its background theories, at least from bottom-up perspective. In particular, as addressed in \cite{Ovalle:2020kpd}, $Q$ can be an electric charge, or a tidal charge of extradimensional origin or any other source depending on the specific background theory. When $Q$ represents an electric charge, the electrovacuum of the RN geometry also contains a tensor-vacuum whose components are those explicitly proportional to $\alpha$. In this paper,  taking  our motivations mentioned in the Introduction into account, we will treat $Q$ as the electric charge differentiating from the additional hairy charge $l_o$, and instead of clarifying its corresponding background theory,  we shall discuss their possible probe from black hole shadow and images. So we denote the above geometry \eqref{eq:metric} as the hairy RN spacetime. Thus,  in the following studies, $M$, $Q$, and $l_o$ are the parameters related with the mass, electric charge and hairy charge of the black hole, respectively. In addition, $l_o$ should satisfy $l_o\leq2M$ to ensure the asymptotic flatness of spacetime.

\begin{figure} [H]
{\centering
\includegraphics[width=5cm]{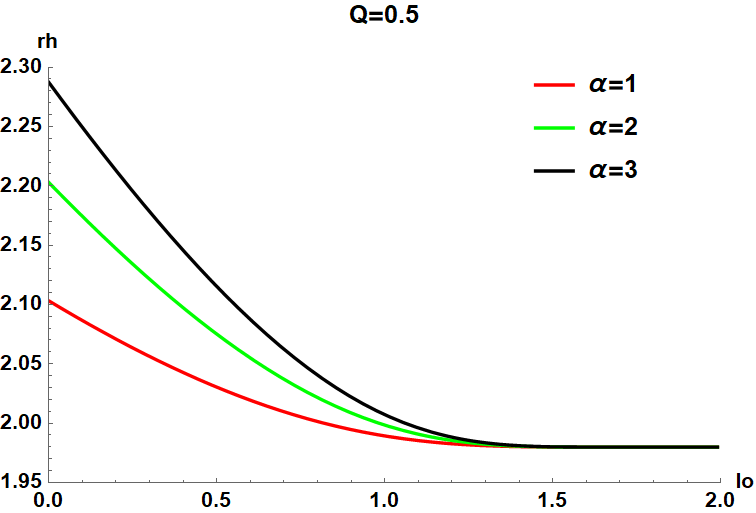}\hspace{5mm}
\includegraphics[width=5cm]{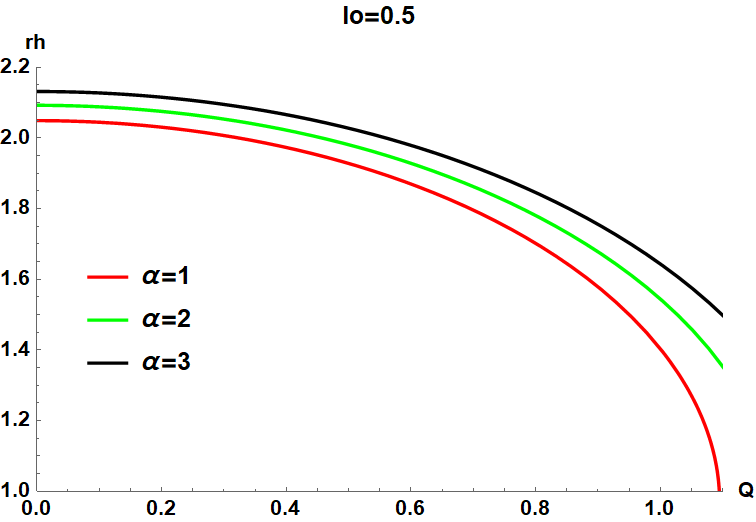}\hspace{5mm}
\includegraphics[width=5cm]{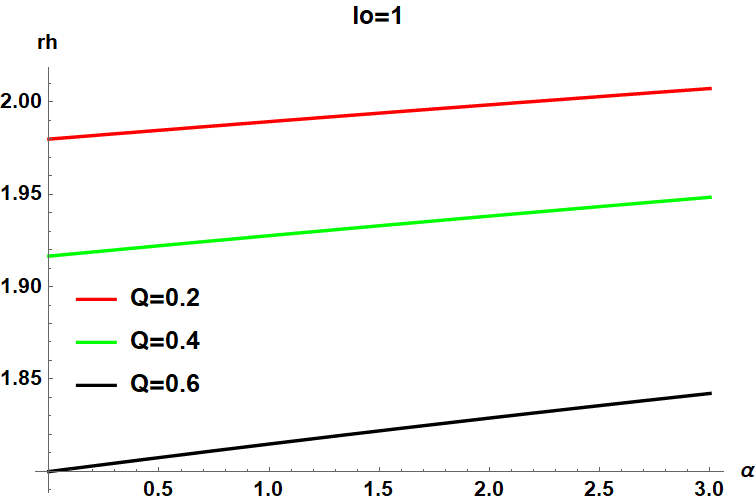}
\caption{The event horizon $r_h$ of the hairy RN black hole as functions of the parameters $l_o$, $Q$ and $\alpha$ of the hairy RN black hole. Here we have set $M=1$.}   \label{fig:rh}}
\end{figure}

Here we focus on the properties of motions of photon in the vicinity of the hairy RN black hole. Before proceeding, we briefly discuss
the event horizon of hairy RN black hole, which can be solved out from
\begin{eqnarray}
f(r)|_{r=r_h}=1-\frac{2M}{r_h}+\frac{Q^2}{r_h^2}-\frac{\alpha}{r_h}\left(M-\frac{l_o}{2}\right)e^{-r_h/(M-l_o/2)}=0.
\label{eventhorizon}
\end{eqnarray}
After numerical calculation, we show how the event horizon $r_h$ is affected by the parameters $\alpha$, $Q$, and $l_o$ with the others fixed in FIG.\ref{fig:rh}. Obviously, with fixed $\alpha$, both the charge parameters $l_o$ and $Q$ will suppress the event horizon $r_h$, but as the deformation parameter $\alpha$ increases, the event horizon  of the hairy RN black hole becomes larger.

\begin{figure} [H]
{\centering
\includegraphics[width=5cm]{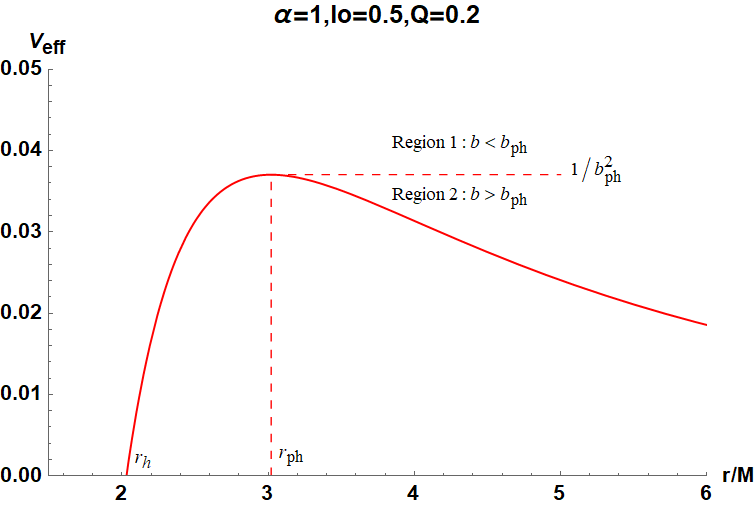}\hspace{5mm}
\includegraphics[width=5cm]{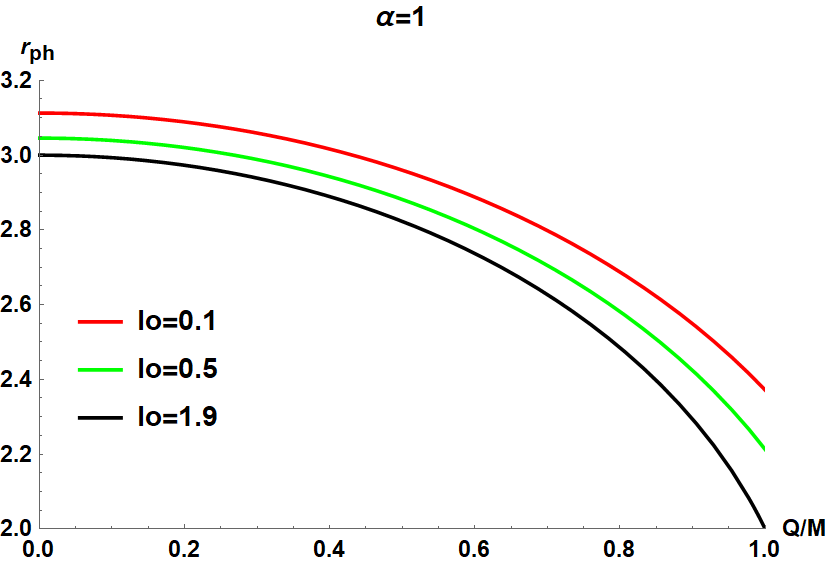}\hspace{5mm}
\includegraphics[width=5cm]{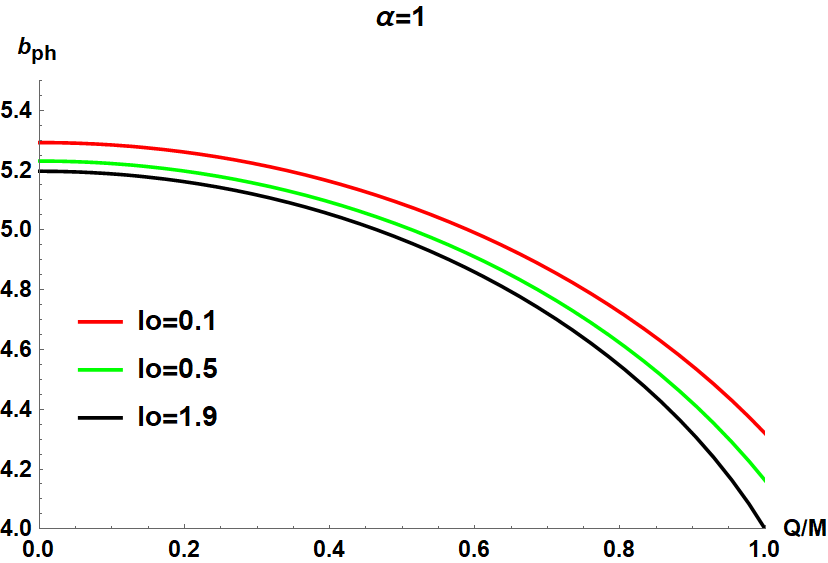}
\caption{Left: a sample of  effective potential $V_\text{{eff}}$ as function of radius $r$. Middle and Right: the radius 
$r_{ph}$ of photon sphere and critical impact parameter $b_{ph}$ for selected parameters.}   \label{fig:Veff} }
\end{figure}

Next, we will investigate the null geodesic motion in the hairy RN black holes. The motions of photons are described by the Euler-Lagrange equation,
\begin{eqnarray}
\frac{d}{d\tau}\Big(\frac{\partial \mathcal{L}}{\partial \dot{x}^\mu}\Big)-\frac{\partial \mathcal{L}}{\partial x^\mu}=0,
\end{eqnarray}
where $\dot{x}^\mu=\frac{dx^\mu}{d\tau}$ represents the four-velocity of the photon and $\tau$ is the affine parameter. In the static spherically symmetric hairy RN black hole, the Lagrangian of the photon takes the form
\begin{eqnarray}
\mathcal{L}=\frac{1}{2}g_{\mu\nu}\dot{x}^\mu\dot{x}^\nu=\frac{1}{2}\big[-f(r)\dot{t}^2+\frac{1}{f(r)}\dot{r}^2+r^2(\dot{\theta}^2+\sin^2\theta\dot{\phi}^2)\big].
\end{eqnarray}
Since $\partial_t$ and $\partial_\phi$ are Killing vector fields of the hairy RN spacetime, we can define the conserved energy, $E$, and the $z$ component of the angular momentum, $L_z$, of the photon as
\begin{eqnarray}
E\equiv-\frac{\partial \mathcal{L}}{\partial \dot{t}}=f(r)\dot{t},~~~~~~~~~~L_z=\frac{\partial \mathcal{L}}{\partial \mathcal{\dot{\phi}}}=r^2 \sin^2\theta \dot{\phi}.
\end{eqnarray}
Furthermore, due to the spherical symmetry of the spacetime, we can focus on the motions of photons on the equatorial plane $(\theta=\pi/2)$ without losing generality. Moreover, by considering $\mathcal{L}=0$ for photon and defining the impact parameter $b=L_z/E$, we can obtain three first-order differential equations of motion for the photons
\begin{eqnarray}
\dot{t}=\frac{1}{bf(r)},~~~\dot{\phi}=\pm\frac{1}{r^2}
\label{eq:phi},~~~
\dot{r}^2=\frac{1}{b^2}-V_{\text{eff}}(r),
\label{eq:r-motion}
\end{eqnarray}
where the effective potential is given as
\begin{eqnarray}
V_{\text{eff}}(r)=\frac{f(r)}{r^2}.
\label{eq:Veff}
\end{eqnarray}
The $\pm$ sign in \eqref{eq:phi} represent the clockwise and counterclockwise direction of photons' trajectories, respectively. Then the fate of a photon is determined by the radial geodesic equation. In particular, the circular orbit should satisfy $\dot{r}=0$ and $\ddot{r}=0$, which will be translated to
\begin{eqnarray}
b_{ph}=\frac{1}{\sqrt{V_{\text{eff}}(r_{ph})}},~~~~~~~~~V'_{\text{eff}}(r_{ph})=0,
\label{eq:rph}
\end{eqnarray}
where $r_{ph}$ is the radius of the photon sphere satisfying $V''_{\text{eff}}(r_{ph})<0$ , and $b_{ph}$ is the critical impact parameter which is related with the shadow radius of the black hole. As shown in the leftmost plot of FIG. \ref{fig:Veff}, the light rays with $b>b_{ph}$, will be deflected by the black hole and escape to infinity{;} on the contrary, the light rays with $b<b_{ph}$ will be captured into the event horizon of the black hole{. The} light rays with $b=b_{ph}$ will circle around the black hole {all the time} with the radius $r_{ph}$.
The effects of the two charges on the  $r_{ph}$ and $b_{ph}$ are shown in the middle and rightmost plot of FIG. \ref{fig:Veff}. It shows that as both charge parameters increase, both $r_{ph}$ and $b_{ph}$ become smaller, similar to {what} occurs for the event horizon{, as shown} in FIG. \ref{fig:rh}, as expected.

Based on the preliminary analysis, we shall study the optical appearance of the hairy RN black hole in various illumination
conditions. It is known that the hot, optically thin accretion flows are surrounding M87*, Sgr A* and
many other supermassive black holes in our Universe \cite{Yuan:2014gma}. Here we will consider the hairy RN black hole illuminated by statically thin disk accretions and spherical accretions, respectively,
which are simple toy accretion models but enough for the purpose of this paper.

\section{Images of hairy RN black hole illuminated by thin  accretions disk}\label{sec:ring and images}

In this section, we shall use the ray-tracing method to analyze the distribution of light rays in the vicinity of the hairy RN black hole and further investigate the optical appearance of the black hole illuminated by an optically and geometrically thin accretion disk. For simplicity, we consider that the accretion disk is fixed at the equatorial plane and the observer is located at the North pole. Due to the strong gravitational lensing effect, the light rays around the black hole may intersect with the accretion disk arbitrary times before falling into the black hole or escaping to infinity, which brings different contributions to the total intensity received by the observer. For this purpose, we should first classify the light rays in terms of the number of times they intersect with the thin accretion disk\del{,} and then figure out the images of the hairy RN black hole.

\subsection{Classification of the light rays}

The motion trajectories of photons around the hairy RN black holes {are} controlled by the trajectory equation 
\begin{eqnarray}
\frac{dr}{d\phi}=\frac{\dot{r}}{\dot{\phi}}=\pm\frac{1}{r^2}\sqrt{\frac{1}{b^2}-V_{\text{eff}}(r)}.
\label{trajectory-light-ray}
\end{eqnarray}
derived from \eqref{eq:r-motion}. It is obvious that the trajectory of light ray is closely related to the impact parameter $b$. As addressed in \cite{Gralla:2019xty}, the regions of impact parameters are divided into three types based on the total number of orbits $n=\phi/(2\pi)$, where $\phi$ is the total change in azimuthal angle. More specifically, the light ray from the North pole only intersects with the equatorial plane once, which corresponds to direct emission with $1/4<n<3/4$. For the second type with $3/4<n<5/4$, the light ray intersects with the equatorial plane twice, which corresponds to the lensed ring emission. The third type is photon ring emission with $n>5/4$, where the light rays intersect with the equatorial plane at least three times.

\begin{figure}[h]
\centering
\subfigure[\, fixing $\alpha=5\,\&\,l_o=0.1$ and changing $Q$]
{\includegraphics[width=4.5cm]{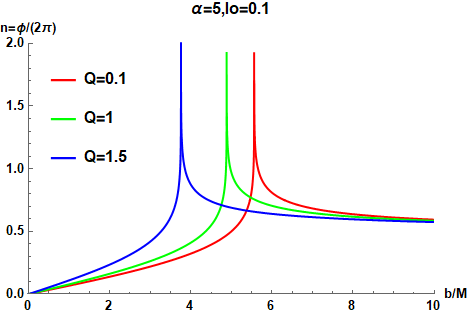} \label{}}\hspace{2mm}
\subfigure[\, $Q=0.1$]
{\includegraphics[width=4cm]{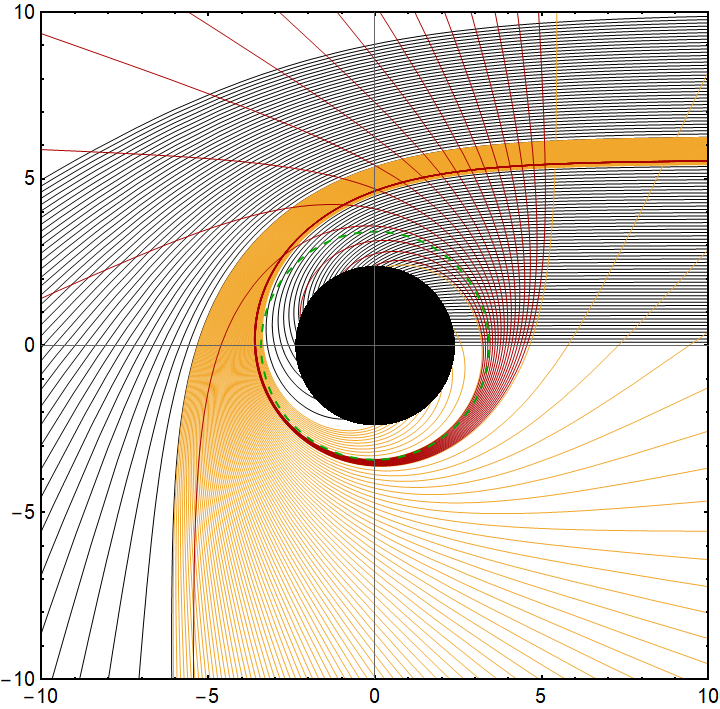}}\hspace{2mm}
\subfigure[\, $Q=1$]
{\includegraphics[width=4cm]{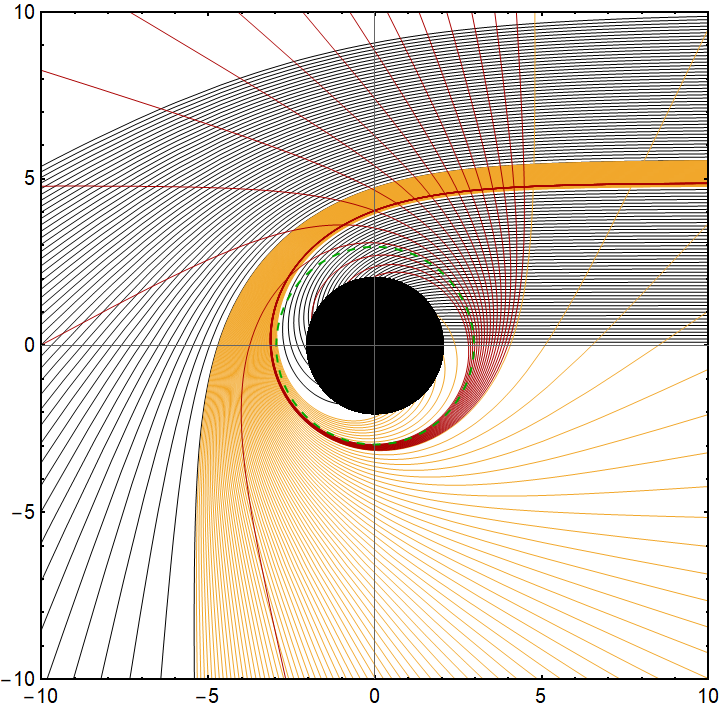}}\hspace{2mm}
\subfigure[\, $Q=1.5$]
{\includegraphics[width=4cm]{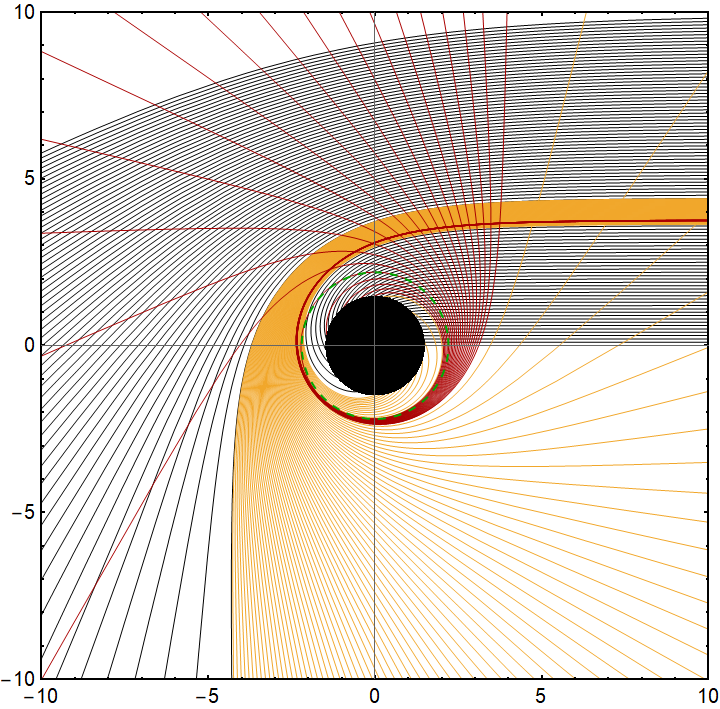}}\\
\subfigure[\, fixing $\alpha=5\,\&\,Q=0.1$ and changing $l_o$]
{\includegraphics[width=4.5cm]{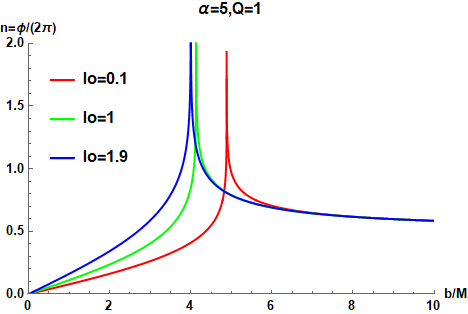} \label{}}\hspace{2mm}
\subfigure[\, $l_o=0.1$]
{\includegraphics[width=4cm]{traject-a-5-lo-01-Q-1}}\hspace{2mm}
\subfigure[\, $l_o=1$]
{\includegraphics[width=4cm]{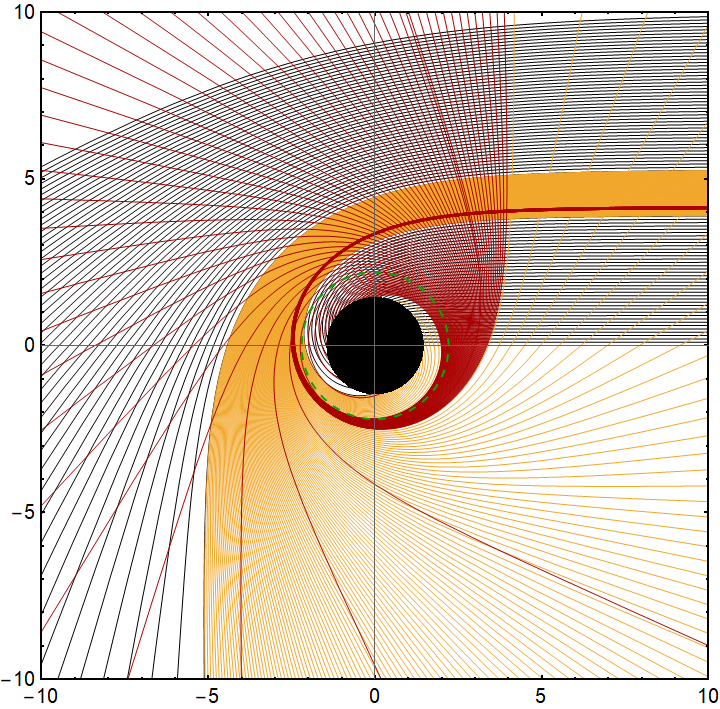}}\hspace{2mm}
\subfigure[\, $l_o=1.9$]
{\includegraphics[width=4cm]{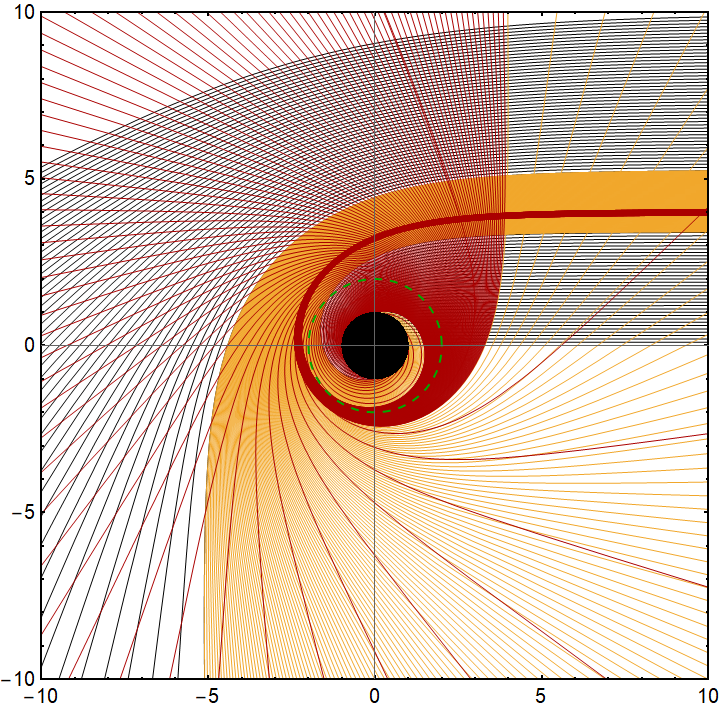}}\\
\subfigure[\, fixing $Q=0.1\,\&\,l_o=0.1$ and changing $\alpha$]
{\includegraphics[width=4.5cm]{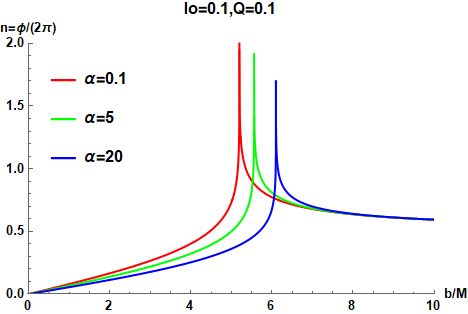} \label{}}\hspace{2mm}
\subfigure[\, $\alpha=0.1$]
{\includegraphics[width=4cm]{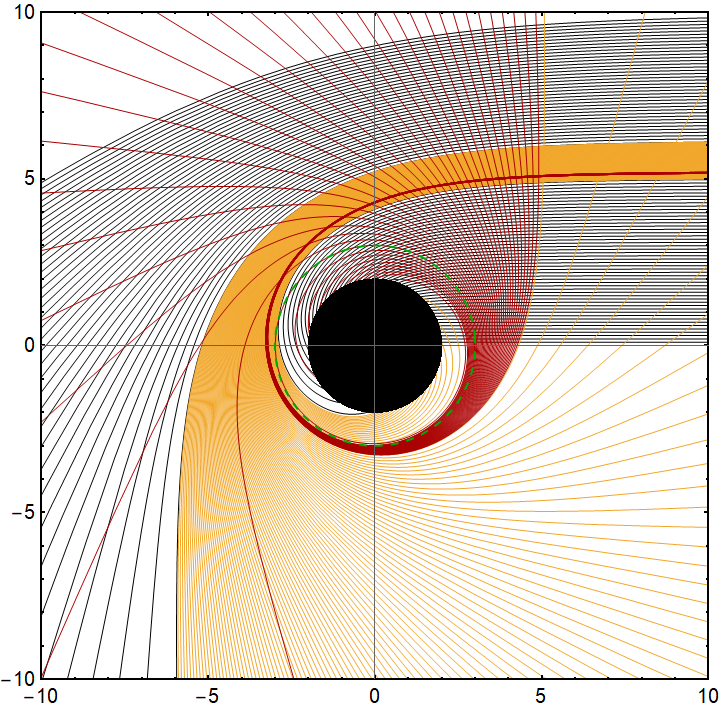}}\hspace{2mm}
\subfigure[\, $\alpha=5$]
{\includegraphics[width=4cm]{traject-a-5-lo-01-Q-01}}\hspace{2mm}
\subfigure[\, $\alpha=20$]
{\includegraphics[width=4cm]{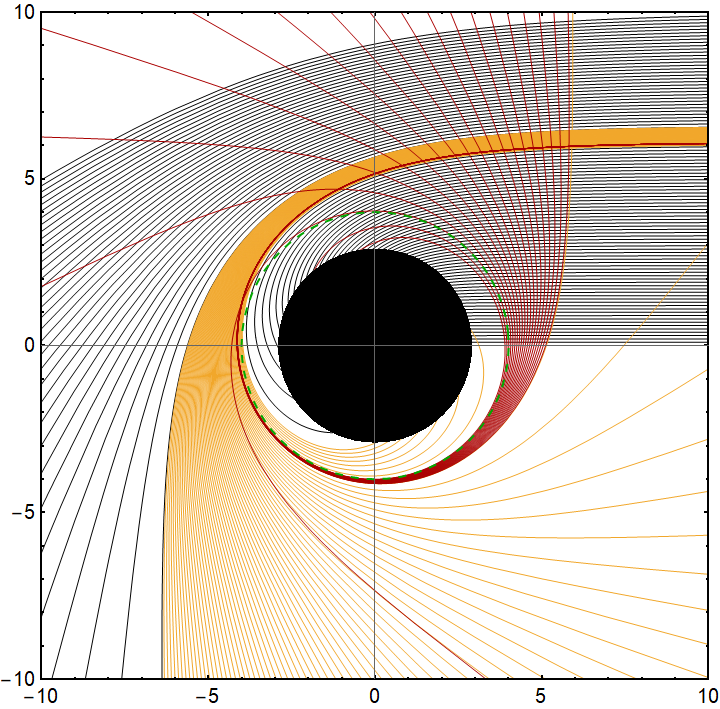}}\\
\caption{(a, e, i): the number of photon orbits $n$ as a function of the impact parameter $b$ for the selected values of $\alpha$, $Q$ and $l_o$. (b-d): the photon trajectories in the Euclidean polar coordinates ($r,\phi$) for different $Q$ with fixed $\alpha=5\,\&\,l_o=0.1$.
(f-h): the photon trajectories in the Euclidean polar coordinates ($r,\phi$) for different $l_o$ with fixed $\alpha=5\,\&\,Q=1$.
(j-l):  the photon trajectories in the Euclidean polar coordinates ($r,\phi$) for different $\alpha$ with fixed $Q=0.1\,\&\,l_o=0.1$. Here
the black, gold, and red curves correspond to the direct emissions ($n<3/4$), lensed ring emissions ($3/4<n<5/4$), and photon ring emissions ($n>5/4$), respectively. The black disk and the green dashed curves denote the {black hole} and photon sphere, respectively.}
\label{fig:orbits-traject}
\end{figure}

In order to better understand how the charges affect the photon trajectories and their classification, we use the ray-tracing method to solve the trajectory equation \eqref{trajectory-light-ray} with selected parameters of the hairy RN black hole. The total number $n$ {of orbits} as function of impact parameters $b$, and the photon trajectories in $(r,\phi)$ panel are shown in FIG. \ref{fig:orbits-traject}. We see from the plots (FIG. \ref{fig:orbits-traject}a, \ref{fig:orbits-traject}e, \ref{fig:orbits-traject}i) that in all cases, as the impact parameter $b$ increases, the total number of photon orbits first increases to be {an infinity} at $b=b_{ph}$, and then gradually decreases. Moreover, the parameters $(\alpha,Q,l_o)$ all affect the width of various emissions we defined previously. 

By further careful calculation, we determine the borders of $b$ for each emissions, and list the results in Table \ref{table01}-Table \ref{table03} in which we also show the results of black hole's event horizon $r_h$, photon sphere radius $r_{ph}$, and critical impact parameters $b_{ph}$. As both charge parameters $Q$ and $l_o$ increase,  $r_h$, $r_{ph}$, $b_{ph}$ all decrease, but the widths of the lensed ring and photon ring emissions both increase. However, the effects of $\alpha$ on all those quantities obey the opposite rules. Furthermore,  we mark the light rays from the direct, lensed ring, and photon ring emissions into black, gold, and red, respectively and figure out photon trajectories around the hairy RN black hole into FIG. \ref{fig:orbits-traject}, where the green dashed circle and black disk represent the photon sphere and the black hole, respectively. The effects of  $Q$, $l_o$ and $\alpha$ on all physical quantities listed in Table \ref{table01}-Table \ref{table03} are explicit if we compare the trajectories of photons in each row of FIG. \ref{fig:orbits-traject}.
In particular, larger both charges enlarge the width of lensed ring and photon ring structure of the black hole in observations. On the contrary, for larger $\alpha$, the lensed ring and photon ring emissions may be less likely to be observed. We expect to verify these phenomena in the optical appearance of the black hole in the next subsection.

\begin{table}[htbp]
{\centering
\begin{tabular}{|c|c|c|c|c|c|c|c|c|}
  \hline
  $\alpha$ & $Q$ & $l_o$& $r_h$& $r_{ph}$& $b_{ph}$&Direct emission&Lensed ring emission& Photon ring emission\\
  \hline
  5 &0.1 &0.1 &2.38260 &3.42178 &5.56182 &$b<5.45742$; $b>6.30779$ &$5.45742<b<5.55909$; $5.57624<b<6.30779$ &$5.55909<b<5.57624$\\
  \hline
  5 &1   &0.1 &2.05833 &2.96629 &4.88403 &$b<4.78411$; $b>5.60076$ &$4.78411<b<4.88124$; $4.89835<b<5.60076$ &$4.88124<b<4.89835$\\
  \hline
  5 &1.5 &0.1 &1.47998 &2.20043 &3.75803 &$b<3.64748$; $b>4.43593$ &$3.64748<b<3.75388$; $3.77501<b<4.43593$ &$3.75388<b<3.77501$\\
  \hline
\end{tabular}
\caption{The event horizon $r_h$, photon sphere radius $r_{ph}$ and the critical impact parameters $b_{ph}$  for different $Q$ with fixed $\alpha=5\,\&\,l_o=0.1$, along with the ranges of impact parameter $b$ corresponding to the direct, lensed ring, and photon ring emissions of the hairy RN black holes. Here we fix $M=1$.
\label{table01} }}
\end{table}
\begin{table}[htbp]
{\centering
\begin{tabular}{|c|c|c|c|c|c|c|c|c|}
  \hline
  $\alpha$ & $Q$ & $l_o$& $r_h$& $r_{ph}$& $b_{ph}$&Direct emission&Lensed ring emission& Photon ring emission\\
   \hline
  5 &1   &0.1 &2.05833 &2.96629 &4.88403 &$b<4.78411$; $b>5.60076$ &$4.78411<b<4.88124$; $4.89835<b<5.60076$ &$4.88124<b<4.89835$\\
  \hline
  5 &1   &1   &1.44737 &2.02571 &4.13132 &$b<3.90957$; $b>5.29458$ &$3.90957<b<4.11637$; $4.18386<b<5.29458$ &$4.11637<b<4.18386$\\
  \hline
  5 &1  &1.9  &1.00002 &2.00000 &4.00000 &$b<3.41818$; $b>5.28835$ &$3.41818<b<3.91948$; $4.09030<b<5.28835$ &$3.91948<b<4.09303$\\
  \hline
\end{tabular}
\caption{The event horizon $r_h$, photon sphere radius $r_{ph}$ and the critical impact parameters $b_{ph}$  for different $l_o$ with fixed $\alpha=5\,\&\,Q=1$, along with the ranges of impact parameter $b$ corresponding to the direct, lensed ring, and photon ring emissions of the hairy RN black holes. Here we fix $M=1$.
\label{table02} }}
\end{table}
\begin{table}[htbp]
{\centering
\begin{tabular}{|c|c|c|c|c|c|c|c|c|}
  \hline
  $\alpha$ & $Q$ & $l_o$& $r_h$& $r_{ph}$& $b_{ph}$&Direct emission&Lensed ring emission& Photon ring emission\\
  \hline
  0.1&0.1&0.1 &2.00651 &3.00572 &5.19796 &$b<5.01959$; $b>6.16345$ &$5.01959<b<5.18985$; $5.22913<b<6.16345$ &$5.18958<b<5.22913$\\
  \hline
  5 &0.1 &0.1 &2.38260 &3.42178 &5.56182 &$b<5.45742$; $b>6.30779$ &$5.45742<b<5.55909$; $5.57624<b<6.30779$ &$5.55909<b<5.57624$\\
  \hline
  20&0.1 &0.1 &2.89692 &4.00597 &6.09891 &$b<6.04478$; $b>6.61119$ &$6.04478<b<6.09818$; $6.10423<b<6.61119$ &$6.09818<b<6.10423$\\
  \hline
\end{tabular}
\caption{The event horizon $r_h$, photon sphere radius $r_{ph}$ and the critical impact parameters $b_{ph}$  for different $\alpha$ with fixed $Q=0.1\,\&\,l_o=0.1$, along with the ranges of impact parameter $b$ corresponding to the direct, lensed ring, and photon ring emissions of the hairy RN black holes. Here we fix $M=1$.
\label{table03} }}
\end{table}

\subsection{Observed intensities and optical appearances}

\begin{figure} [H]
{\centering
\subfigure[]
{\includegraphics[width=5.5cm]{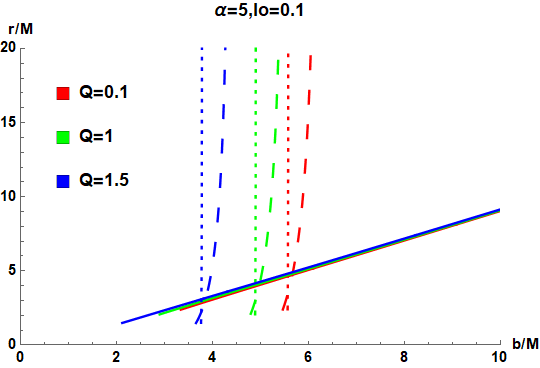}}\hspace{5mm}
\subfigure[]
{\includegraphics[width=5.5cm]{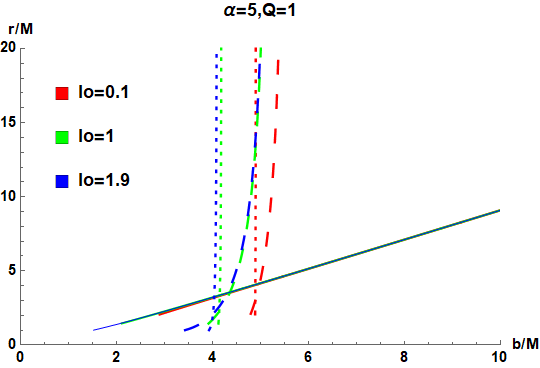}}\hspace{5mm}
\subfigure[]
{\includegraphics[width=5.5cm]{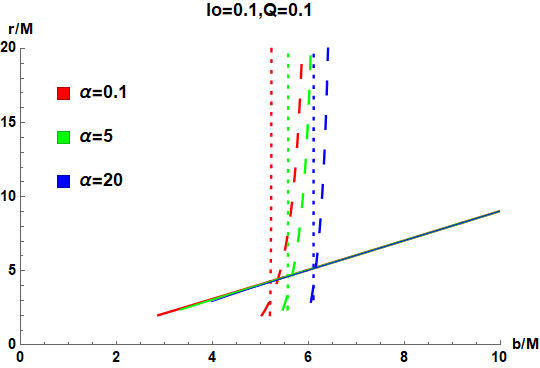}}
   \caption{The first three transfer functions of the hairy RN black holes. In each plot, $r_1(b)$,
$r_2(b)$ and $r_3(b)$ are denoted by  the solid, dashed, and dotted curves, respectively.  From left to right, we check the effects of $Q$, $l_o$ and $\alpha$, respectively, with other two parameters fixed. Here we have set $M=1$.}   \label{fig:transfer}}
\end{figure}

The light rays will extract energy when they intersect with the accretion disk each time, which will directly affect the contribution of the direct, lensed ring, and photon ring emissions to the observed brightness. Since the parameters $Q, l_o$ and $\alpha$ have obvious influences on the emission widths of the three types. Therefore, it is interesting to examine their effects on the optical appearance of the hairy RN black hole. Considering that the thin accretion disk emits isotropically in the rest frame of static worldlines, the specific intensity received by the observer with emission frequency $\nu_e$ is
\begin{eqnarray}
I_o(r,\nu_o)=g^3I_e(r,\nu_e),
\end{eqnarray}
where $g=\nu_o/\nu_e=\sqrt{f(r)}$ is the redshift factor, $\nu_o$ and $I_e(r,\nu_e)$ are the frequency of the observed light and the specific intensity of the accretion disk, respectively \cite{Bromley:1996wb}. The total observed intensity $I_{obs}(r)$ is then evaluated by integrating over all the observed frequencies {of} $I_o(r,\nu_o)$ as
\begin{eqnarray}
I_{obs}(r)=\int I_o(r,\nu_o)d\nu_o=\int g^4I_e(r,\nu_e)d\nu_e=f(r)^2I_{em}(r),
\end{eqnarray}
where we denote $I_{em}(r)=\int I_e(r,\nu_e)d\nu_e$ as the total emitted intensity.  Therefore, since the reverse light ray from the observer may intersect the accretion disk multiple times, depending on the type of emission, so the total intensity received by the observer is the sum of the intensities from each intersection \cite{Gralla:2019xty}
\begin{eqnarray}
I_{obs}(b)=\sum_m f(r)^2I_{em}(r)\mid_{r=r_m(b)},
\label{eq:observed}
\end{eqnarray}
where $r_m(b)$ is known as the transfer function, which represents the radial coordinate of the $m-$th intersection of the light ray with impact parameter $b$ and the accretion disk.
Thus, the slope $dr/db$ describes the demagnification factor, and  the case with larger $m$ corresponds to strong  demagnification so as to contribute much less to the total luminosity.
Moreover, as we  illustrated in \cite{Wang:2023vcv}, the first transfer function corresponds to the direct image
originating from direct, lensed and photon ring emission; the second transfer function can origin from lensed ring and photon ring emission; while the third transfer function can
only origin from photon ring emission.  We show the first three transfer functions for the selected parameters in  FIG. \ref{fig:transfer}, where the solid, dashed, and dotted curves denote $r_1(b)$, 
$r_2(b)$ and $r_3(b)$, respectively. We see that as the charge parameters $Q$ or $l_o$ increases (FIG. \ref{fig:transfer}a and FIG. \ref{fig:transfer}b), the three transfer functions areas become wider and move toward the smaller $b$, while the parameter $\alpha$ gives opposite influence (FIG. \ref{fig:transfer}c). However, there are still some common properties in all cases: (i) the slope of $r_1(b)$ is almost 1, because the direct image corresponds to the source profile after redshift. (ii)  $r_2(b)$ and $r_3(b)$ are highly steeper comparing to $r_1(b)$, and  $r_3(b)$ is steeper than $r_2(b)$. This feature means that $r_1(b)$ will dominate in the total luminosity and other transfer functions contribute very little. 

We proceed to  investigate the optical appearance of the hairy RN black hole by considering a specific toy model emission function of the accretion disk, which gradually decays from the event horizon $r_h$ in the form \cite{Wang:2022yvi,Yang:2022btw}
\begin{align}
    \begin{split}
        I_{em}(r)=\left \{
        \begin{array}{ll}
            I_o\frac{\frac{\pi}{2}-\arctan(r-r_{isco}+1)}{\frac{\pi}{2}-\arctan(r_{h}-r_{isco}+1)}                 ~~~~~~~~~~~~& r>r_{h}\\\\
            0,                                                                                        ~~~~~~~~~~~~& r\leq r_{h}
        \end{array}
        \right.
    \end{split}.
    \label{obser3}
\end{align}
Here, $I_o$ is the maximum intensity and the innermost stable circular orbit $r_{isco}$  can be computed by \cite{Wang:2023vcv}
\begin{eqnarray}
r_{isco}=\frac{3f(r_{isco})f'(r_{isco})}{2f'(r_{isco})^2-f(r_{isco})f''(r_{isco})},
\end{eqnarray}
in which $f$ is the metric function of the hairy RN black hole, and  the prime represents the derivative  with respect to the radial coordinate $r$.
\begin{figure}[h]
\centering
\subfigure[\, fixing $\alpha=5\,\&\, l_o=0.1$ and changing $Q$]
{\includegraphics[width=4.5cm]{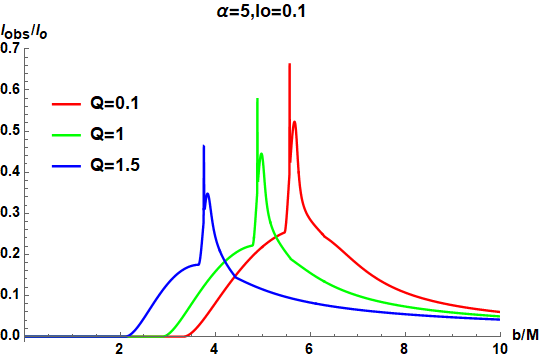} \label{}}\hspace{2mm}
\subfigure[\, $Q=0.1$]
{\includegraphics[width=4cm]{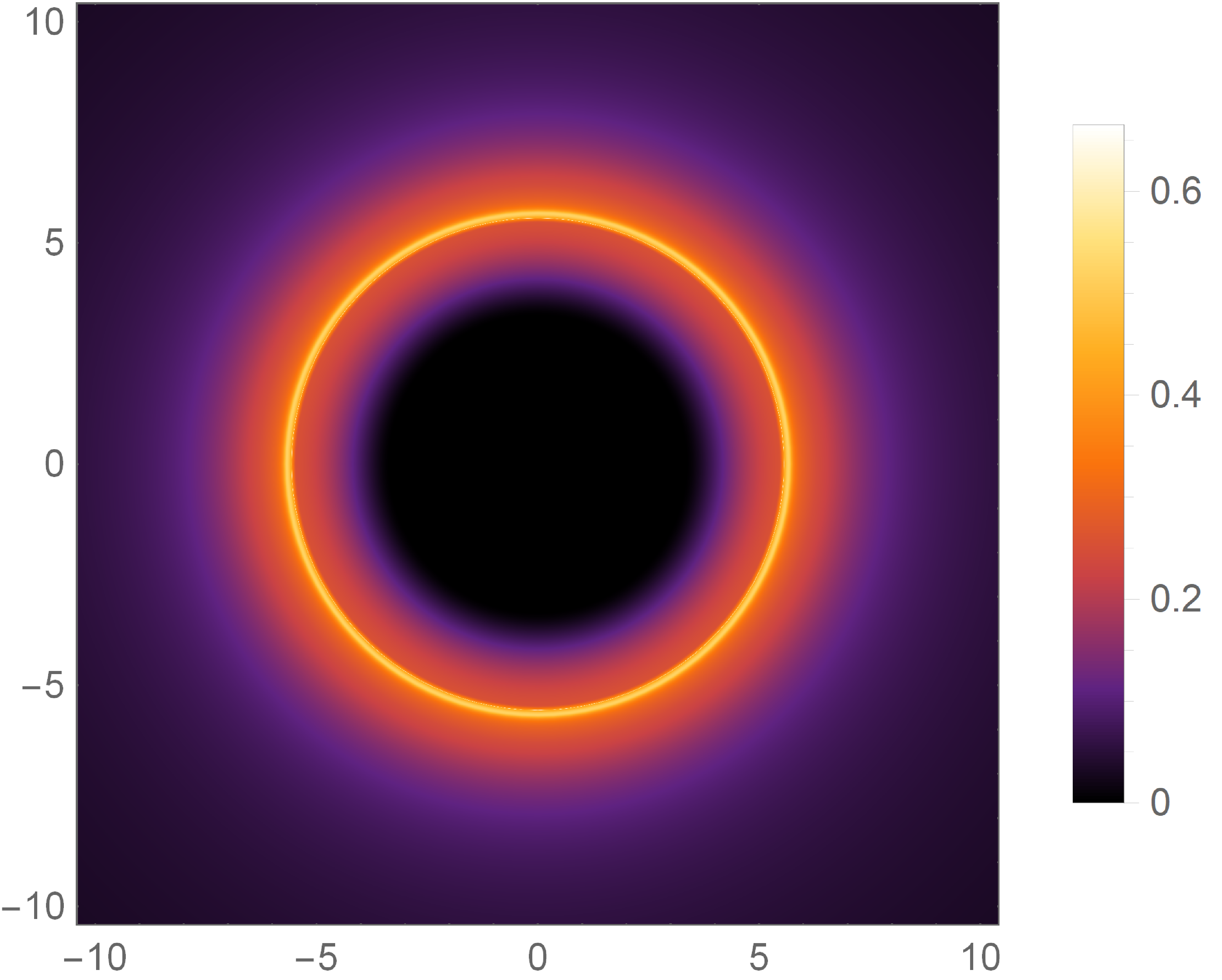}}\hspace{2mm}
\subfigure[\, $Q=1$]
{\includegraphics[width=4cm]{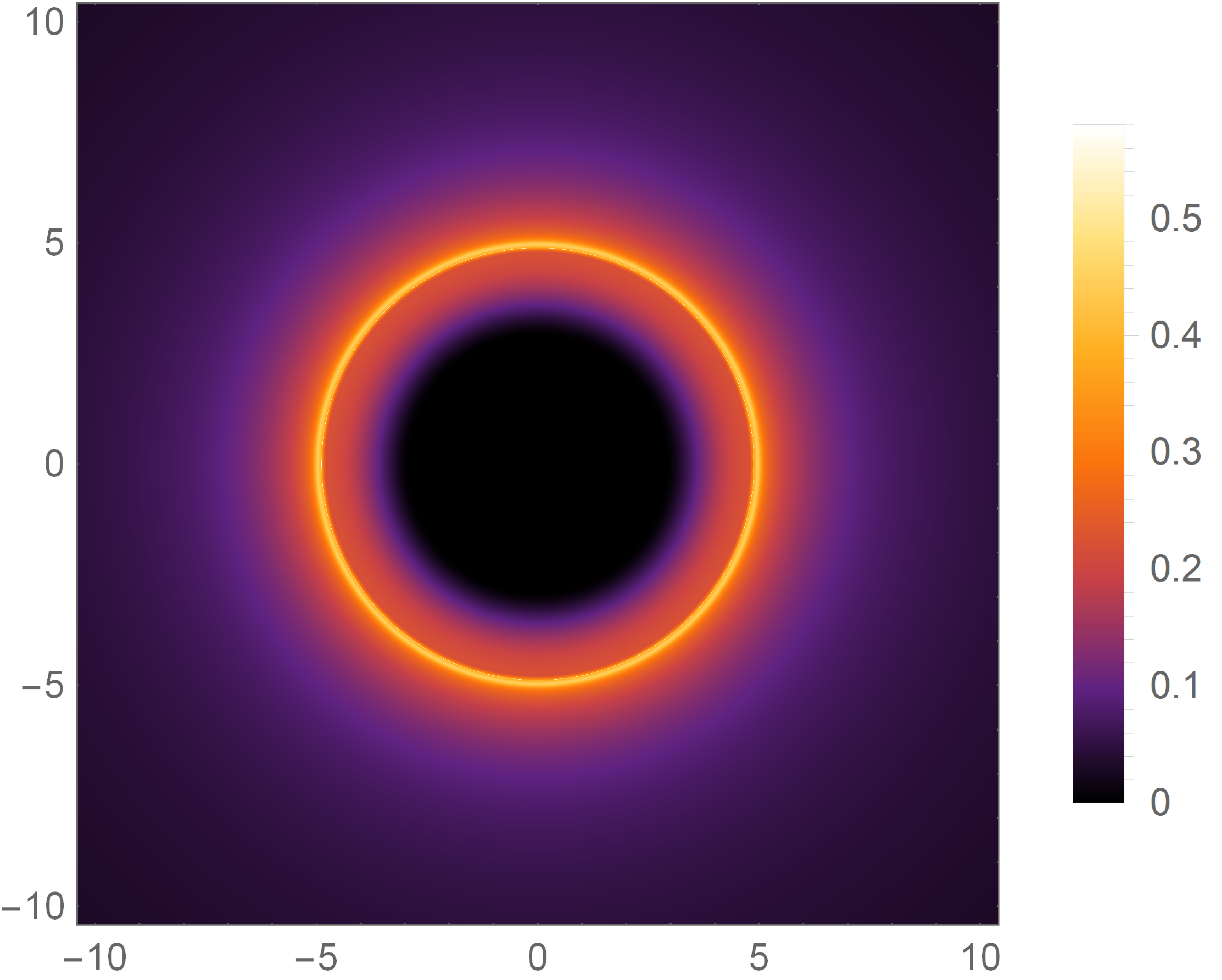}}\hspace{2mm}
\subfigure[\, $Q=1.5$]
{\includegraphics[width=4cm]{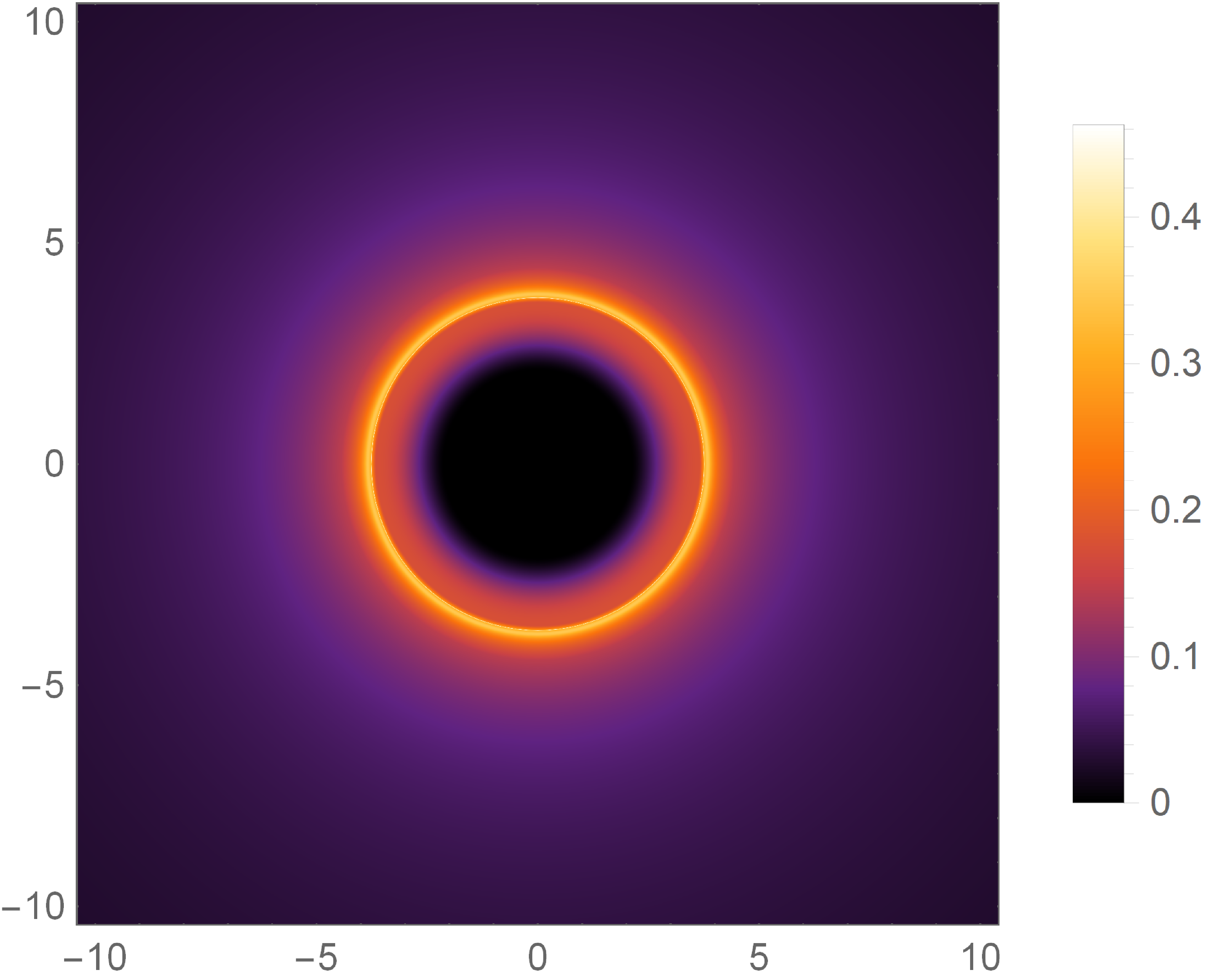}}\\
\subfigure[\, fixing $\alpha=5\,\&\,Q=0.1$ and changing $l_o$]
{\includegraphics[width=4.5cm]{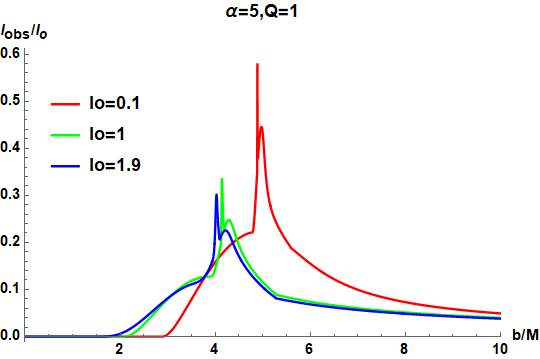} \label{}}\hspace{2mm}
\subfigure[\, $l_o=0.1$]
{\includegraphics[width=4cm]{image-a-5-lo-01-Q-1}}\hspace{2mm}
\subfigure[\, $l_o=1$]
{\includegraphics[width=4cm]{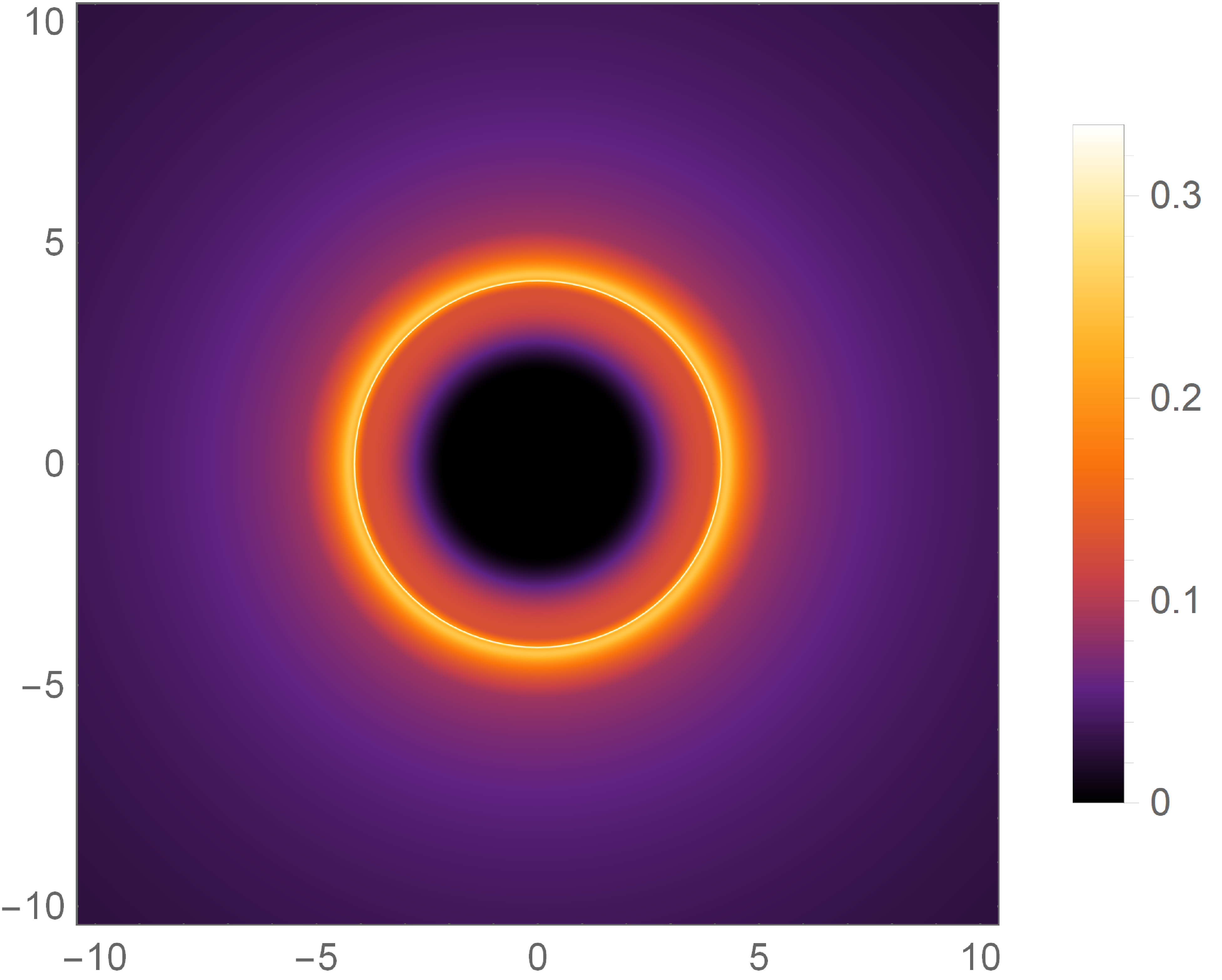}}\hspace{2mm}
\subfigure[\, $l_o=1.9$]
{\includegraphics[width=4cm]{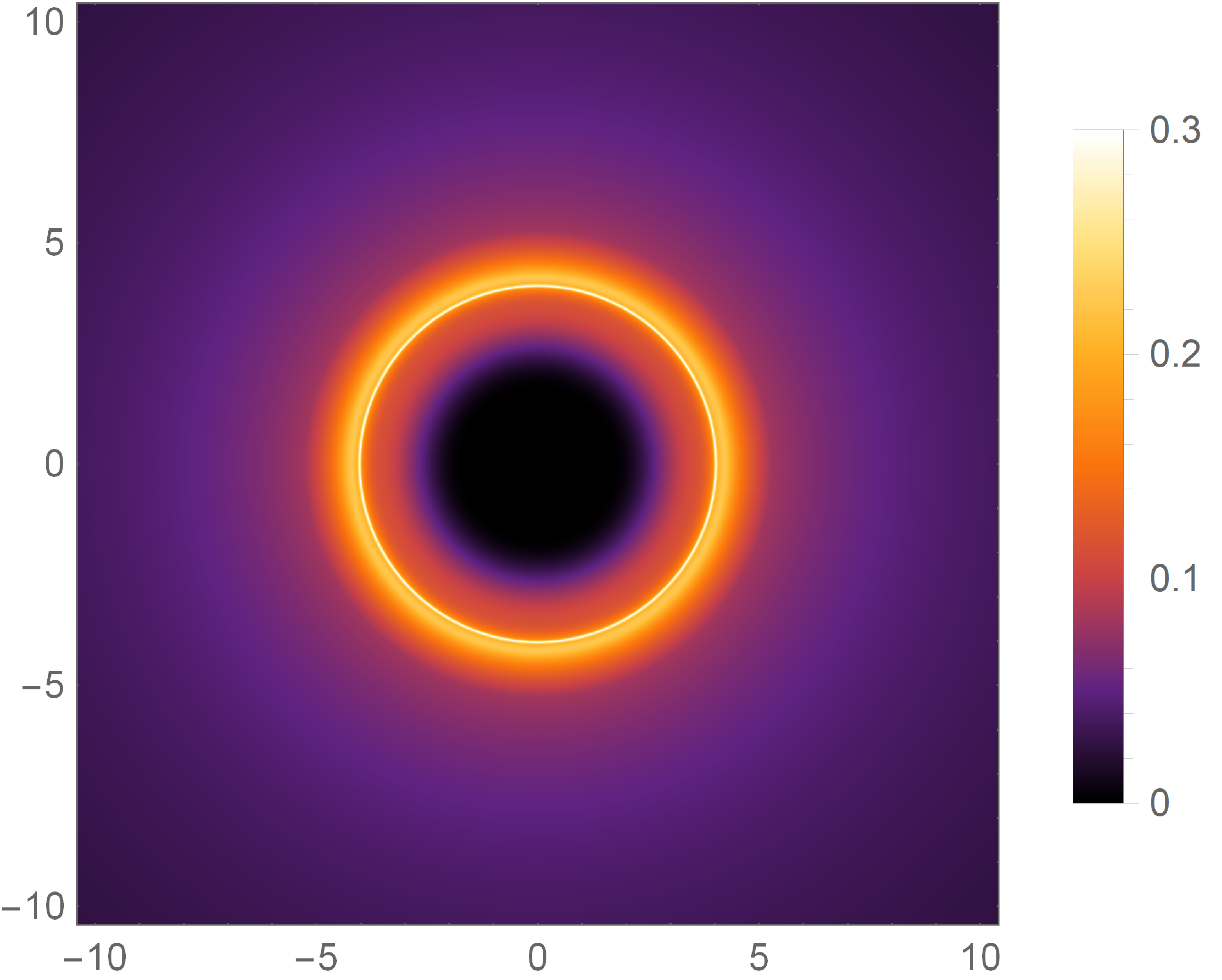}}\\
\subfigure[\, fixing $Q=0.1\,\&\,l_o=0.1$ and changing $\alpha$]
{\includegraphics[width=4.5cm]{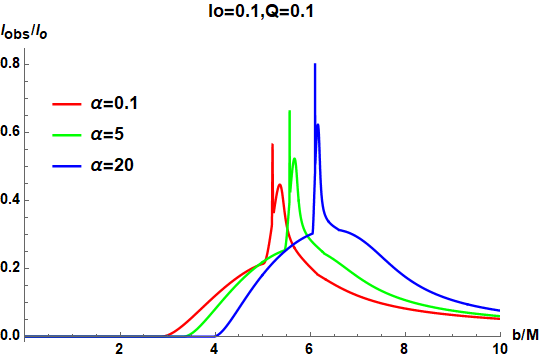} \label{}}\hspace{2mm}
\subfigure[\, $\alpha=0.1$]
{\includegraphics[width=4cm]{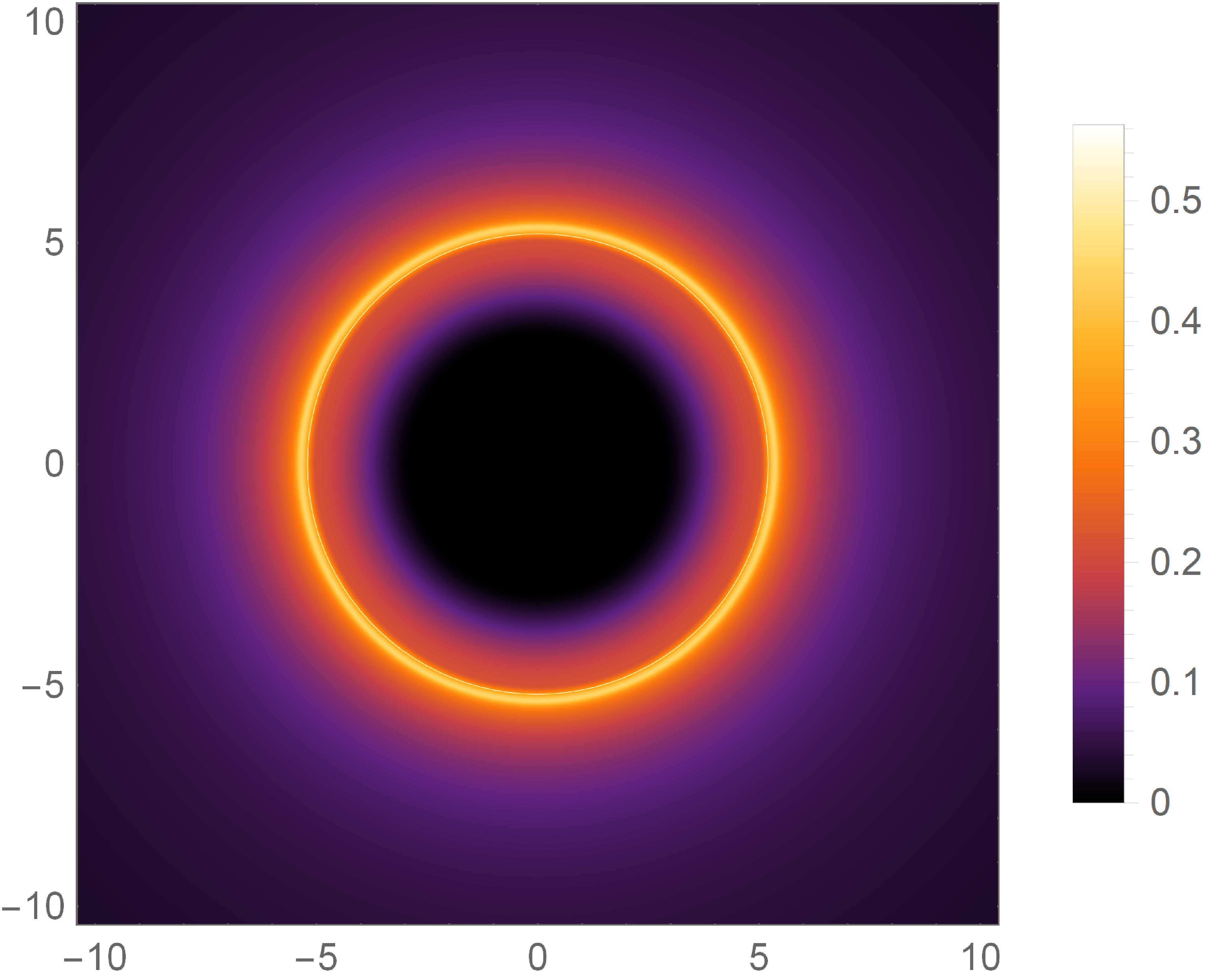}}\hspace{2mm}
\subfigure[\, $\alpha=5$]
{\includegraphics[width=4cm]{image-a-5-lo-01-Q-01}}\hspace{2mm}
\subfigure[\, $\alpha=20$]
{\includegraphics[width=4cm]{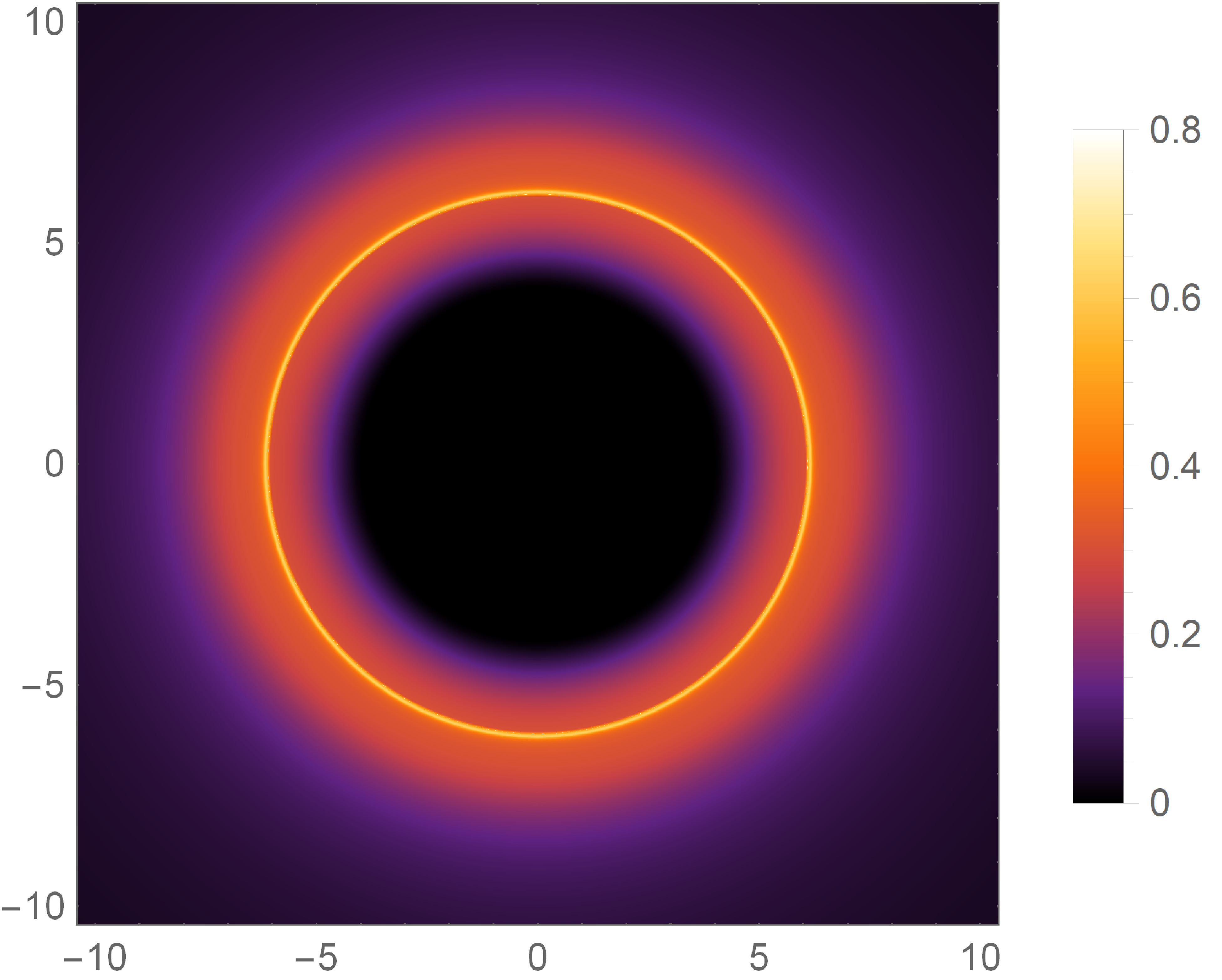}}\\
\caption{The total observed intensities \eqref{eq:observed} radiated from the emission function \eqref{obser3} of the accretion disk, as a function of impact parameter, and the optical appearances of the hairy RN black hole for selected parameters. From upper to bottom rows, we respectively check the effects of $Q$, $l_o$ and $\alpha$, with other two parameters fixed. }
\label{Iemit-observed}
\end{figure}

Having both the transfer and emission functions of the accretion disk in hands, we can evaluate the observed intensities via \eqref{eq:observed} and figure out the black hole images. 
Our results for the selected parameters are shown in FIG. \ref{Iemit-observed}.
In each row, we first show the total observed intensities as a function of impact parameter for different parameters, and then project them into a two dimensional plane to explicitly present the optical appearance of the accretion disk.  In all cases, we can see two obvious peaks in the total observed intensities, corresponding to two rings in the images. Moreover,
for larger  $Q$ ($l_o$) with fixed $l_o$ ($Q$) and $\alpha$, the two peaks of $I_{obs}$ are both smaller and shift to smaller $b$. On the contrary, for larger $\alpha$ with fixed $Q$ and $l_o$, the peaks are enhanced and occur at larger $b$. In addition, it is obvious from the images  that larger $Q$ ( $l_o$) corresponds to smaller shadow region while larger deviation parameter $\alpha$ gives larger shadow region as we illustrated in section \ref{sec:tragectories and photons}.
Consequently, our results imply that one may not be able to tell which kind of charge the hairy RN black hole carry  through the rings and images of the accretion disk, since the two charge parameters, $Q$ and  $l_o$, have similar influence on the observed intensities of the black hole {surrounded} by the accretion disk.
 Moreover, there are potential degeneracies in the hairy RN black hole images because $\alpha$ and $(Q, l_o)$ may counteract their effects. This may also result in the optical appearance of the hairy RN black hole  being undistinguished from that of the RN black hole.

\section{Images of hairy RN black hole illuminated by static spherical accretions} \label{sec:spherical accretions}
In this section, we will investigate the images of hairy RN black holes surrounded by a static spherical accretion. 
When the materials in the Universe are trapped by a black hole, the disk-shaped accretion flow  usually forms around the black hole and rotates with a large angular momentum, but the matter will flow radially to the black hole and form a spherically symmetric accretion when the angular momentum is extremely small \cite{Yuan:2014gma}.  For an optically and geometrically thin static accretion with spherical symmetry, 
the observed specific intensity $I(\nu_o)$ detected by an observer at infinity $r=\infty$ (measured in erg s$^{-1}$ cm$^{-2}$ str$^{-1}$ Hz$^{-1}$) is evaluated by integrating the specific emissivity along the photon path $\gamma$ \cite{Bambi:2013nla}
\begin{eqnarray}
I(\nu_o)=\int_\gamma g^3j_e(\nu_e)dl_{prop}.
\label{specific-intensity1}
\end{eqnarray}
Here, $\nu_e$ and $\nu_o$ are the emitted photon frequency and the observed photon frequency, respectively. $g=\nu_o/\nu_e=f(r)^{1/2}$ is the redshift factor. $j(\nu_e)$ is the emissivity per unit volume in the rest frame and usually taken the form $j_e(\nu_e)\propto \delta(\nu_r-\nu_e)/r^2$ with $\nu_r$ the emitter's rest-frame frequency \cite{Bambi:2013nla}. And $dl_{prop}$ is the infinitesimal proper length given by
\begin{eqnarray}
dl_{prop}=\sqrt{\frac{1}{f(r)}dr^2+r^2d\phi^2}=\sqrt{\frac{1}{f(r)}+r^2\big(\frac{d\phi}{dr}\big)^2}dr,
\label{Eq:intensity-1}
\end{eqnarray}
where the $d\phi/dr$ is expressed in \eqref{trajectory-light-ray}. Subsequently, by further integrating \eqref{specific-intensity1} over all the observed frequencies, we obtain the total observed intensity as
\begin{eqnarray}
I_{obs}=\int_{\nu o}I(\nu_o)d\nu_o=\int_{\nu_e}\int_{\gamma}g^4j_e(\nu_e)dl_{prop}d\nu_e=\int_\gamma\frac{f(r)^2}{r^2}\sqrt{\frac{1}{f(r)}+r^2\Big(\frac{d\phi}{dr}\Big)^2}dr.
\label{observed-intensity}
\end{eqnarray}

\begin{figure}[h]
\centering
\subfigure[\, fixing $\alpha=5\,\&\,l_o=0.1$ and changing $Q$]
{\includegraphics[width=4.5cm]{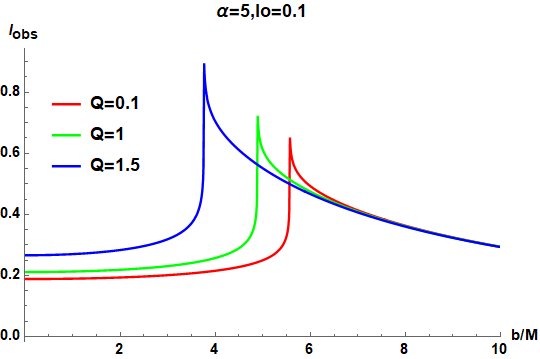} \label{}}\hspace{2mm}
\subfigure[\, $Q=0.1$]
{\includegraphics[width=4cm]{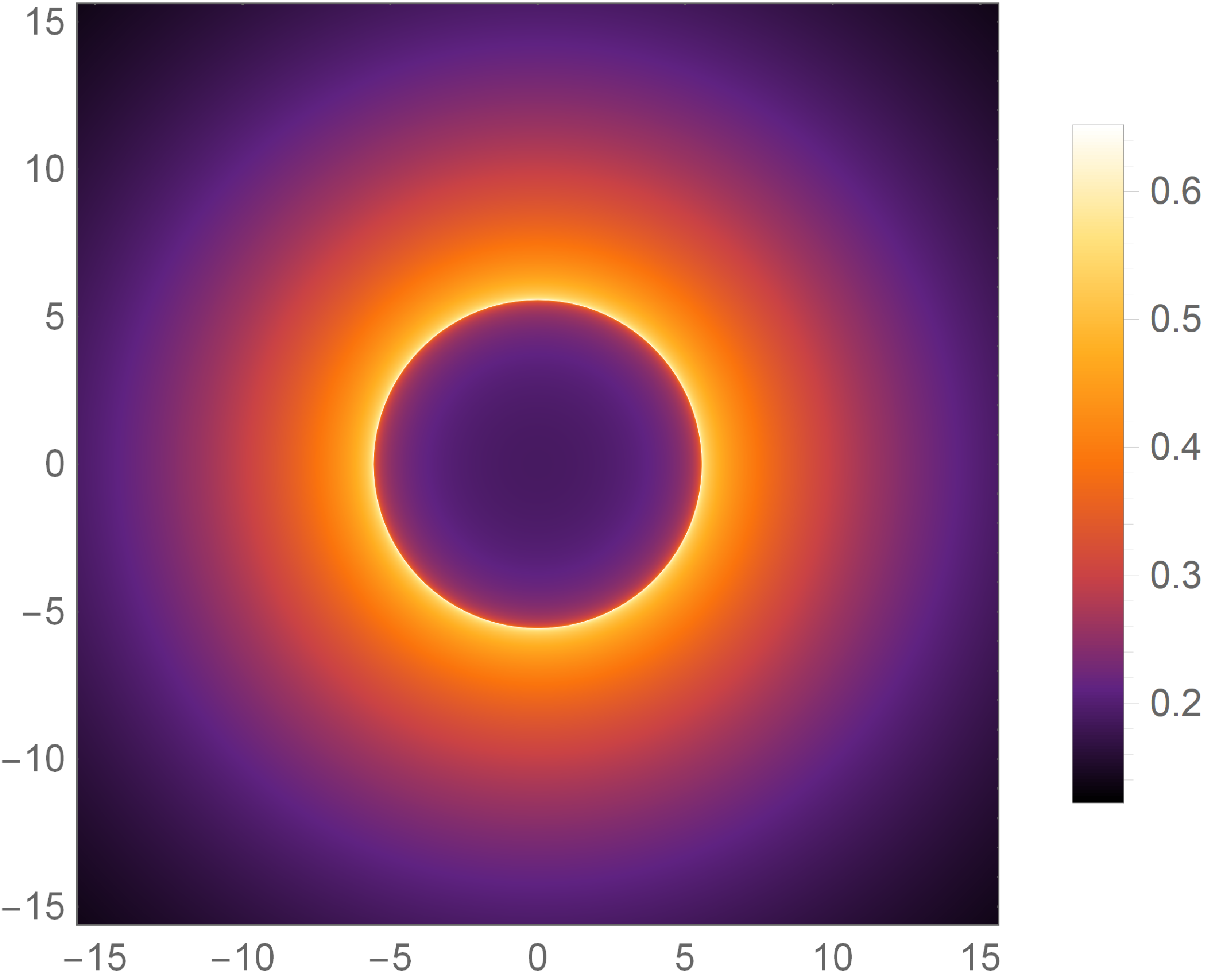}}\hspace{2mm}
\subfigure[\, $Q=1$]
{\includegraphics[width=4cm]{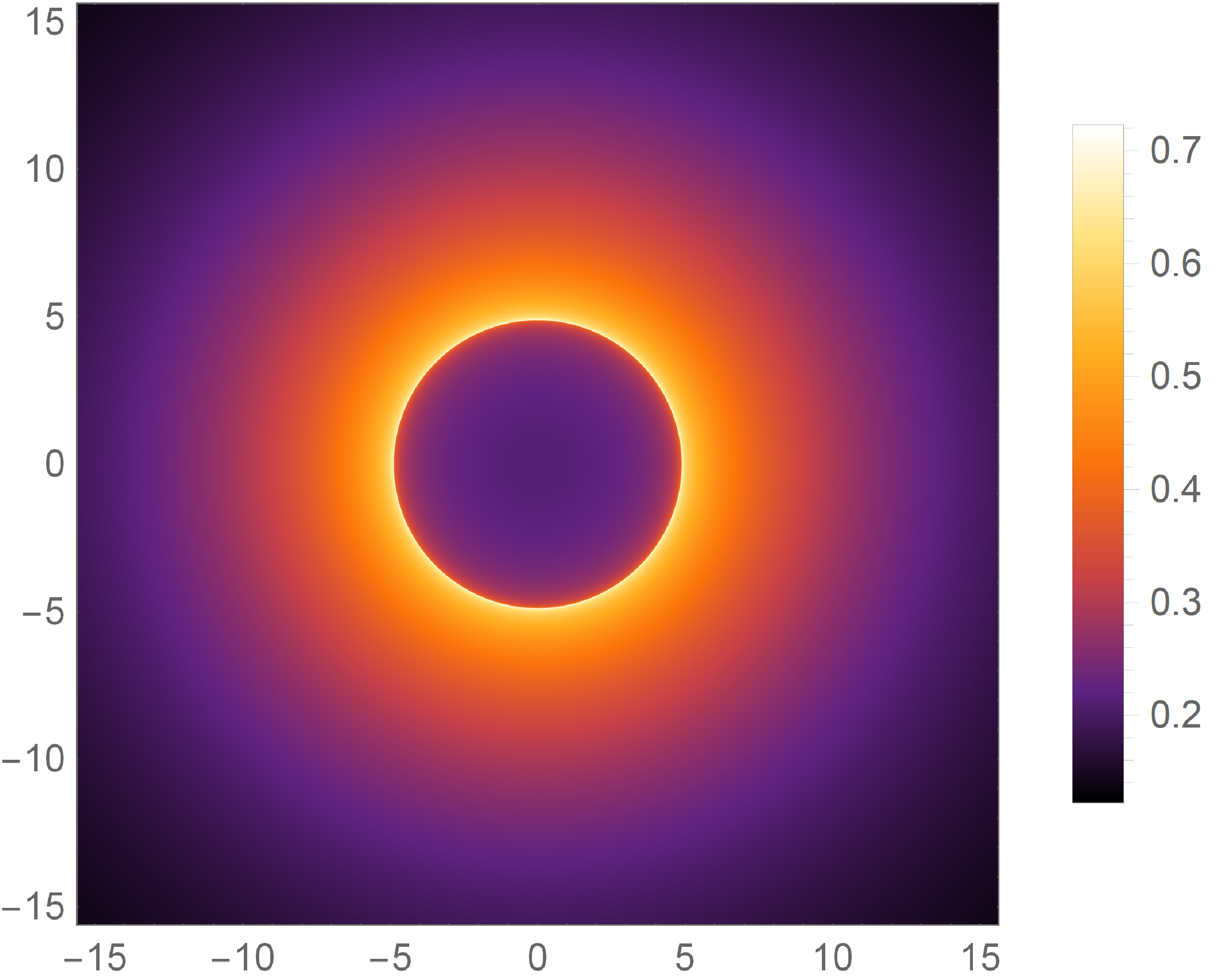}}\hspace{2mm}
\subfigure[\, $Q=1.5$]
{\includegraphics[width=4cm]{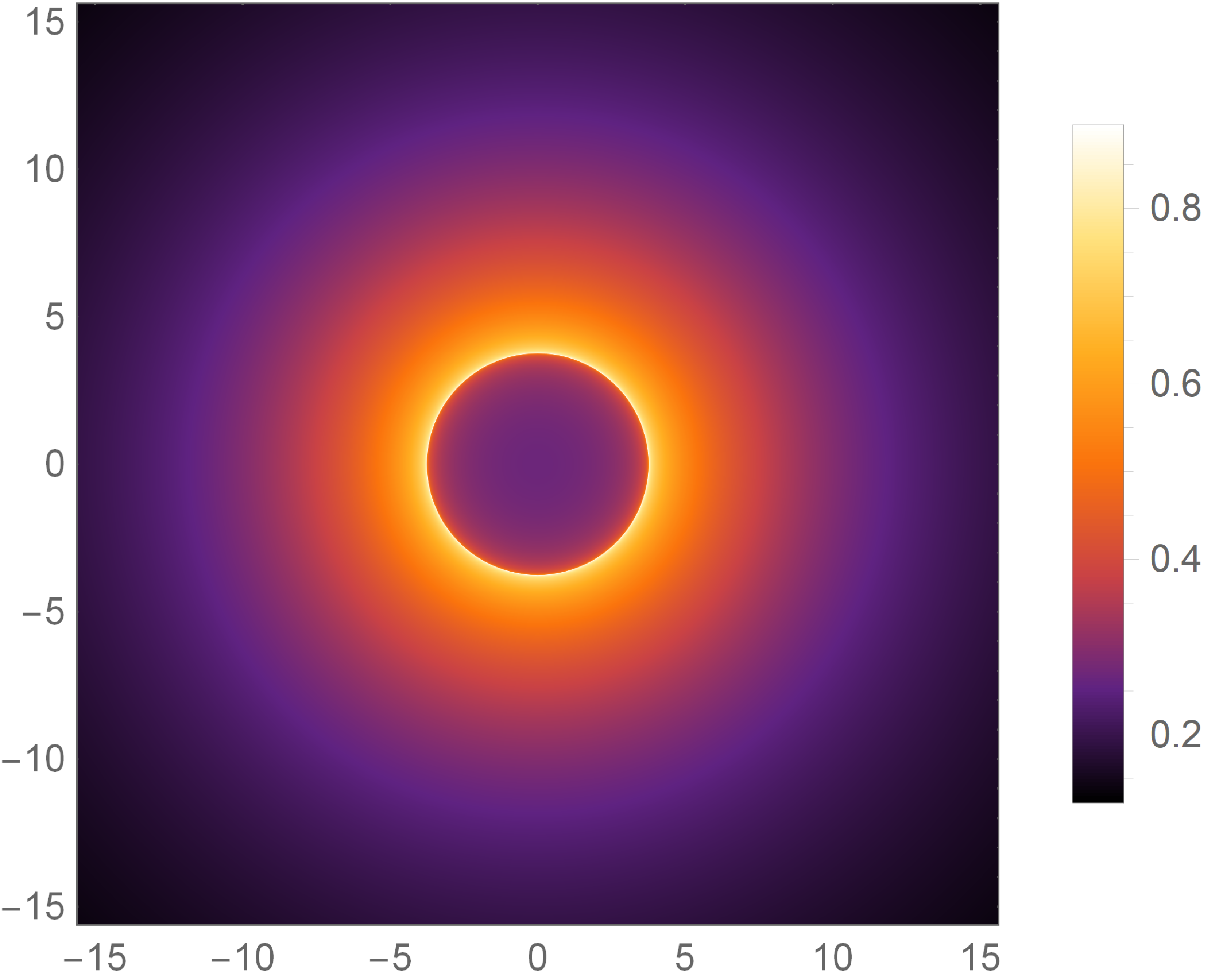}}\\
\subfigure[\, fixing $\alpha=5\,\&\,Q=1$ and changing $l_o$]
{\includegraphics[width=4.5cm]{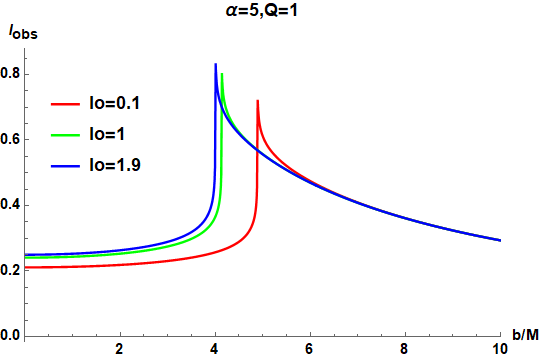} \label{}}\hspace{2mm}
\subfigure[\, $l_o=0.1$]
{\includegraphics[width=4cm]{QX-a-5-lo-01-Q-1}}\hspace{2mm}
\subfigure[\, $l_o=1$]
{\includegraphics[width=4cm]{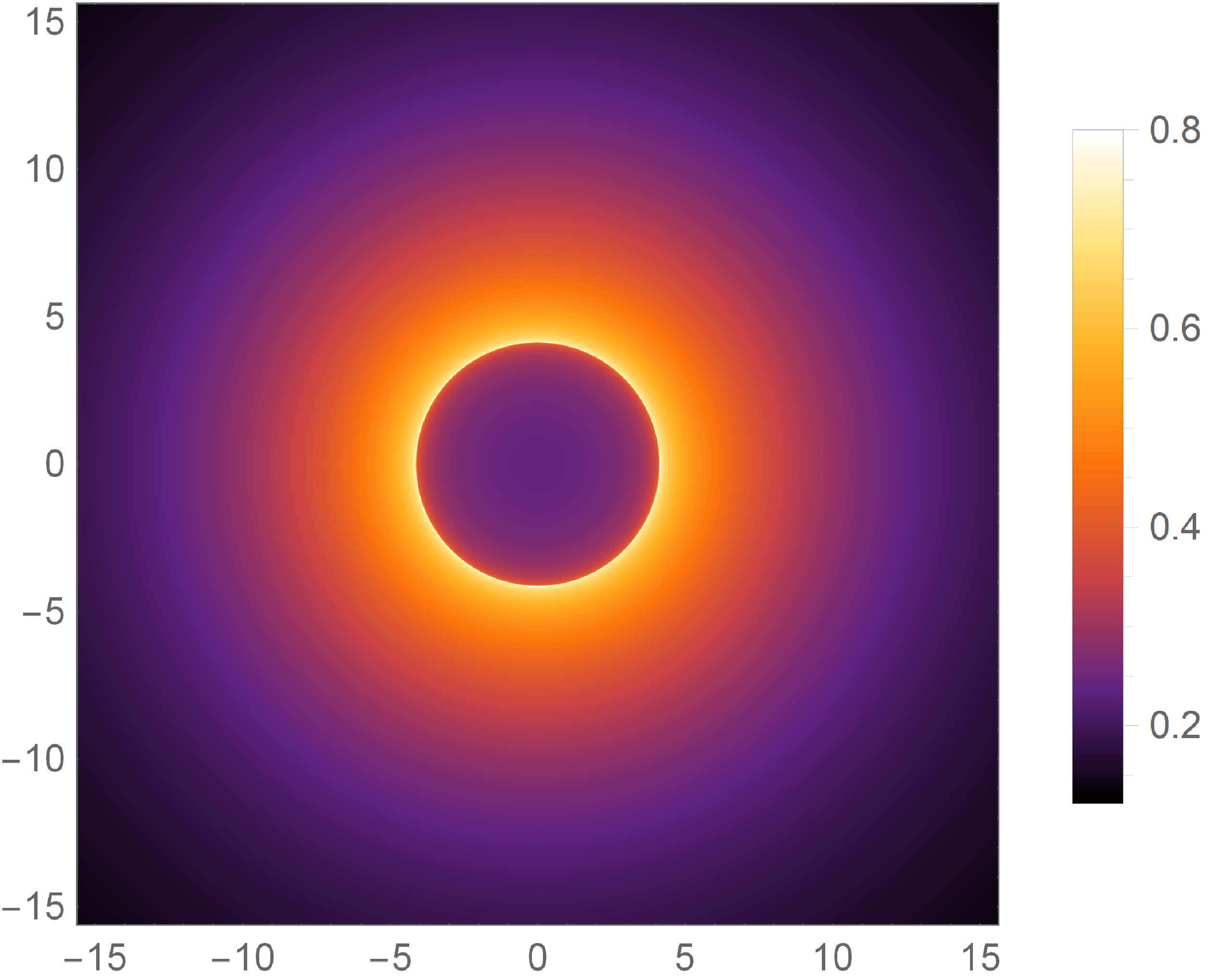}}\hspace{2mm}
\subfigure[\, $l_o=1.9$]
{\includegraphics[width=4cm]{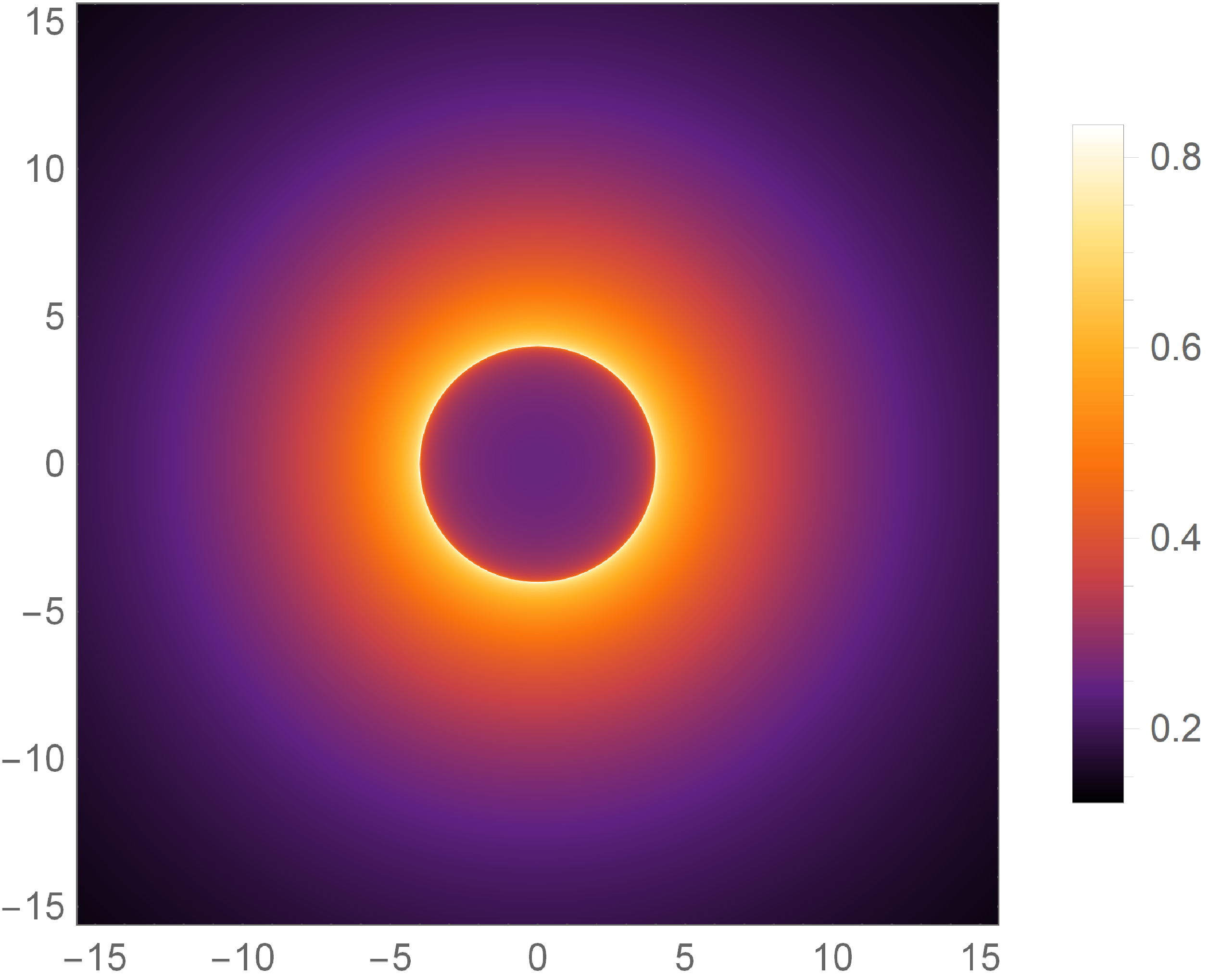}}\\
\subfigure[\, fixing $Q=0.1\,\&\,l_o=0.1$ and changing $\alpha$]
{\includegraphics[width=4.5cm]{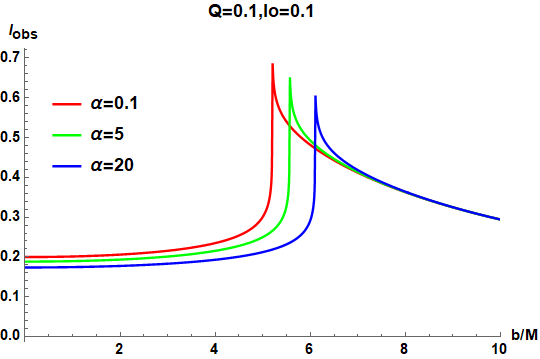} \label{}}\hspace{2mm}
\subfigure[\, $\alpha=0.1$]
{\includegraphics[width=4cm]{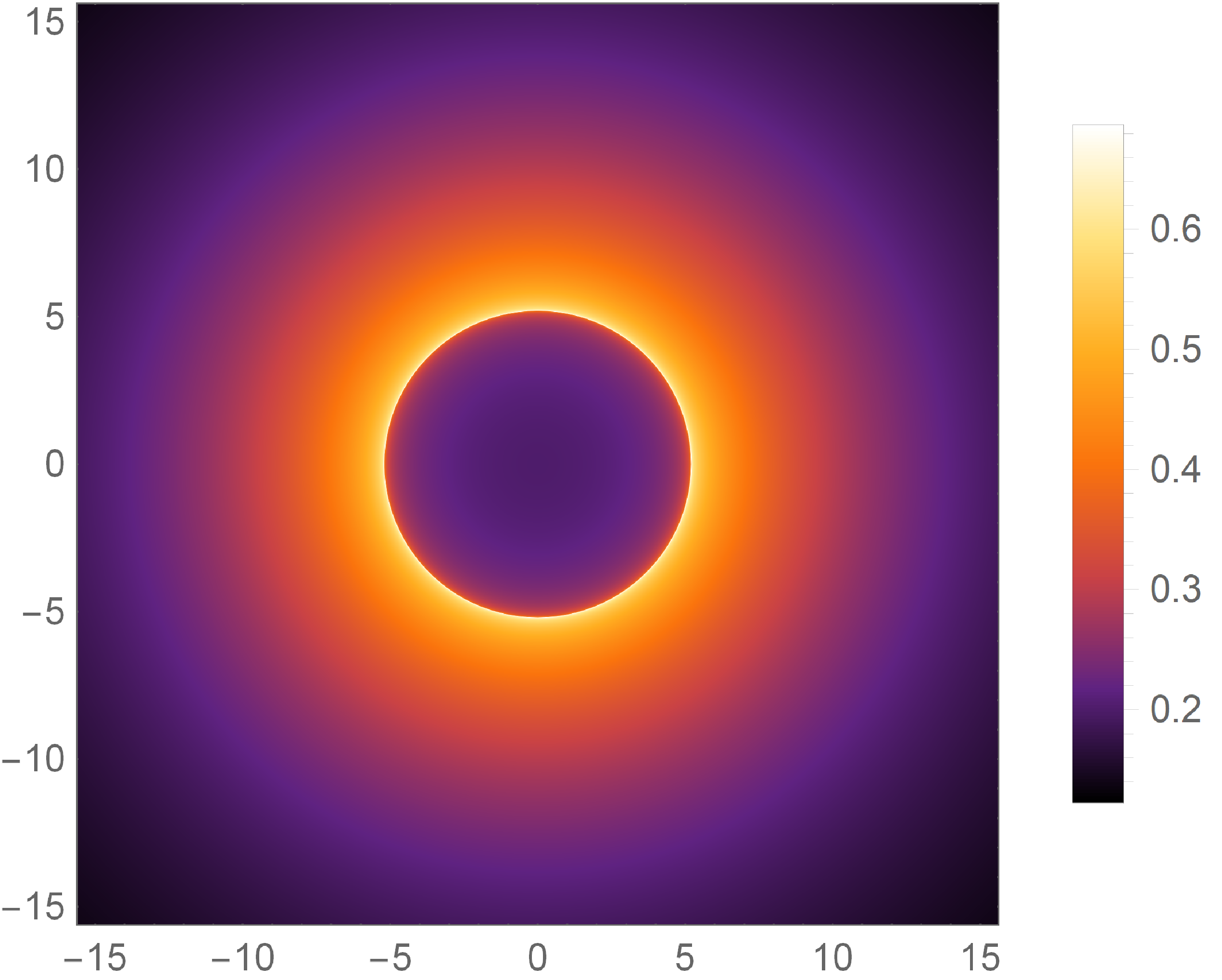}}\hspace{2mm}
\subfigure[\, $\alpha=5$]
{\includegraphics[width=4cm]{QX-a-5-lo-01-Q-01}}\hspace{2mm}
\subfigure[\, $\alpha=20$]
{\includegraphics[width=4cm]{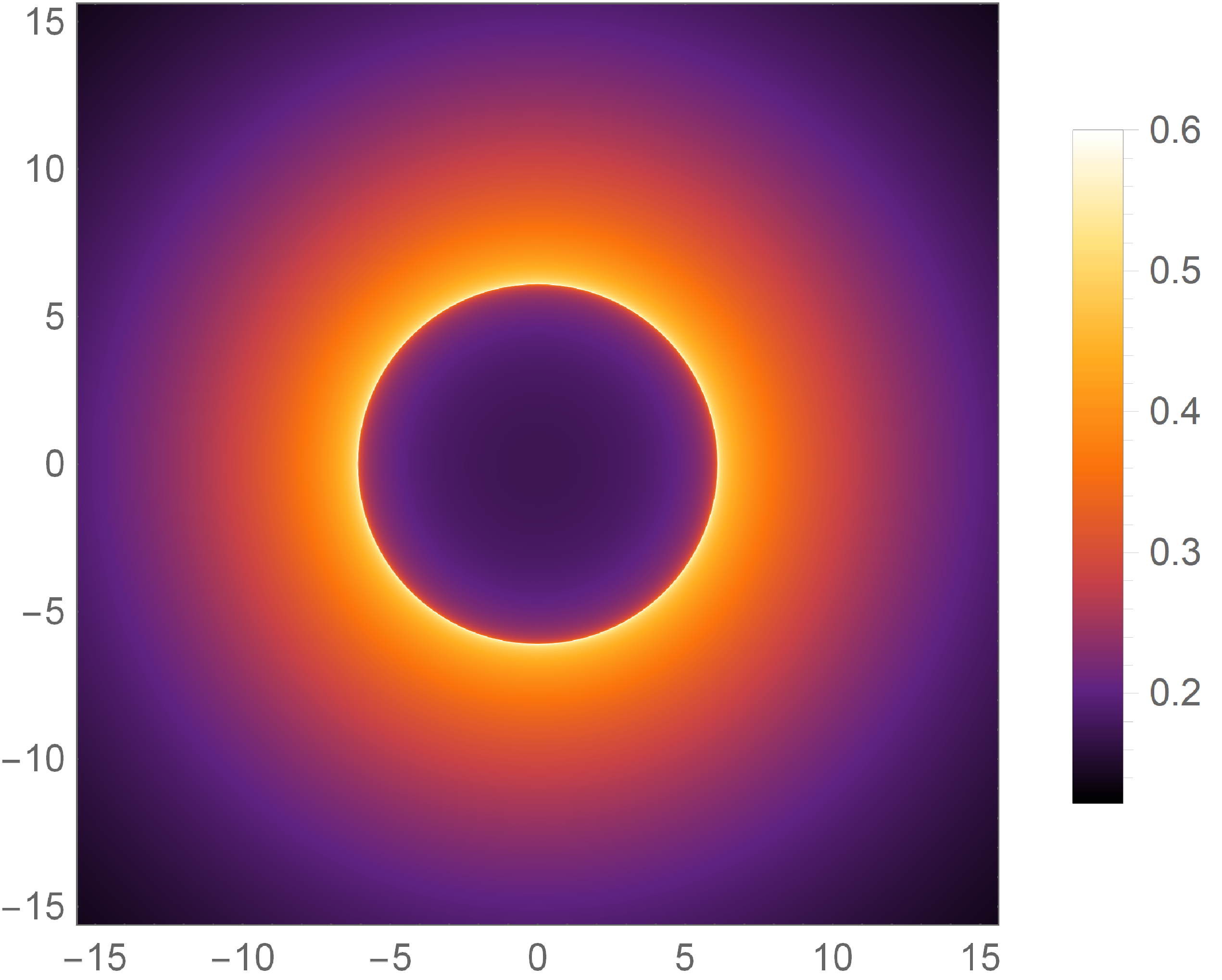}}\\
\caption{The total observed intensities \eqref{observed-intensity} radiated from the static spherical accretion as a function of impact parameter, and the optical appearances of the hairy RN black hole for selected parameters. From upper to bottom rows, we respectively check the effects of $Q$, $l_o$ and $\alpha$, with other two parameters fixed.}
\label{fig:QX-images}
\end{figure}

Then we calculate the total observed intensities, and depict them into a two dimensional plane as the images of hairy RN black hole with the same selected parameters as previous sections. The results are presented in FIG. \ref{fig:QX-images}. For the total observed intensities in each case, there exists a peak which corresponds to the bright ring in the images of the black hole. As the increasing of both $Q$ and $l_o$ (FIG.\ref{fig:QX-images}a and \ref{fig:QX-images}e), the peak is enhanced and shifts to smaller impact parameters, indicating that the ring becomes brighter but the faint illuminating region in the center is smaller for larger $Q$ (Fig. \ref{fig:QX-images}b-\ref{fig:QX-images}d) or $l_o$ (Fig. \ref{fig:QX-images}f-\ref{fig:QX-images}h). In addition, FIG. \ref{fig:QX-images}i-\ref{fig:QX-images}l show that the effect of increasing of $\alpha$ obeys a completely different rule from that of the charges. Namely, increasing $\alpha$ suppresses the peak but enlarges the corresponding impact parameters, thus, we see larger faint illuminating region in the images of black hole with larger $\alpha$.
Thus, comparing to the RN black hole, the introduction of additional field can enlarge/suppress or keep the brightness as well as the size of photon ring, depending on the competition between the parameters $l_o$ and $\alpha$. This implies the possible degeneracies in the optical appearances of the hairy RN black hole and RN black hole surrounded by the static spherical accretion.

\section{Conclusion and discussion}\label{sec:conclusion}

The black hole images released by the EHT collaborators open a new window for testing gravity in strong field regime and provide new inspiration for testing possible hairs around the black hole. In this paper, taking the hairy RN black hole constructed by GD proposal \cite{Ovalle:2020kpd}, which could contain electric charge and additional arbitrary hairy charge, as
the central object, we investigated its optical appearance illuminated by some toy accretion models to discuss the possible probes of the two charges. 

As the preparation, we first analyzed the effects of electric charge, additional hairy charge and the deviation parameter on the photon sphere and critical impact parameters. We found that  as both charges increase, the photon sphere and critical impact parameters of the hairy RN black hole  both decrease, but the deviation parameter enlarges them. Then, using the back ray-tracing method, we studied the photon trajectories around the hairy RN black hole and found that the charges have a significant influence on the distributions and classification of light rays.
In particular, for the hairy RN black hole with larger electric charge or additional hairy charge,  the widths of the lensed ring and photon ring emissions of light rays is larger, on the contrary, the one with larger deviation parameter $\alpha$  corresponds to narrower  widths of the lensed ring and photon ring emissions. This phenomena would indeed have {been printed}  on the black hole images.

We then investigated the optical appearance of the hairy RN black holes illuminated by an optically and geometrically thin accretion {disk}, which is radiated in the form \eqref{obser3}. There always exist two explicit peaks in the total observed intensity,  which correspond to two bright rings in the image of the  hairy RN black hole as we project the total observed intensity into a two dimensional plane. We found that hairy RN black hole with larger charge parameter $Q$ ($l_o$)  gives rings with weaker brightness,  while that with larger $\alpha$ has brighter rings. In addition, for larger charge parameter $Q$ ($l_o$), the central black hole shadow is smaller, but the larger deviation parameter $\alpha$ tends to enlarge the central shadow size. 
Finally, we figured out the optical appearances of hairy RN black holes illuminated by a static spherical accretion. The observed intensity shows a peak at critical impact parameter, indicating that in the optical appearance of the hairy RN black hole, a bright ring enclosing a shaded region appears and the brightness becomes weaker as the impact parameter increase due to the redshift effect. Moreover, we found that as the charge parameter $Q$ ($l_o$) increases, the ring becomes brighter but  the enclosed shaded region in the center is smaller. While the stronger deviation parameter $\alpha$ corresponds to a darker ring but larger faint illuminated region in the central of black hole images, which are completely opposite to the influence of charge parameter $Q$ or $l_o$. 

In conclusion, we investigated the effects of electric charge $Q$,  additional hairy charge $l_o$ and the deviation parameter $\alpha$ on the accretion images of the hairy RN black holes in the scenario of GD proposal. Our results show that the electric charge and additional hairy  
charge have similar effects on the rings and images of the hairy RN black hole, while the deviation parameters show completely different influences. Thus, we discussed that one may not be able to distinguish the electric charge and additional hairy charge via the rings and images of the black hole in this scenario. Besides, there could be degeneracies in the optical appearances of the hairy RN black hole, caused by the competing effects of the charge parameters and the deviation parameter. This could also produce the undistinguished images between the hairy RN black holes and RN black holes. 

The significance of the current study could exist in two aspects. From observable  point,  our study chooses the hairy RN black hole as a first attempt to utilize  ring and image of black holes as potential probes to seek the possible distinguishing of electric charge and other charges. We hope that our study can inspire more related researches in this direction. From theoretical point, as previously mentioned, the complete background theory that generates the hairy RN black hole is unclear now, so our results may give some beneficial hints for building the background theory of hairy RN black hole from the phenomena besides for the existed thermodynamics \cite{Mahapatra:2022xea}, quasinormal modes \cite{Cavalcanti:2022cga,Yang:2022ifo,Li:2022hkq}, strong gravitational lensing \cite{Afrin:2021imp,Islam:2021dyk,Meng:2023htc}, precession and Lense-Thirring effect \cite{Wu:2023wld} and gravitational waves \cite{Zi:2023omh}.

\begin{acknowledgments}
This work is partly supported by the Natural Science Foundation of China under Grants Nos. 12375054 and 12375055,  Natural Science Foundation of Jiangsu Province under Grant No. BK20211601, and  the Postgraduate Research $\&$ Practice Innovation Program of Jiangsu Province under Grants Nos. KYCX22$_{-}$3452 and KYCX21$_{-}$3192.
\end{acknowledgments}

\begin{appendices}
\section{Brief derivation of the hairy RN black hole via GD approach}\label{appendix}
 Recently, Ovalle {et al.}  used the GD approach to obtain a spherically symmetric metric with electric charge and additional hairy charge \cite{Ovalle:2020kpd},
in which the corresponding Einstein equation is expressed by
\begin{equation}\label{eq-EE}
G_{\mu\nu}\equiv R_{\mu\nu}-\frac{1}{2}Rg_{\mu\nu}=8\pi\Tilde{T}_{\mu\nu}.
\end{equation}
The total energy momentum tensor $\Tilde{T}_{\mu\nu}=T_{\mu\nu}+\vartheta_{\mu\nu}$ contains two parts: $T_{\mu\nu}$ is the energy momentum tensor associated with a known solution of  GR and $\vartheta_{\mu\nu}$ is  introduced by new matter fields or a new gravitational sector.  The Bianchi identity requires $\nabla^\mu \Tilde{T}_{\mu\nu}=0$. In the GD approach proposed in  \cite{Ovalle:2017fgl}, $\vartheta_{\mu\nu}$ is always assumed to be decoupled from  $T_{\mu\nu}$ \cite{Contreras:2021yxe,Ovalle:2020kpd}. We will review the main steps to better understand how the GD approach works in the construction of a deformed solution from a seed metric. As we will show that under the decoupling assumption, we can decouple the equations of motion for the two sectors.

Firstly, we write the static spherically symmetric solution ${g}_{\mu\nu}$ to \eqref{eq-EE} as
\begin{equation}\label{eq-swx0}
ds^2=-e^{\nu(r)}dt^2+e^{\lambda(r)}dr^2+r^2(d\theta^2+\sin^2\theta d\phi^2),
\end{equation}
and the corresponding  Einstein tensor is denoted as $G_{\mu}^{~\nu}(\nu(r),\lambda(r))$. Secondly, we consider that the above solution is generated by the seed metric 
 \begin{equation}\label{eq-swx1}
ds^2=-e^{\xi(r)}dt^2+e^{\mu(r)}dr^2+r^2(d\theta^2+\sin^2\theta d\phi^2),
\end{equation}
which only sources from $T_{\mu\nu}$ (i.e. $\vartheta_{\mu\nu}=0$),
and the introduction of additional source $\vartheta_{\mu\nu}$ is equivalent to deforming the seed metric by
\begin{gather}\label{eq-deswx}
\xi(r)\rightarrow\nu(r)= \xi(r)+\alpha~k(r), ~~~~~~~e^{-\mu(r)}\rightarrow e^{-\lambda(r)}= e^{-\mu(r)}+\alpha~h(r).
\end{gather}
Here the parameter $\alpha$ is introduced to keep track of the deformations. Subsequently, we can check that the Einstein equation \eqref{eq-EE} can be decomposed into the standard Einstein equation {${G}_{\mu}^{~\nu}(\xi(r),\mu(r))=8\pi T_{\mu}^{~\nu}$ and an additional equation $\alpha~\mathcal{G}_{\mu}^{~\nu}(\xi(r),\mu(r);k(r),h(r))$=$8\pi\vartheta_{\mu}^{~\nu}$}, respectively.
It is obvious that the tensor $\vartheta_{\mu\nu}$ will vanish for the vanishing of the metric deformations. Also, under the transformation \eqref{eq-deswx}, the Einstein tensor will be transformed by
\begin{gather}
{G}_{\mu}^{~\nu}(\xi(r),\mu(r))\to  G_{\mu}^{~\nu}(\nu(r),\lambda(r)) ={G}_{\mu}^{~\nu}(\xi(r),\mu(r))+\alpha~\mathcal{G}_{\mu}^{~\nu}(\nu(r),\lambda(r)),
\end{gather}
which is a linear combination, originated from two sources added linearly in the r.h.s of \eqref{eq-EE}. In fact, this is the key point existing in GD approach to analytically construct the deforming metric from the seed solution.

Then, the authors of \cite{Ovalle:2020kpd} consider the seed metric \eqref{eq-swx1} as the Schwarzschild one (with  $T_{\mu\nu}=0$), and  treat the additional source as the anisotropic fluid satisfying dominant energy condition. They solve out the Einstein equation and obtain the hairy charged solution deformed from the Schwarzschild metric, and the metric for the hairy charged black hole reads as 
\begin{equation}\label{eq-static}
ds^2=-f(r)dt^2+\frac{dr^2}{f(r)}+r^2(d\theta^2+\sin^2\theta d\phi^2)
~~\mathrm{with}~~ f(r)=1-\frac{2M}{r}+\frac{Q^2}{r^2}-\frac{\alpha}{r}\left(M-\frac{l_o}{2}\right)e^{-r/(M-l_o/2)}.
\end{equation}
This metric describes certain deformation of the Schwarzschild solution due to the introduction of additional material sources.

\end{appendices}

\bibliography{ref}

\begin{thebibliography}{121}
\expandafter\ifx\csname natexlab\endcsname\relax\def\natexlab#1{#1}\fi
\expandafter\ifx\csname bibnamefont\endcsname\relax
  \def\bibnamefont#1{#1}\fi
\expandafter\ifx\csname bibfnamefont\endcsname\relax
  \def\bibfnamefont#1{#1}\fi
\expandafter\ifx\csname citenamefont\endcsname\relax
  \def\citenamefont#1{#1}\fi
\expandafter\ifx\csname url\endcsname\relax
  \def\url#1{\texttt{#1}}\fi
\expandafter\ifx\csname urlprefix\endcsname\relax\def\urlprefix{URL }\fi
\providecommand{\bibinfo}[2]{#2}
\providecommand{\eprint}[2][]{\url{#2}}

\bibitem[{\citenamefont{Akiyama
  et~al.}(2019{\natexlab{a}})}]{EventHorizonTelescope:2019dse}
\bibinfo{author}{\bibfnamefont{K.}~\bibnamefont{Akiyama}} \bibnamefont{et~al.}
  (\bibinfo{collaboration}{Event Horizon Telescope}),
  \bibinfo{journal}{Astrophys. J. Lett.} \textbf{\bibinfo{volume}{875}},
  \bibinfo{pages}{L1} (\bibinfo{year}{2019}{\natexlab{a}}),
  \eprint{1906.11238}.

\bibitem[{\citenamefont{Akiyama
  et~al.}(2019{\natexlab{b}})}]{EventHorizonTelescope:2019uob}
\bibinfo{author}{\bibfnamefont{K.}~\bibnamefont{Akiyama}} \bibnamefont{et~al.}
  (\bibinfo{collaboration}{Event Horizon Telescope}),
  \bibinfo{journal}{Astrophys. J. Lett.} \textbf{\bibinfo{volume}{875}},
  \bibinfo{pages}{L2} (\bibinfo{year}{2019}{\natexlab{b}}),
  \eprint{1906.11239}.

\bibitem[{\citenamefont{Akiyama
  et~al.}(2019{\natexlab{c}})}]{EventHorizonTelescope:2019jan}
\bibinfo{author}{\bibfnamefont{K.}~\bibnamefont{Akiyama}} \bibnamefont{et~al.}
  (\bibinfo{collaboration}{Event Horizon Telescope}),
  \bibinfo{journal}{Astrophys. J. Lett.} \textbf{\bibinfo{volume}{875}},
  \bibinfo{pages}{L3} (\bibinfo{year}{2019}{\natexlab{c}}),
  \eprint{1906.11240}.

\bibitem[{\citenamefont{Akiyama
  et~al.}(2019{\natexlab{d}})}]{EventHorizonTelescope:2019ths}
\bibinfo{author}{\bibfnamefont{K.}~\bibnamefont{Akiyama}} \bibnamefont{et~al.}
  (\bibinfo{collaboration}{Event Horizon Telescope}),
  \bibinfo{journal}{Astrophys. J. Lett.} \textbf{\bibinfo{volume}{875}},
  \bibinfo{pages}{L4} (\bibinfo{year}{2019}{\natexlab{d}}),
  \eprint{1906.11241}.

\bibitem[{\citenamefont{Akiyama
  et~al.}(2019{\natexlab{e}})}]{EventHorizonTelescope:2019pgp}
\bibinfo{author}{\bibfnamefont{K.}~\bibnamefont{Akiyama}} \bibnamefont{et~al.}
  (\bibinfo{collaboration}{Event Horizon Telescope}),
  \bibinfo{journal}{Astrophys. J. Lett.} \textbf{\bibinfo{volume}{875}},
  \bibinfo{pages}{L5} (\bibinfo{year}{2019}{\natexlab{e}}),
  \eprint{1906.11242}.

\bibitem[{\citenamefont{Akiyama
  et~al.}(2019{\natexlab{f}})}]{EventHorizonTelescope:2019ggy}
\bibinfo{author}{\bibfnamefont{K.}~\bibnamefont{Akiyama}} \bibnamefont{et~al.}
  (\bibinfo{collaboration}{Event Horizon Telescope}),
  \bibinfo{journal}{Astrophys. J. Lett.} \textbf{\bibinfo{volume}{875}},
  \bibinfo{pages}{L6} (\bibinfo{year}{2019}{\natexlab{f}}),
  \eprint{1906.11243}.

\bibitem[{\citenamefont{Akiyama
  et~al.}(2022{\natexlab{a}})}]{EventHorizonTelescope:2022wkp}
\bibinfo{author}{\bibfnamefont{K.}~\bibnamefont{Akiyama}} \bibnamefont{et~al.}
  (\bibinfo{collaboration}{Event Horizon Telescope}),
  \bibinfo{journal}{Astrophys. J. Lett.} \textbf{\bibinfo{volume}{930}},
  \bibinfo{pages}{L12} (\bibinfo{year}{2022}{\natexlab{a}}).

\bibitem[{\citenamefont{Akiyama
  et~al.}(2022{\natexlab{b}})}]{EventHorizonTelescope:2022apq}
\bibinfo{author}{\bibfnamefont{K.}~\bibnamefont{Akiyama}} \bibnamefont{et~al.}
  (\bibinfo{collaboration}{Event Horizon Telescope}),
  \bibinfo{journal}{Astrophys. J. Lett.} \textbf{\bibinfo{volume}{930}},
  \bibinfo{pages}{L13} (\bibinfo{year}{2022}{\natexlab{b}}).

\bibitem[{\citenamefont{Akiyama
  et~al.}(2022{\natexlab{c}})}]{EventHorizonTelescope:2022wok}
\bibinfo{author}{\bibfnamefont{K.}~\bibnamefont{Akiyama}} \bibnamefont{et~al.}
  (\bibinfo{collaboration}{Event Horizon Telescope}),
  \bibinfo{journal}{Astrophys. J. Lett.} \textbf{\bibinfo{volume}{930}},
  \bibinfo{pages}{L14} (\bibinfo{year}{2022}{\natexlab{c}}).

\bibitem[{\citenamefont{Akiyama
  et~al.}(2022{\natexlab{d}})}]{EventHorizonTelescope:2022exc}
\bibinfo{author}{\bibfnamefont{K.}~\bibnamefont{Akiyama}} \bibnamefont{et~al.}
  (\bibinfo{collaboration}{Event Horizon Telescope}),
  \bibinfo{journal}{Astrophys. J. Lett.} \textbf{\bibinfo{volume}{930}},
  \bibinfo{pages}{L15} (\bibinfo{year}{2022}{\natexlab{d}}).

\bibitem[{\citenamefont{Akiyama
  et~al.}(2022{\natexlab{e}})}]{EventHorizonTelescope:2022urf}
\bibinfo{author}{\bibfnamefont{K.}~\bibnamefont{Akiyama}} \bibnamefont{et~al.}
  (\bibinfo{collaboration}{Event Horizon Telescope}),
  \bibinfo{journal}{Astrophys. J. Lett.} \textbf{\bibinfo{volume}{930}},
  \bibinfo{pages}{L16} (\bibinfo{year}{2022}{\natexlab{e}}).

\bibitem[{\citenamefont{Akiyama
  et~al.}(2022{\natexlab{f}})}]{EventHorizonTelescope:2022xqj}
\bibinfo{author}{\bibfnamefont{K.}~\bibnamefont{Akiyama}} \bibnamefont{et~al.}
  (\bibinfo{collaboration}{Event Horizon Telescope}),
  \bibinfo{journal}{Astrophys. J. Lett.} \textbf{\bibinfo{volume}{930}},
  \bibinfo{pages}{L17} (\bibinfo{year}{2022}{\natexlab{f}}).

\bibitem[{\citenamefont{Virbhadra and Ellis}(2000)}]{Virbhadra:1999nm}
\bibinfo{author}{\bibfnamefont{K.~S.} \bibnamefont{Virbhadra}}
  \bibnamefont{and} \bibinfo{author}{\bibfnamefont{G.~F.~R.}
  \bibnamefont{Ellis}}, \bibinfo{journal}{Phys. Rev. D}
  \textbf{\bibinfo{volume}{62}}, \bibinfo{pages}{084003}
  (\bibinfo{year}{2000}), \eprint{astro-ph/9904193}.

\bibitem[{\citenamefont{Virbhadra and Ellis}(2002)}]{Virbhadra:2002ju}
\bibinfo{author}{\bibfnamefont{K.~S.} \bibnamefont{Virbhadra}}
  \bibnamefont{and} \bibinfo{author}{\bibfnamefont{G.~F.~R.}
  \bibnamefont{Ellis}}, \bibinfo{journal}{Phys. Rev. D}
  \textbf{\bibinfo{volume}{65}}, \bibinfo{pages}{103004}
  (\bibinfo{year}{2002}).

\bibitem[{\citenamefont{Virbhadra and Keeton}(2008)}]{Virbhadra:2007kw}
\bibinfo{author}{\bibfnamefont{K.~S.} \bibnamefont{Virbhadra}}
  \bibnamefont{and} \bibinfo{author}{\bibfnamefont{C.~R.}
  \bibnamefont{Keeton}}, \bibinfo{journal}{Phys. Rev. D}
  \textbf{\bibinfo{volume}{77}}, \bibinfo{pages}{124014}
  (\bibinfo{year}{2008}), \eprint{0710.2333}.

\bibitem[{\citenamefont{Synge}(1966)}]{Synge:1966okc}
\bibinfo{author}{\bibfnamefont{J.~L.} \bibnamefont{Synge}},
  \bibinfo{journal}{Mon. Not. Roy. Astron. Soc.}
  \textbf{\bibinfo{volume}{131}}, \bibinfo{pages}{463} (\bibinfo{year}{1966}).

\bibitem[{\citenamefont{Bardeen et~al.}(1972)\citenamefont{Bardeen, Press, and
  Teukolsky}}]{Bardeen:1972fi}
\bibinfo{author}{\bibfnamefont{J.~M.} \bibnamefont{Bardeen}},
  \bibinfo{author}{\bibfnamefont{W.~H.} \bibnamefont{Press}}, \bibnamefont{and}
  \bibinfo{author}{\bibfnamefont{S.~A.} \bibnamefont{Teukolsky}},
  \bibinfo{journal}{Astrophys. J.} \textbf{\bibinfo{volume}{178}},
  \bibinfo{pages}{347} (\bibinfo{year}{1972}).

\bibitem[{\citenamefont{Bozza}(2010)}]{Bozza:2010xqn}
\bibinfo{author}{\bibfnamefont{V.}~\bibnamefont{Bozza}}, \bibinfo{journal}{Gen.
  Rel. Grav.} \textbf{\bibinfo{volume}{42}}, \bibinfo{pages}{2269}
  (\bibinfo{year}{2010}), \eprint{0911.2187}.

\bibitem[{\citenamefont{Luminet}(1979)}]{Luminet:1979nyg}
\bibinfo{author}{\bibfnamefont{J.~P.} \bibnamefont{Luminet}},
  \bibinfo{journal}{Astron. Astrophys.} \textbf{\bibinfo{volume}{75}},
  \bibinfo{pages}{228} (\bibinfo{year}{1979}).

\bibitem[{\citenamefont{Shen et~al.}(2005)\citenamefont{Shen, Lo, Liang, Ho,
  and Zhao}}]{Shen:2005cw}
\bibinfo{author}{\bibfnamefont{Z.-Q.} \bibnamefont{Shen}},
  \bibinfo{author}{\bibfnamefont{K.~Y.} \bibnamefont{Lo}},
  \bibinfo{author}{\bibfnamefont{M.~C.} \bibnamefont{Liang}},
  \bibinfo{author}{\bibfnamefont{P.~T.~P.} \bibnamefont{Ho}}, \bibnamefont{and}
  \bibinfo{author}{\bibfnamefont{J.~H.} \bibnamefont{Zhao}},
  \bibinfo{journal}{Nature} \textbf{\bibinfo{volume}{438}}, \bibinfo{pages}{62}
  (\bibinfo{year}{2005}), \eprint{astro-ph/0512515}.

\bibitem[{\citenamefont{Yumoto et~al.}(2012)\citenamefont{Yumoto, Nitta, Chiba,
  and Sugiyama}}]{Yumoto:2012kz}
\bibinfo{author}{\bibfnamefont{A.}~\bibnamefont{Yumoto}},
  \bibinfo{author}{\bibfnamefont{D.}~\bibnamefont{Nitta}},
  \bibinfo{author}{\bibfnamefont{T.}~\bibnamefont{Chiba}}, \bibnamefont{and}
  \bibinfo{author}{\bibfnamefont{N.}~\bibnamefont{Sugiyama}},
  \bibinfo{journal}{Phys. Rev. D} \textbf{\bibinfo{volume}{86}},
  \bibinfo{pages}{103001} (\bibinfo{year}{2012}), \eprint{1208.0635}.

\bibitem[{\citenamefont{Atamurotov et~al.}(2013)\citenamefont{Atamurotov,
  Abdujabbarov, and Ahmedov}}]{Atamurotov:2013sca}
\bibinfo{author}{\bibfnamefont{F.}~\bibnamefont{Atamurotov}},
  \bibinfo{author}{\bibfnamefont{A.}~\bibnamefont{Abdujabbarov}},
  \bibnamefont{and} \bibinfo{author}{\bibfnamefont{B.}~\bibnamefont{Ahmedov}},
  \bibinfo{journal}{Phys. Rev. D} \textbf{\bibinfo{volume}{88}},
  \bibinfo{pages}{064004} (\bibinfo{year}{2013}).

\bibitem[{\citenamefont{Papnoi et~al.}(2014)\citenamefont{Papnoi, Atamurotov,
  Ghosh, and Ahmedov}}]{Papnoi:2014aaa}
\bibinfo{author}{\bibfnamefont{U.}~\bibnamefont{Papnoi}},
  \bibinfo{author}{\bibfnamefont{F.}~\bibnamefont{Atamurotov}},
  \bibinfo{author}{\bibfnamefont{S.~G.} \bibnamefont{Ghosh}}, \bibnamefont{and}
  \bibinfo{author}{\bibfnamefont{B.}~\bibnamefont{Ahmedov}},
  \bibinfo{journal}{Phys. Rev. D} \textbf{\bibinfo{volume}{90}},
  \bibinfo{pages}{024073} (\bibinfo{year}{2014}), \eprint{1407.0834}.

\bibitem[{\citenamefont{Abdujabbarov et~al.}(2015)\citenamefont{Abdujabbarov,
  Rezzolla, and Ahmedov}}]{Abdujabbarov:2015xqa}
\bibinfo{author}{\bibfnamefont{A.~A.} \bibnamefont{Abdujabbarov}},
  \bibinfo{author}{\bibfnamefont{L.}~\bibnamefont{Rezzolla}}, \bibnamefont{and}
  \bibinfo{author}{\bibfnamefont{B.~J.} \bibnamefont{Ahmedov}},
  \bibinfo{journal}{Mon. Not. Roy. Astron. Soc.}
  \textbf{\bibinfo{volume}{454}}, \bibinfo{pages}{2423} (\bibinfo{year}{2015}),
  \eprint{1503.09054}.

\bibitem[{\citenamefont{Kumar and Ghosh}(2020)}]{Kumar:2018ple}
\bibinfo{author}{\bibfnamefont{R.}~\bibnamefont{Kumar}} \bibnamefont{and}
  \bibinfo{author}{\bibfnamefont{S.~G.} \bibnamefont{Ghosh}},
  \bibinfo{journal}{Astrophys. J.} \textbf{\bibinfo{volume}{892}},
  \bibinfo{pages}{78} (\bibinfo{year}{2020}), \eprint{1811.01260}.

\bibitem[{\citenamefont{Meng et~al.}(2022)\citenamefont{Meng, Kuang, and
  Tang}}]{Meng:2022kjs}
\bibinfo{author}{\bibfnamefont{Y.}~\bibnamefont{Meng}},
  \bibinfo{author}{\bibfnamefont{X.-M.} \bibnamefont{Kuang}}, \bibnamefont{and}
  \bibinfo{author}{\bibfnamefont{Z.-Y.} \bibnamefont{Tang}},
  \bibinfo{journal}{Phys. Rev. D} \textbf{\bibinfo{volume}{106}},
  \bibinfo{pages}{064006} (\bibinfo{year}{2022}), \eprint{2204.00897}.

\bibitem[{\citenamefont{Ma and Lu}(2020)}]{Ma:2019ybz}
\bibinfo{author}{\bibfnamefont{L.}~\bibnamefont{Ma}} \bibnamefont{and}
  \bibinfo{author}{\bibfnamefont{H.}~\bibnamefont{Lu}}, \bibinfo{journal}{Phys.
  Lett. B} \textbf{\bibinfo{volume}{807}}, \bibinfo{pages}{135535}
  (\bibinfo{year}{2020}), \eprint{1912.05569}.

\bibitem[{\citenamefont{Guo and Li}(2020)}]{Guo:2020zmf}
\bibinfo{author}{\bibfnamefont{M.}~\bibnamefont{Guo}} \bibnamefont{and}
  \bibinfo{author}{\bibfnamefont{P.-C.} \bibnamefont{Li}},
  \bibinfo{journal}{Eur. Phys. J. C} \textbf{\bibinfo{volume}{80}},
  \bibinfo{pages}{588} (\bibinfo{year}{2020}), \eprint{2003.02523}.

\bibitem[{\citenamefont{Meng et~al.}(2023{\natexlab{a}})\citenamefont{Meng,
  Kuang, Wang, and Wu}}]{Meng:2023wgi}
\bibinfo{author}{\bibfnamefont{Y.}~\bibnamefont{Meng}},
  \bibinfo{author}{\bibfnamefont{X.-M.} \bibnamefont{Kuang}},
  \bibinfo{author}{\bibfnamefont{X.-J.} \bibnamefont{Wang}}, \bibnamefont{and}
  \bibinfo{author}{\bibfnamefont{J.-P.} \bibnamefont{Wu}},
  \bibinfo{journal}{Phys. Lett. B} \textbf{\bibinfo{volume}{841}},
  \bibinfo{pages}{137940} (\bibinfo{year}{2023}{\natexlab{a}}),
  \eprint{2305.04210}.

\bibitem[{\citenamefont{Ayzenberg and Yunes}(2018)}]{Ayzenberg:2018jip}
\bibinfo{author}{\bibfnamefont{D.}~\bibnamefont{Ayzenberg}} \bibnamefont{and}
  \bibinfo{author}{\bibfnamefont{N.}~\bibnamefont{Yunes}},
  \bibinfo{journal}{Class. Quant. Grav.} \textbf{\bibinfo{volume}{35}},
  \bibinfo{pages}{235002} (\bibinfo{year}{2018}), \eprint{1807.08422}.

\bibitem[{\citenamefont{Amarilla et~al.}(2010)\citenamefont{Amarilla, Eiroa,
  and Giribet}}]{Amarilla:2010zq}
\bibinfo{author}{\bibfnamefont{L.}~\bibnamefont{Amarilla}},
  \bibinfo{author}{\bibfnamefont{E.~F.} \bibnamefont{Eiroa}}, \bibnamefont{and}
  \bibinfo{author}{\bibfnamefont{G.}~\bibnamefont{Giribet}},
  \bibinfo{journal}{Phys. Rev. D} \textbf{\bibinfo{volume}{81}},
  \bibinfo{pages}{124045} (\bibinfo{year}{2010}), \eprint{1005.0607}.

\bibitem[{\citenamefont{Addazi et~al.}(2021)\citenamefont{Addazi, Capozziello,
  and Odintsov}}]{Addazi:2021pty}
\bibinfo{author}{\bibfnamefont{A.}~\bibnamefont{Addazi}},
  \bibinfo{author}{\bibfnamefont{S.}~\bibnamefont{Capozziello}},
  \bibnamefont{and} \bibinfo{author}{\bibfnamefont{S.}~\bibnamefont{Odintsov}},
  \bibinfo{journal}{Phys. Lett. B} \textbf{\bibinfo{volume}{816}},
  \bibinfo{pages}{136257} (\bibinfo{year}{2021}), \eprint{2103.16856}.

\bibitem[{\citenamefont{Dastan et~al.}(2022)\citenamefont{Dastan, Saffari, and
  Soroushfar}}]{Dastan:2016vhb}
\bibinfo{author}{\bibfnamefont{S.}~\bibnamefont{Dastan}},
  \bibinfo{author}{\bibfnamefont{R.}~\bibnamefont{Saffari}}, \bibnamefont{and}
  \bibinfo{author}{\bibfnamefont{S.}~\bibnamefont{Soroushfar}},
  \bibinfo{journal}{Eur. Phys. J. Plus} \textbf{\bibinfo{volume}{137}},
  \bibinfo{pages}{1002} (\bibinfo{year}{2022}), \eprint{1606.06994}.

\bibitem[{\citenamefont{Amarilla and Eiroa}(2012)}]{Amarilla:2011fx}
\bibinfo{author}{\bibfnamefont{L.}~\bibnamefont{Amarilla}} \bibnamefont{and}
  \bibinfo{author}{\bibfnamefont{E.~F.} \bibnamefont{Eiroa}},
  \bibinfo{journal}{Phys. Rev. D} \textbf{\bibinfo{volume}{85}},
  \bibinfo{pages}{064019} (\bibinfo{year}{2012}), \eprint{1112.6349}.

\bibitem[{\citenamefont{Amarilla and Eiroa}(2013)}]{Amarilla:2013sj}
\bibinfo{author}{\bibfnamefont{L.}~\bibnamefont{Amarilla}} \bibnamefont{and}
  \bibinfo{author}{\bibfnamefont{E.~F.} \bibnamefont{Eiroa}},
  \bibinfo{journal}{Phys. Rev. D} \textbf{\bibinfo{volume}{87}},
  \bibinfo{pages}{044057} (\bibinfo{year}{2013}), \eprint{1301.0532}.

\bibitem[{\citenamefont{Amir et~al.}(2018)\citenamefont{Amir, Singh, and
  Ghosh}}]{Amir:2017slq}
\bibinfo{author}{\bibfnamefont{M.}~\bibnamefont{Amir}},
  \bibinfo{author}{\bibfnamefont{B.~P.} \bibnamefont{Singh}}, \bibnamefont{and}
  \bibinfo{author}{\bibfnamefont{S.~G.} \bibnamefont{Ghosh}},
  \bibinfo{journal}{Eur. Phys. J. C} \textbf{\bibinfo{volume}{78}},
  \bibinfo{pages}{399} (\bibinfo{year}{2018}), \eprint{1707.09521}.

\bibitem[{\citenamefont{Mizuno et~al.}(2018)\citenamefont{Mizuno, Younsi,
  Fromm, Porth, De~Laurentis, Olivares, Falcke, Kramer, and
  Rezzolla}}]{Mizuno:2018lxz}
\bibinfo{author}{\bibfnamefont{Y.}~\bibnamefont{Mizuno}},
  \bibinfo{author}{\bibfnamefont{Z.}~\bibnamefont{Younsi}},
  \bibinfo{author}{\bibfnamefont{C.~M.} \bibnamefont{Fromm}},
  \bibinfo{author}{\bibfnamefont{O.}~\bibnamefont{Porth}},
  \bibinfo{author}{\bibfnamefont{M.}~\bibnamefont{De~Laurentis}},
  \bibinfo{author}{\bibfnamefont{H.}~\bibnamefont{Olivares}},
  \bibinfo{author}{\bibfnamefont{H.}~\bibnamefont{Falcke}},
  \bibinfo{author}{\bibfnamefont{M.}~\bibnamefont{Kramer}}, \bibnamefont{and}
  \bibinfo{author}{\bibfnamefont{L.}~\bibnamefont{Rezzolla}},
  \bibinfo{journal}{Nature Astron.} \textbf{\bibinfo{volume}{2}},
  \bibinfo{pages}{585} (\bibinfo{year}{2018}), \eprint{1804.05812}.

\bibitem[{\citenamefont{Eiroa and Sendra}(2018)}]{Eiroa:2017uuq}
\bibinfo{author}{\bibfnamefont{E.~F.} \bibnamefont{Eiroa}} \bibnamefont{and}
  \bibinfo{author}{\bibfnamefont{C.~M.} \bibnamefont{Sendra}},
  \bibinfo{journal}{Eur. Phys. J. C} \textbf{\bibinfo{volume}{78}},
  \bibinfo{pages}{91} (\bibinfo{year}{2018}), \eprint{1711.08380}.

\bibitem[{\citenamefont{Vagnozzi and Visinelli}(2019)}]{Vagnozzi:2019apd}
\bibinfo{author}{\bibfnamefont{S.}~\bibnamefont{Vagnozzi}} \bibnamefont{and}
  \bibinfo{author}{\bibfnamefont{L.}~\bibnamefont{Visinelli}},
  \bibinfo{journal}{Phys. Rev. D} \textbf{\bibinfo{volume}{100}},
  \bibinfo{pages}{024020} (\bibinfo{year}{2019}), \eprint{1905.12421}.

\bibitem[{\citenamefont{Banerjee et~al.}(2020)\citenamefont{Banerjee,
  Chakraborty, and SenGupta}}]{Banerjee:2019nnj}
\bibinfo{author}{\bibfnamefont{I.}~\bibnamefont{Banerjee}},
  \bibinfo{author}{\bibfnamefont{S.}~\bibnamefont{Chakraborty}},
  \bibnamefont{and} \bibinfo{author}{\bibfnamefont{S.}~\bibnamefont{SenGupta}},
  \bibinfo{journal}{Phys. Rev. D} \textbf{\bibinfo{volume}{101}},
  \bibinfo{pages}{041301} (\bibinfo{year}{2020}), \eprint{1909.09385}.

\bibitem[{\citenamefont{Chowdhuri and Bhattacharyya}(2021)}]{Chowdhuri:2020ipb}
\bibinfo{author}{\bibfnamefont{A.}~\bibnamefont{Chowdhuri}} \bibnamefont{and}
  \bibinfo{author}{\bibfnamefont{A.}~\bibnamefont{Bhattacharyya}},
  \bibinfo{journal}{Phys. Rev. D} \textbf{\bibinfo{volume}{104}},
  \bibinfo{pages}{064039} (\bibinfo{year}{2021}), \eprint{2012.12914}.

\bibitem[{\citenamefont{Konoplya and Zhidenko}(2019)}]{Konoplya:2019goy}
\bibinfo{author}{\bibfnamefont{R.~A.} \bibnamefont{Konoplya}} \bibnamefont{and}
  \bibinfo{author}{\bibfnamefont{A.}~\bibnamefont{Zhidenko}},
  \bibinfo{journal}{Phys. Rev. D} \textbf{\bibinfo{volume}{100}},
  \bibinfo{pages}{044015} (\bibinfo{year}{2019}), \eprint{1907.05551}.

\bibitem[{\citenamefont{Younsi et~al.}(2016)\citenamefont{Younsi, Zhidenko,
  Rezzolla, Konoplya, and Mizuno}}]{Younsi:2016azx}
\bibinfo{author}{\bibfnamefont{Z.}~\bibnamefont{Younsi}},
  \bibinfo{author}{\bibfnamefont{A.}~\bibnamefont{Zhidenko}},
  \bibinfo{author}{\bibfnamefont{L.}~\bibnamefont{Rezzolla}},
  \bibinfo{author}{\bibfnamefont{R.}~\bibnamefont{Konoplya}}, \bibnamefont{and}
  \bibinfo{author}{\bibfnamefont{Y.}~\bibnamefont{Mizuno}},
  \bibinfo{journal}{Phys. Rev. D} \textbf{\bibinfo{volume}{94}},
  \bibinfo{pages}{084025} (\bibinfo{year}{2016}), \eprint{1607.05767}.

\bibitem[{\citenamefont{Olmo et~al.}(2023)\citenamefont{Olmo, Rosa,
  Rubiera-Garcia, and Saez-Chillon~Gomez}}]{Olmo:2023lil}
\bibinfo{author}{\bibfnamefont{G.~J.} \bibnamefont{Olmo}},
  \bibinfo{author}{\bibfnamefont{J.~L.} \bibnamefont{Rosa}},
  \bibinfo{author}{\bibfnamefont{D.}~\bibnamefont{Rubiera-Garcia}},
  \bibnamefont{and}
  \bibinfo{author}{\bibfnamefont{D.}~\bibnamefont{Saez-Chillon~Gomez}},
  \bibinfo{journal}{Class. Quant. Grav.} \textbf{\bibinfo{volume}{40}},
  \bibinfo{pages}{174002} (\bibinfo{year}{2023}), \eprint{2302.12064}.

\bibitem[{\citenamefont{Shaikh et~al.}(2019{\natexlab{a}})\citenamefont{Shaikh,
  Kocherlakota, Narayan, and Joshi}}]{Shaikh:2018lcc}
\bibinfo{author}{\bibfnamefont{R.}~\bibnamefont{Shaikh}},
  \bibinfo{author}{\bibfnamefont{P.}~\bibnamefont{Kocherlakota}},
  \bibinfo{author}{\bibfnamefont{R.}~\bibnamefont{Narayan}}, \bibnamefont{and}
  \bibinfo{author}{\bibfnamefont{P.~S.} \bibnamefont{Joshi}},
  \bibinfo{journal}{Mon. Not. Roy. Astron. Soc.}
  \textbf{\bibinfo{volume}{482}}, \bibinfo{pages}{52}
  (\bibinfo{year}{2019}{\natexlab{a}}), \eprint{1802.08060}.

\bibitem[{\citenamefont{Joshi et~al.}(2020)\citenamefont{Joshi, Dey, Joshi, and
  Bambhaniya}}]{Joshi:2020tlq}
\bibinfo{author}{\bibfnamefont{A.~B.} \bibnamefont{Joshi}},
  \bibinfo{author}{\bibfnamefont{D.}~\bibnamefont{Dey}},
  \bibinfo{author}{\bibfnamefont{P.~S.} \bibnamefont{Joshi}}, \bibnamefont{and}
  \bibinfo{author}{\bibfnamefont{P.}~\bibnamefont{Bambhaniya}},
  \bibinfo{journal}{Phys. Rev. D} \textbf{\bibinfo{volume}{102}},
  \bibinfo{pages}{024022} (\bibinfo{year}{2020}), \eprint{2004.06525}.

\bibitem[{\citenamefont{Dey et~al.}(2021)\citenamefont{Dey, Shaikh, and
  Joshi}}]{Dey:2020bgo}
\bibinfo{author}{\bibfnamefont{D.}~\bibnamefont{Dey}},
  \bibinfo{author}{\bibfnamefont{R.}~\bibnamefont{Shaikh}}, \bibnamefont{and}
  \bibinfo{author}{\bibfnamefont{P.~S.} \bibnamefont{Joshi}},
  \bibinfo{journal}{Phys. Rev. D} \textbf{\bibinfo{volume}{103}},
  \bibinfo{pages}{024015} (\bibinfo{year}{2021}), \eprint{2009.07487}.

\bibitem[{\citenamefont{Rahaman et~al.}(2021)\citenamefont{Rahaman, Singh,
  Shaikh, Manna, and Aktar}}]{Rahaman:2021web}
\bibinfo{author}{\bibfnamefont{F.}~\bibnamefont{Rahaman}},
  \bibinfo{author}{\bibfnamefont{K.~N.} \bibnamefont{Singh}},
  \bibinfo{author}{\bibfnamefont{R.}~\bibnamefont{Shaikh}},
  \bibinfo{author}{\bibfnamefont{T.}~\bibnamefont{Manna}}, \bibnamefont{and}
  \bibinfo{author}{\bibfnamefont{S.}~\bibnamefont{Aktar}},
  \bibinfo{journal}{Class. Quant. Grav.} \textbf{\bibinfo{volume}{38}},
  \bibinfo{pages}{215007} (\bibinfo{year}{2021}), \eprint{2108.09930}.

\bibitem[{\citenamefont{Kasuya and Kobayashi}(2021)}]{Kasuya:2021cpk}
\bibinfo{author}{\bibfnamefont{S.}~\bibnamefont{Kasuya}} \bibnamefont{and}
  \bibinfo{author}{\bibfnamefont{M.}~\bibnamefont{Kobayashi}},
  \bibinfo{journal}{Phys. Rev. D} \textbf{\bibinfo{volume}{103}},
  \bibinfo{pages}{104050} (\bibinfo{year}{2021}), \eprint{2103.13086}.

\bibitem[{\citenamefont{Shaikh et~al.}(2019{\natexlab{b}})\citenamefont{Shaikh,
  Banerjee, Paul, and Sarkar}}]{Shaikh:2018oul}
\bibinfo{author}{\bibfnamefont{R.}~\bibnamefont{Shaikh}},
  \bibinfo{author}{\bibfnamefont{P.}~\bibnamefont{Banerjee}},
  \bibinfo{author}{\bibfnamefont{S.}~\bibnamefont{Paul}}, \bibnamefont{and}
  \bibinfo{author}{\bibfnamefont{T.}~\bibnamefont{Sarkar}},
  \bibinfo{journal}{Phys. Lett. B} \textbf{\bibinfo{volume}{789}},
  \bibinfo{pages}{270} (\bibinfo{year}{2019}{\natexlab{b}}),
  \bibinfo{note}{[Erratum: Phys.Lett.B 791, 422--423 (2019)]},
  \eprint{1811.08245}.

\bibitem[{\citenamefont{Wielgus et~al.}(2020)\citenamefont{Wielgus, Horak,
  Vincent, and Abramowicz}}]{Wielgus:2020uqz}
\bibinfo{author}{\bibfnamefont{M.}~\bibnamefont{Wielgus}},
  \bibinfo{author}{\bibfnamefont{J.}~\bibnamefont{Horak}},
  \bibinfo{author}{\bibfnamefont{F.}~\bibnamefont{Vincent}}, \bibnamefont{and}
  \bibinfo{author}{\bibfnamefont{M.}~\bibnamefont{Abramowicz}},
  \bibinfo{journal}{Phys. Rev. D} \textbf{\bibinfo{volume}{102}},
  \bibinfo{pages}{084044} (\bibinfo{year}{2020}), \eprint{2008.10130}.

\bibitem[{\citenamefont{Peng et~al.}(2021{\natexlab{a}})\citenamefont{Peng,
  Guo, and Feng}}]{Peng:2021osd}
\bibinfo{author}{\bibfnamefont{J.}~\bibnamefont{Peng}},
  \bibinfo{author}{\bibfnamefont{M.}~\bibnamefont{Guo}}, \bibnamefont{and}
  \bibinfo{author}{\bibfnamefont{X.-H.} \bibnamefont{Feng}},
  \bibinfo{journal}{Phys. Rev. D} \textbf{\bibinfo{volume}{104}},
  \bibinfo{pages}{124010} (\bibinfo{year}{2021}{\natexlab{a}}),
  \eprint{2102.05488}.

\bibitem[{\citenamefont{Neto et~al.}(2023)\citenamefont{Neto, P\'erez, and
  Pelle}}]{Neto:2022pmu}
\bibinfo{author}{\bibfnamefont{M.~R.} \bibnamefont{Neto}},
  \bibinfo{author}{\bibfnamefont{D.}~\bibnamefont{P\'erez}}, \bibnamefont{and}
  \bibinfo{author}{\bibfnamefont{J.}~\bibnamefont{Pelle}},
  \bibinfo{journal}{Int. J. Mod. Phys. D} \textbf{\bibinfo{volume}{32}},
  \bibinfo{pages}{2250137} (\bibinfo{year}{2023}), \eprint{2210.14106}.

\bibitem[{\citenamefont{Tsukamoto et~al.}(2012)\citenamefont{Tsukamoto, Harada,
  and Yajima}}]{Tsukamoto:2012xs}
\bibinfo{author}{\bibfnamefont{N.}~\bibnamefont{Tsukamoto}},
  \bibinfo{author}{\bibfnamefont{T.}~\bibnamefont{Harada}}, \bibnamefont{and}
  \bibinfo{author}{\bibfnamefont{K.}~\bibnamefont{Yajima}},
  \bibinfo{journal}{Phys. Rev. D} \textbf{\bibinfo{volume}{86}},
  \bibinfo{pages}{104062} (\bibinfo{year}{2012}), \eprint{1207.0047}.

\bibitem[{\citenamefont{Vagnozzi et~al.}(2023)}]{Vagnozzi:2022moj}
\bibinfo{author}{\bibfnamefont{S.}~\bibnamefont{Vagnozzi}}
  \bibnamefont{et~al.}, \bibinfo{journal}{Class. Quant. Grav.}
  \textbf{\bibinfo{volume}{40}}, \bibinfo{pages}{165007}
  (\bibinfo{year}{2023}), \eprint{2205.07787}.

\bibitem[{\citenamefont{Afrin et~al.}(2021)\citenamefont{Afrin, Kumar, and
  Ghosh}}]{Afrin:2021imp}
\bibinfo{author}{\bibfnamefont{M.}~\bibnamefont{Afrin}},
  \bibinfo{author}{\bibfnamefont{R.}~\bibnamefont{Kumar}}, \bibnamefont{and}
  \bibinfo{author}{\bibfnamefont{S.~G.} \bibnamefont{Ghosh}},
  \bibinfo{journal}{Mon. Not. Roy. Astron. Soc.}
  \textbf{\bibinfo{volume}{504}}, \bibinfo{pages}{5927} (\bibinfo{year}{2021}),
  \eprint{2103.11417}.

\bibitem[{\citenamefont{Kuang et~al.}(2022)\citenamefont{Kuang, Tang, Wang, and
  Wang}}]{Kuang:2022ojj}
\bibinfo{author}{\bibfnamefont{X.-M.} \bibnamefont{Kuang}},
  \bibinfo{author}{\bibfnamefont{Z.-Y.} \bibnamefont{Tang}},
  \bibinfo{author}{\bibfnamefont{B.}~\bibnamefont{Wang}}, \bibnamefont{and}
  \bibinfo{author}{\bibfnamefont{A.}~\bibnamefont{Wang}},
  \bibinfo{journal}{Phys. Rev. D} \textbf{\bibinfo{volume}{106}},
  \bibinfo{pages}{064012} (\bibinfo{year}{2022}), \eprint{2206.05878}.

\bibitem[{\citenamefont{Tang et~al.}(2022)\citenamefont{Tang, Kuang, Wang, and
  Qian}}]{Tang:2022hsu}
\bibinfo{author}{\bibfnamefont{Z.-Y.} \bibnamefont{Tang}},
  \bibinfo{author}{\bibfnamefont{X.-M.} \bibnamefont{Kuang}},
  \bibinfo{author}{\bibfnamefont{B.}~\bibnamefont{Wang}}, \bibnamefont{and}
  \bibinfo{author}{\bibfnamefont{W.-L.} \bibnamefont{Qian}},
  \bibinfo{journal}{Sci. Bull.} \textbf{\bibinfo{volume}{67}},
  \bibinfo{pages}{2272} (\bibinfo{year}{2022}), \eprint{2206.08608}.

\bibitem[{\citenamefont{Tang et~al.}(2023)\citenamefont{Tang, Kuang, Wang, and
  Qian}}]{Tang:2022bcm}
\bibinfo{author}{\bibfnamefont{Z.-Y.} \bibnamefont{Tang}},
  \bibinfo{author}{\bibfnamefont{X.-M.} \bibnamefont{Kuang}},
  \bibinfo{author}{\bibfnamefont{B.}~\bibnamefont{Wang}}, \bibnamefont{and}
  \bibinfo{author}{\bibfnamefont{W.-L.} \bibnamefont{Qian}},
  \bibinfo{journal}{Eur. Phys. J. C} \textbf{\bibinfo{volume}{83}},
  \bibinfo{pages}{837} (\bibinfo{year}{2023}), \eprint{2211.08137}.

\bibitem[{\citenamefont{Kuang and \"Ovg\"un}(2022)}]{Kuang:2022xjp}
\bibinfo{author}{\bibfnamefont{X.-M.} \bibnamefont{Kuang}} \bibnamefont{and}
  \bibinfo{author}{\bibfnamefont{A.}~\bibnamefont{\"Ovg\"un}},
  \bibinfo{journal}{Annals Phys.} \textbf{\bibinfo{volume}{447}},
  \bibinfo{pages}{169147} (\bibinfo{year}{2022}), \eprint{2205.11003}.

\bibitem[{\citenamefont{Kumar et~al.}(2019)\citenamefont{Kumar, Ghosh, and
  Wang}}]{Kumar:2019pjp}
\bibinfo{author}{\bibfnamefont{R.}~\bibnamefont{Kumar}},
  \bibinfo{author}{\bibfnamefont{S.~G.} \bibnamefont{Ghosh}}, \bibnamefont{and}
  \bibinfo{author}{\bibfnamefont{A.}~\bibnamefont{Wang}},
  \bibinfo{journal}{Phys. Rev. D} \textbf{\bibinfo{volume}{100}},
  \bibinfo{pages}{124024} (\bibinfo{year}{2019}), \eprint{1912.05154}.

\bibitem[{\citenamefont{Shaikh et~al.}(2021)\citenamefont{Shaikh, Pal, Pal, and
  Sarkar}}]{Shaikh:2021yux}
\bibinfo{author}{\bibfnamefont{R.}~\bibnamefont{Shaikh}},
  \bibinfo{author}{\bibfnamefont{K.}~\bibnamefont{Pal}},
  \bibinfo{author}{\bibfnamefont{K.}~\bibnamefont{Pal}}, \bibnamefont{and}
  \bibinfo{author}{\bibfnamefont{T.}~\bibnamefont{Sarkar}},
  \bibinfo{journal}{Mon. Not. Roy. Astron. Soc.}
  \textbf{\bibinfo{volume}{506}}, \bibinfo{pages}{1229} (\bibinfo{year}{2021}),
  \eprint{2102.04299}.

\bibitem[{\citenamefont{Wu et~al.}(2023{\natexlab{a}})\citenamefont{Wu, Zhang,
  Shao, and Qian}}]{Wu:2023yhp}
\bibinfo{author}{\bibfnamefont{W.-H.} \bibnamefont{Wu}},
  \bibinfo{author}{\bibfnamefont{C.-Y.} \bibnamefont{Zhang}},
  \bibinfo{author}{\bibfnamefont{C.-G.} \bibnamefont{Shao}}, \bibnamefont{and}
  \bibinfo{author}{\bibfnamefont{W.-L.} \bibnamefont{Qian}},
  \bibinfo{journal}{Chin. Phys. C} \textbf{\bibinfo{volume}{47}},
  \bibinfo{pages}{085102} (\bibinfo{year}{2023}{\natexlab{a}}),
  \eprint{2303.04907}.

\bibitem[{\citenamefont{Capozziello et~al.}(2023)\citenamefont{Capozziello,
  Zare, and Hassanabadi}}]{Capozziello:2023tbo}
\bibinfo{author}{\bibfnamefont{S.}~\bibnamefont{Capozziello}},
  \bibinfo{author}{\bibfnamefont{S.}~\bibnamefont{Zare}}, \bibnamefont{and}
  \bibinfo{author}{\bibfnamefont{H.}~\bibnamefont{Hassanabadi}}
  (\bibinfo{year}{2023}), \eprint{2311.12896}.

\bibitem[{\citenamefont{Sui et~al.}(2023)\citenamefont{Sui, Fu, and
  Guo}}]{Sui:2023rfh}
\bibinfo{author}{\bibfnamefont{T.-T.} \bibnamefont{Sui}},
  \bibinfo{author}{\bibfnamefont{Q.-M.} \bibnamefont{Fu}}, \bibnamefont{and}
  \bibinfo{author}{\bibfnamefont{W.-D.} \bibnamefont{Guo}},
  \bibinfo{journal}{Phys. Lett. B} \textbf{\bibinfo{volume}{845}},
  \bibinfo{pages}{138135} (\bibinfo{year}{2023}), \eprint{2311.10930}.

\bibitem[{\citenamefont{Pantig et~al.}(2023)\citenamefont{Pantig, \"Ovg\"un,
  and Demir}}]{Pantig:2022qak}
\bibinfo{author}{\bibfnamefont{R.~C.} \bibnamefont{Pantig}},
  \bibinfo{author}{\bibfnamefont{A.}~\bibnamefont{\"Ovg\"un}},
  \bibnamefont{and} \bibinfo{author}{\bibfnamefont{D.}~\bibnamefont{Demir}},
  \bibinfo{journal}{Eur. Phys. J. C} \textbf{\bibinfo{volume}{83}},
  \bibinfo{pages}{250} (\bibinfo{year}{2023}), \eprint{2208.02969}.

\bibitem[{\citenamefont{Ghosh et~al.}(2023)\citenamefont{Ghosh, Sk, and
  Sarkar}}]{Ghosh:2023kge}
\bibinfo{author}{\bibfnamefont{R.}~\bibnamefont{Ghosh}},
  \bibinfo{author}{\bibfnamefont{S.}~\bibnamefont{Sk}}, \bibnamefont{and}
  \bibinfo{author}{\bibfnamefont{S.}~\bibnamefont{Sarkar}},
  \bibinfo{journal}{Phys. Rev. D} \textbf{\bibinfo{volume}{108}},
  \bibinfo{pages}{L041501} (\bibinfo{year}{2023}), \eprint{2306.14193}.

\bibitem[{\citenamefont{Tsukamoto et~al.}(2014)\citenamefont{Tsukamoto, Li, and
  Bambi}}]{Tsukamoto:2014tja}
\bibinfo{author}{\bibfnamefont{N.}~\bibnamefont{Tsukamoto}},
  \bibinfo{author}{\bibfnamefont{Z.}~\bibnamefont{Li}}, \bibnamefont{and}
  \bibinfo{author}{\bibfnamefont{C.}~\bibnamefont{Bambi}},
  \bibinfo{journal}{JCAP} \textbf{\bibinfo{volume}{06}}, \bibinfo{pages}{043}
  (\bibinfo{year}{2014}), \eprint{1403.0371}.

\bibitem[{\citenamefont{Porth et~al.}(2019)}]{EventHorizonTelescope:2019pcy}
\bibinfo{author}{\bibfnamefont{O.}~\bibnamefont{Porth}} \bibnamefont{et~al.}
  (\bibinfo{collaboration}{Event Horizon Telescope}),
  \bibinfo{journal}{Astrophys. J. Suppl.} \textbf{\bibinfo{volume}{243}},
  \bibinfo{pages}{26} (\bibinfo{year}{2019}), \eprint{1904.04923}.

\bibitem[{\citenamefont{Gralla et~al.}(2019)\citenamefont{Gralla, Holz, and
  Wald}}]{Gralla:2019xty}
\bibinfo{author}{\bibfnamefont{S.~E.} \bibnamefont{Gralla}},
  \bibinfo{author}{\bibfnamefont{D.~E.} \bibnamefont{Holz}}, \bibnamefont{and}
  \bibinfo{author}{\bibfnamefont{R.~M.} \bibnamefont{Wald}},
  \bibinfo{journal}{Phys. Rev. D} \textbf{\bibinfo{volume}{100}},
  \bibinfo{pages}{024018} (\bibinfo{year}{2019}), \eprint{1906.00873}.

\bibitem[{\citenamefont{Dokuchaev and Nazarova}(2019)}]{Dokuchaev:2019pcx}
\bibinfo{author}{\bibfnamefont{V.~I.} \bibnamefont{Dokuchaev}}
  \bibnamefont{and} \bibinfo{author}{\bibfnamefont{N.~O.}
  \bibnamefont{Nazarova}}, \bibinfo{journal}{Universe}
  \textbf{\bibinfo{volume}{5}}, \bibinfo{pages}{183} (\bibinfo{year}{2019}),
  \eprint{1906.07171}.

\bibitem[{\citenamefont{Peng et~al.}(2021{\natexlab{b}})\citenamefont{Peng,
  Guo, and Feng}}]{Peng:2020wun}
\bibinfo{author}{\bibfnamefont{J.}~\bibnamefont{Peng}},
  \bibinfo{author}{\bibfnamefont{M.}~\bibnamefont{Guo}}, \bibnamefont{and}
  \bibinfo{author}{\bibfnamefont{X.-H.} \bibnamefont{Feng}},
  \bibinfo{journal}{Chin. Phys. C} \textbf{\bibinfo{volume}{45}},
  \bibinfo{pages}{085103} (\bibinfo{year}{2021}{\natexlab{b}}),
  \eprint{2008.00657}.

\bibitem[{\citenamefont{He et~al.}(2022)\citenamefont{He, Guo, Tan, and
  Li}}]{He:2021htq}
\bibinfo{author}{\bibfnamefont{K.-J.} \bibnamefont{He}},
  \bibinfo{author}{\bibfnamefont{S.}~\bibnamefont{Guo}},
  \bibinfo{author}{\bibfnamefont{S.-C.} \bibnamefont{Tan}}, \bibnamefont{and}
  \bibinfo{author}{\bibfnamefont{G.-P.} \bibnamefont{Li}},
  \bibinfo{journal}{Chin. Phys. C} \textbf{\bibinfo{volume}{46}},
  \bibinfo{pages}{085106} (\bibinfo{year}{2022}), \eprint{2103.13664}.

\bibitem[{\citenamefont{Eichhorn and Held}(2021)}]{Eichhorn:2021iwq}
\bibinfo{author}{\bibfnamefont{A.}~\bibnamefont{Eichhorn}} \bibnamefont{and}
  \bibinfo{author}{\bibfnamefont{A.}~\bibnamefont{Held}},
  \bibinfo{journal}{JCAP} \textbf{\bibinfo{volume}{05}}, \bibinfo{pages}{073}
  (\bibinfo{year}{2021}), \eprint{2103.13163}.

\bibitem[{\citenamefont{Li and He}(2021)}]{Li:2021riw}
\bibinfo{author}{\bibfnamefont{G.-P.} \bibnamefont{Li}} \bibnamefont{and}
  \bibinfo{author}{\bibfnamefont{K.-J.} \bibnamefont{He}},
  \bibinfo{journal}{JCAP} \textbf{\bibinfo{volume}{06}}, \bibinfo{pages}{037}
  (\bibinfo{year}{2021}), \eprint{2105.08521}.

\bibitem[{\citenamefont{Wang et~al.}(2023)\citenamefont{Wang, Kuang, Meng,
  Wang, and Wu}}]{Wang:2023vcv}
\bibinfo{author}{\bibfnamefont{X.-J.} \bibnamefont{Wang}},
  \bibinfo{author}{\bibfnamefont{X.-M.} \bibnamefont{Kuang}},
  \bibinfo{author}{\bibfnamefont{Y.}~\bibnamefont{Meng}},
  \bibinfo{author}{\bibfnamefont{B.}~\bibnamefont{Wang}}, \bibnamefont{and}
  \bibinfo{author}{\bibfnamefont{J.-P.} \bibnamefont{Wu}},
  \bibinfo{journal}{Phys. Rev. D} \textbf{\bibinfo{volume}{107}},
  \bibinfo{pages}{124052} (\bibinfo{year}{2023}), \eprint{2304.10015}.

\bibitem[{\citenamefont{Zeng et~al.}(2020)\citenamefont{Zeng, Zhang, and
  Zhang}}]{Zeng:2020dco}
\bibinfo{author}{\bibfnamefont{X.-X.} \bibnamefont{Zeng}},
  \bibinfo{author}{\bibfnamefont{H.-Q.} \bibnamefont{Zhang}}, \bibnamefont{and}
  \bibinfo{author}{\bibfnamefont{H.}~\bibnamefont{Zhang}},
  \bibinfo{journal}{Eur. Phys. J. C} \textbf{\bibinfo{volume}{80}},
  \bibinfo{pages}{872} (\bibinfo{year}{2020}), \eprint{2004.12074}.

\bibitem[{\citenamefont{Saurabh and Jusufi}(2021)}]{Saurabh:2020zqg}
\bibinfo{author}{\bibfnamefont{K.}~\bibnamefont{Saurabh}} \bibnamefont{and}
  \bibinfo{author}{\bibfnamefont{K.}~\bibnamefont{Jusufi}},
  \bibinfo{journal}{Eur. Phys. J. C} \textbf{\bibinfo{volume}{81}},
  \bibinfo{pages}{490} (\bibinfo{year}{2021}), \eprint{2009.10599}.

\bibitem[{\citenamefont{Zeng and Zhang}(2020)}]{Zeng:2020vsj}
\bibinfo{author}{\bibfnamefont{X.-X.} \bibnamefont{Zeng}} \bibnamefont{and}
  \bibinfo{author}{\bibfnamefont{H.-Q.} \bibnamefont{Zhang}},
  \bibinfo{journal}{Eur. Phys. J. C} \textbf{\bibinfo{volume}{80}},
  \bibinfo{pages}{1058} (\bibinfo{year}{2020}), \eprint{2007.06333}.

\bibitem[{\citenamefont{Qin et~al.}(2021)\citenamefont{Qin, Chen, and
  Jing}}]{Qin:2020xzu}
\bibinfo{author}{\bibfnamefont{X.}~\bibnamefont{Qin}},
  \bibinfo{author}{\bibfnamefont{S.}~\bibnamefont{Chen}}, \bibnamefont{and}
  \bibinfo{author}{\bibfnamefont{J.}~\bibnamefont{Jing}},
  \bibinfo{journal}{Class. Quant. Grav.} \textbf{\bibinfo{volume}{38}},
  \bibinfo{pages}{115008} (\bibinfo{year}{2021}), \eprint{2011.04310}.

\bibitem[{\citenamefont{Narayan et~al.}(2019)\citenamefont{Narayan, Johnson,
  and Gammie}}]{Narayan:2019imo}
\bibinfo{author}{\bibfnamefont{R.}~\bibnamefont{Narayan}},
  \bibinfo{author}{\bibfnamefont{M.~D.} \bibnamefont{Johnson}},
  \bibnamefont{and} \bibinfo{author}{\bibfnamefont{C.~F.}
  \bibnamefont{Gammie}}, \bibinfo{journal}{Astrophys. J. Lett.}
  \textbf{\bibinfo{volume}{885}}, \bibinfo{pages}{L33} (\bibinfo{year}{2019}),
  \eprint{1910.02957}.

\bibitem[{\citenamefont{Gan et~al.}(2021{\natexlab{a}})\citenamefont{Gan, Wang,
  Wu, and Yang}}]{Gan:2021xdl}
\bibinfo{author}{\bibfnamefont{Q.}~\bibnamefont{Gan}},
  \bibinfo{author}{\bibfnamefont{P.}~\bibnamefont{Wang}},
  \bibinfo{author}{\bibfnamefont{H.}~\bibnamefont{Wu}}, \bibnamefont{and}
  \bibinfo{author}{\bibfnamefont{H.}~\bibnamefont{Yang}},
  \bibinfo{journal}{Phys. Rev. D} \textbf{\bibinfo{volume}{104}},
  \bibinfo{pages}{044049} (\bibinfo{year}{2021}{\natexlab{a}}),
  \eprint{2105.11770}.

\bibitem[{\citenamefont{Gan et~al.}(2021{\natexlab{b}})\citenamefont{Gan, Wang,
  Wu, and Yang}}]{Gan:2021pwu}
\bibinfo{author}{\bibfnamefont{Q.}~\bibnamefont{Gan}},
  \bibinfo{author}{\bibfnamefont{P.}~\bibnamefont{Wang}},
  \bibinfo{author}{\bibfnamefont{H.}~\bibnamefont{Wu}}, \bibnamefont{and}
  \bibinfo{author}{\bibfnamefont{H.}~\bibnamefont{Yang}},
  \bibinfo{journal}{Phys. Rev. D} \textbf{\bibinfo{volume}{104}},
  \bibinfo{pages}{024003} (\bibinfo{year}{2021}{\natexlab{b}}),
  \eprint{2104.08703}.

\bibitem[{\citenamefont{Meng et~al.}(2023{\natexlab{b}})\citenamefont{Meng,
  Kuang, Wang, Wang, and Wu}}]{Meng:2023htc}
\bibinfo{author}{\bibfnamefont{Y.}~\bibnamefont{Meng}},
  \bibinfo{author}{\bibfnamefont{X.-M.} \bibnamefont{Kuang}},
  \bibinfo{author}{\bibfnamefont{X.-J.} \bibnamefont{Wang}},
  \bibinfo{author}{\bibfnamefont{B.}~\bibnamefont{Wang}}, \bibnamefont{and}
  \bibinfo{author}{\bibfnamefont{J.-P.} \bibnamefont{Wu}},
  \bibinfo{journal}{Phys. Rev. D} \textbf{\bibinfo{volume}{108}},
  \bibinfo{pages}{064013} (\bibinfo{year}{2023}{\natexlab{b}}),
  \eprint{2306.10459}.

\bibitem[{\citenamefont{Boshkayev et~al.}(2022)\citenamefont{Boshkayev,
  Konysbayev, Kurmanov, Luongo, and Malafarina}}]{Boshkayev:2022vlv}
\bibinfo{author}{\bibfnamefont{K.}~\bibnamefont{Boshkayev}},
  \bibinfo{author}{\bibfnamefont{T.}~\bibnamefont{Konysbayev}},
  \bibinfo{author}{\bibfnamefont{Y.}~\bibnamefont{Kurmanov}},
  \bibinfo{author}{\bibfnamefont{O.}~\bibnamefont{Luongo}}, \bibnamefont{and}
  \bibinfo{author}{\bibfnamefont{D.}~\bibnamefont{Malafarina}},
  \bibinfo{journal}{Astrophys. J.} \textbf{\bibinfo{volume}{936}},
  \bibinfo{pages}{96} (\bibinfo{year}{2022}), \eprint{2205.04208}.

\bibitem[{\citenamefont{Xavier et~al.}(2023)\citenamefont{Xavier, Lima, and
  Crispino}}]{Xavier:2023exm}
\bibinfo{author}{\bibfnamefont{S.~V. M. C.~B.} \bibnamefont{Xavier}},
  \bibinfo{author}{\bibfnamefont{H.~C.~D.} \bibnamefont{Lima},
  \bibfnamefont{Junior.}}, \bibnamefont{and}
  \bibinfo{author}{\bibfnamefont{L.~C.~B.} \bibnamefont{Crispino}},
  \bibinfo{journal}{Phys. Rev. D} \textbf{\bibinfo{volume}{107}},
  \bibinfo{pages}{064040} (\bibinfo{year}{2023}), \eprint{2303.17666}.

\bibitem[{\citenamefont{Sakai et~al.}(2014)\citenamefont{Sakai, Saida, and
  Tamaki}}]{Sakai:2014pga}
\bibinfo{author}{\bibfnamefont{N.}~\bibnamefont{Sakai}},
  \bibinfo{author}{\bibfnamefont{H.}~\bibnamefont{Saida}}, \bibnamefont{and}
  \bibinfo{author}{\bibfnamefont{T.}~\bibnamefont{Tamaki}},
  \bibinfo{journal}{Phys. Rev. D} \textbf{\bibinfo{volume}{90}},
  \bibinfo{pages}{104013} (\bibinfo{year}{2014}), \eprint{1408.6929}.

\bibitem[{\citenamefont{Bacchini et~al.}(2021)\citenamefont{Bacchini, Mayerson,
  Ripperda, Davelaar, Olivares, Hertog, and Vercnocke}}]{Bacchini:2021fig}
\bibinfo{author}{\bibfnamefont{F.}~\bibnamefont{Bacchini}},
  \bibinfo{author}{\bibfnamefont{D.~R.} \bibnamefont{Mayerson}},
  \bibinfo{author}{\bibfnamefont{B.}~\bibnamefont{Ripperda}},
  \bibinfo{author}{\bibfnamefont{J.}~\bibnamefont{Davelaar}},
  \bibinfo{author}{\bibfnamefont{H.}~\bibnamefont{Olivares}},
  \bibinfo{author}{\bibfnamefont{T.}~\bibnamefont{Hertog}}, \bibnamefont{and}
  \bibinfo{author}{\bibfnamefont{B.}~\bibnamefont{Vercnocke}},
  \bibinfo{journal}{Phys. Rev. Lett.} \textbf{\bibinfo{volume}{127}},
  \bibinfo{pages}{171601} (\bibinfo{year}{2021}), \eprint{2103.12075}.

\bibitem[{\citenamefont{Destounis et~al.}(2023)\citenamefont{Destounis,
  Angeloni, Vaglio, and Pani}}]{Destounis:2023khj}
\bibinfo{author}{\bibfnamefont{K.}~\bibnamefont{Destounis}},
  \bibinfo{author}{\bibfnamefont{F.}~\bibnamefont{Angeloni}},
  \bibinfo{author}{\bibfnamefont{M.}~\bibnamefont{Vaglio}}, \bibnamefont{and}
  \bibinfo{author}{\bibfnamefont{P.}~\bibnamefont{Pani}},
  \bibinfo{journal}{Phys. Rev. D} \textbf{\bibinfo{volume}{108}},
  \bibinfo{pages}{084062} (\bibinfo{year}{2023}), \eprint{2305.05691}.

\bibitem[{\citenamefont{Guo et~al.}(2023{\natexlab{a}})\citenamefont{Guo, Li,
  and Liang}}]{Guo:2022iiy}
\bibinfo{author}{\bibfnamefont{S.}~\bibnamefont{Guo}},
  \bibinfo{author}{\bibfnamefont{G.-R.} \bibnamefont{Li}}, \bibnamefont{and}
  \bibinfo{author}{\bibfnamefont{E.-W.} \bibnamefont{Liang}},
  \bibinfo{journal}{Eur. Phys. J. C} \textbf{\bibinfo{volume}{83}},
  \bibinfo{pages}{663} (\bibinfo{year}{2023}{\natexlab{a}}),
  \eprint{2210.03010}.

\bibitem[{\citenamefont{Guo et~al.}(2023{\natexlab{b}})\citenamefont{Guo, Lu,
  Wang, Wu, and Yang}}]{Guo:2022ghl}
\bibinfo{author}{\bibfnamefont{G.}~\bibnamefont{Guo}},
  \bibinfo{author}{\bibfnamefont{Y.}~\bibnamefont{Lu}},
  \bibinfo{author}{\bibfnamefont{P.}~\bibnamefont{Wang}},
  \bibinfo{author}{\bibfnamefont{H.}~\bibnamefont{Wu}}, \bibnamefont{and}
  \bibinfo{author}{\bibfnamefont{H.}~\bibnamefont{Yang}},
  \bibinfo{journal}{Phys. Rev. D} \textbf{\bibinfo{volume}{107}},
  \bibinfo{pages}{124037} (\bibinfo{year}{2023}{\natexlab{b}}),
  \eprint{2212.12901}.

\bibitem[{\citenamefont{Archer-Smith and Zhang}(2021)}]{Archer-Smith:2020hqq}
\bibinfo{author}{\bibfnamefont{P.}~\bibnamefont{Archer-Smith}}
  \bibnamefont{and} \bibinfo{author}{\bibfnamefont{Y.}~\bibnamefont{Zhang}},
  \bibinfo{journal}{Phys. Lett. B} \textbf{\bibinfo{volume}{817}},
  \bibinfo{pages}{136309} (\bibinfo{year}{2021}), \eprint{2005.08980}.

\bibitem[{\citenamefont{Okyay and \"Ovg\"un}(2022)}]{Okyay:2021nnh}
\bibinfo{author}{\bibfnamefont{M.}~\bibnamefont{Okyay}} \bibnamefont{and}
  \bibinfo{author}{\bibfnamefont{A.}~\bibnamefont{\"Ovg\"un}},
  \bibinfo{journal}{JCAP} \textbf{\bibinfo{volume}{01}}, \bibinfo{pages}{009}
  (\bibinfo{year}{2022}), \eprint{2108.07766}.

\bibitem[{\citenamefont{Uniyal et~al.}(2023{\natexlab{a}})\citenamefont{Uniyal,
  Pantig, and \"Ovg\"un}}]{Uniyal:2022vdu}
\bibinfo{author}{\bibfnamefont{A.}~\bibnamefont{Uniyal}},
  \bibinfo{author}{\bibfnamefont{R.~C.} \bibnamefont{Pantig}},
  \bibnamefont{and}
  \bibinfo{author}{\bibfnamefont{A.}~\bibnamefont{\"Ovg\"un}},
  \bibinfo{journal}{Phys. Dark Univ.} \textbf{\bibinfo{volume}{40}},
  \bibinfo{pages}{101178} (\bibinfo{year}{2023}{\natexlab{a}}),
  \eprint{2205.11072}.

\bibitem[{\citenamefont{Hou et~al.}(2022)\citenamefont{Hou, Zhang, Yan, Guo,
  and Chen}}]{Hou:2022eev}
\bibinfo{author}{\bibfnamefont{Y.}~\bibnamefont{Hou}},
  \bibinfo{author}{\bibfnamefont{Z.}~\bibnamefont{Zhang}},
  \bibinfo{author}{\bibfnamefont{H.}~\bibnamefont{Yan}},
  \bibinfo{author}{\bibfnamefont{M.}~\bibnamefont{Guo}}, \bibnamefont{and}
  \bibinfo{author}{\bibfnamefont{B.}~\bibnamefont{Chen}},
  \bibinfo{journal}{Phys. Rev. D} \textbf{\bibinfo{volume}{106}},
  \bibinfo{pages}{064058} (\bibinfo{year}{2022}), \eprint{2206.13744}.

\bibitem[{\citenamefont{Uniyal et~al.}(2023{\natexlab{b}})\citenamefont{Uniyal,
  Chakrabarti, Pantig, and \"Ovg\"un}}]{Uniyal:2023inx}
\bibinfo{author}{\bibfnamefont{A.}~\bibnamefont{Uniyal}},
  \bibinfo{author}{\bibfnamefont{S.}~\bibnamefont{Chakrabarti}},
  \bibinfo{author}{\bibfnamefont{R.~C.} \bibnamefont{Pantig}},
  \bibnamefont{and} \bibinfo{author}{\bibfnamefont{A.}~\bibnamefont{\"Ovg\"un}}
  (\bibinfo{year}{2023}{\natexlab{b}}), \eprint{2303.07174}.

\bibitem[{\citenamefont{Akbarieh et~al.}(2023)\citenamefont{Akbarieh,
  Khoshragbaf, and Atazadeh}}]{Akbarieh:2023kjv}
\bibinfo{author}{\bibfnamefont{A.~R.} \bibnamefont{Akbarieh}},
  \bibinfo{author}{\bibfnamefont{M.}~\bibnamefont{Khoshragbaf}},
  \bibnamefont{and} \bibinfo{author}{\bibfnamefont{M.}~\bibnamefont{Atazadeh}}
  (\bibinfo{year}{2023}), \eprint{2302.02784}.

\bibitem[{\citenamefont{Gao et~al.}(2023)\citenamefont{Gao, Sui, Zeng, An, and
  Hu}}]{Gao:2023mjb}
\bibinfo{author}{\bibfnamefont{X.-J.} \bibnamefont{Gao}},
  \bibinfo{author}{\bibfnamefont{T.-T.} \bibnamefont{Sui}},
  \bibinfo{author}{\bibfnamefont{X.-X.} \bibnamefont{Zeng}},
  \bibinfo{author}{\bibfnamefont{Y.-S.} \bibnamefont{An}}, \bibnamefont{and}
  \bibinfo{author}{\bibfnamefont{Y.-P.} \bibnamefont{Hu}},
  \bibinfo{journal}{Eur. Phys. J. C} \textbf{\bibinfo{volume}{83}},
  \bibinfo{pages}{1052} (\bibinfo{year}{2023}), \eprint{2311.11780}.

\bibitem[{\citenamefont{Theodosopoulos
  et~al.}(2023)\citenamefont{Theodosopoulos, Karakasis, Koutsoumbas, and
  Papantonopoulos}}]{Theodosopoulos:2023ice}
\bibinfo{author}{\bibfnamefont{D.~P.} \bibnamefont{Theodosopoulos}},
  \bibinfo{author}{\bibfnamefont{T.}~\bibnamefont{Karakasis}},
  \bibinfo{author}{\bibfnamefont{G.}~\bibnamefont{Koutsoumbas}},
  \bibnamefont{and}
  \bibinfo{author}{\bibfnamefont{E.}~\bibnamefont{Papantonopoulos}}
  (\bibinfo{year}{2023}), \eprint{2311.02740}.

\bibitem[{\citenamefont{Zeng et~al.}(2023)\citenamefont{Zeng, Ling, Jiang, and
  Li}}]{Zeng:2023fqy}
\bibinfo{author}{\bibfnamefont{W.}~\bibnamefont{Zeng}},
  \bibinfo{author}{\bibfnamefont{Y.}~\bibnamefont{Ling}},
  \bibinfo{author}{\bibfnamefont{Q.-Q.} \bibnamefont{Jiang}}, \bibnamefont{and}
  \bibinfo{author}{\bibfnamefont{G.-P.} \bibnamefont{Li}},
  \bibinfo{journal}{Phys. Rev. D} \textbf{\bibinfo{volume}{108}},
  \bibinfo{pages}{104072} (\bibinfo{year}{2023}), \eprint{2308.00976}.

\bibitem[{\citenamefont{Meng et~al.}(2023{\natexlab{c}})\citenamefont{Meng,
  Fan, Li, Han, and Zhang}}]{Meng:2023uws}
\bibinfo{author}{\bibfnamefont{K.}~\bibnamefont{Meng}},
  \bibinfo{author}{\bibfnamefont{X.-L.} \bibnamefont{Fan}},
  \bibinfo{author}{\bibfnamefont{S.}~\bibnamefont{Li}},
  \bibinfo{author}{\bibfnamefont{W.-B.} \bibnamefont{Han}}, \bibnamefont{and}
  \bibinfo{author}{\bibfnamefont{H.}~\bibnamefont{Zhang}},
  \bibinfo{journal}{JHEP} \textbf{\bibinfo{volume}{11}}, \bibinfo{pages}{141}
  (\bibinfo{year}{2023}{\natexlab{c}}), \eprint{2307.08953}.

\bibitem[{\citenamefont{Liebling and Palenzuela}(2023)}]{Liebling:2012fv}
\bibinfo{author}{\bibfnamefont{S.~L.} \bibnamefont{Liebling}} \bibnamefont{and}
  \bibinfo{author}{\bibfnamefont{C.}~\bibnamefont{Palenzuela}},
  \bibinfo{journal}{Living Rev. Rel.} \textbf{\bibinfo{volume}{26}},
  \bibinfo{pages}{1} (\bibinfo{year}{2023}), \eprint{1202.5809}.

\bibitem[{\citenamefont{Rosa et~al.}(2023)\citenamefont{Rosa, Macedo, and
  Rubiera-Garcia}}]{Rosa:2023qcv}
\bibinfo{author}{\bibfnamefont{J.~a.~L.} \bibnamefont{Rosa}},
  \bibinfo{author}{\bibfnamefont{C.~F.~B.} \bibnamefont{Macedo}},
  \bibnamefont{and}
  \bibinfo{author}{\bibfnamefont{D.}~\bibnamefont{Rubiera-Garcia}},
  \bibinfo{journal}{Phys. Rev. D} \textbf{\bibinfo{volume}{108}},
  \bibinfo{pages}{044021} (\bibinfo{year}{2023}), \eprint{2303.17296}.

\bibitem[{\citenamefont{Rosa and Rubiera-Garcia}(2022)}]{Rosa:2022tfv}
\bibinfo{author}{\bibfnamefont{J.~a.~L.} \bibnamefont{Rosa}} \bibnamefont{and}
  \bibinfo{author}{\bibfnamefont{D.}~\bibnamefont{Rubiera-Garcia}},
  \bibinfo{journal}{Phys. Rev. D} \textbf{\bibinfo{volume}{106}},
  \bibinfo{pages}{084004} (\bibinfo{year}{2022}), \eprint{2204.12949}.

\bibitem[{\citenamefont{Vincent et~al.}(2016)\citenamefont{Vincent, Meliani,
  Grandclement, Gourgoulhon, and Straub}}]{Vincent:2015xta}
\bibinfo{author}{\bibfnamefont{F.~H.} \bibnamefont{Vincent}},
  \bibinfo{author}{\bibfnamefont{Z.}~\bibnamefont{Meliani}},
  \bibinfo{author}{\bibfnamefont{P.}~\bibnamefont{Grandclement}},
  \bibinfo{author}{\bibfnamefont{E.}~\bibnamefont{Gourgoulhon}},
  \bibnamefont{and} \bibinfo{author}{\bibfnamefont{O.}~\bibnamefont{Straub}},
  \bibinfo{journal}{Class. Quant. Grav.} \textbf{\bibinfo{volume}{33}},
  \bibinfo{pages}{105015} (\bibinfo{year}{2016}), \eprint{1510.04170}.

\bibitem[{\citenamefont{Ovalle et~al.}(2021)\citenamefont{Ovalle, Casadio,
  Contreras, and Sotomayor}}]{Ovalle:2020kpd}
\bibinfo{author}{\bibfnamefont{J.}~\bibnamefont{Ovalle}},
  \bibinfo{author}{\bibfnamefont{R.}~\bibnamefont{Casadio}},
  \bibinfo{author}{\bibfnamefont{E.}~\bibnamefont{Contreras}},
  \bibnamefont{and}
  \bibinfo{author}{\bibfnamefont{A.}~\bibnamefont{Sotomayor}},
  \bibinfo{journal}{Phys. Dark Univ.} \textbf{\bibinfo{volume}{31}},
  \bibinfo{pages}{100744} (\bibinfo{year}{2021}), \eprint{2006.06735}.

\bibitem[{\citenamefont{Contreras et~al.}(2021)\citenamefont{Contreras, Ovalle,
  and Casadio}}]{Contreras:2021yxe}
\bibinfo{author}{\bibfnamefont{E.}~\bibnamefont{Contreras}},
  \bibinfo{author}{\bibfnamefont{J.}~\bibnamefont{Ovalle}}, \bibnamefont{and}
  \bibinfo{author}{\bibfnamefont{R.}~\bibnamefont{Casadio}},
  \bibinfo{journal}{Phys. Rev. D} \textbf{\bibinfo{volume}{103}},
  \bibinfo{pages}{044020} (\bibinfo{year}{2021}), \eprint{2101.08569}.

\bibitem[{\citenamefont{Mahapatra and Banerjee}(2023)}]{Mahapatra:2022xea}
\bibinfo{author}{\bibfnamefont{S.}~\bibnamefont{Mahapatra}} \bibnamefont{and}
  \bibinfo{author}{\bibfnamefont{I.}~\bibnamefont{Banerjee}},
  \bibinfo{journal}{Phys. Dark Univ.} \textbf{\bibinfo{volume}{39}},
  \bibinfo{pages}{101172} (\bibinfo{year}{2023}), \eprint{2208.05796}.

\bibitem[{\citenamefont{Cavalcanti et~al.}(2022)\citenamefont{Cavalcanti,
  de~Paiva, and da~Rocha}}]{Cavalcanti:2022cga}
\bibinfo{author}{\bibfnamefont{R.~T.} \bibnamefont{Cavalcanti}},
  \bibinfo{author}{\bibfnamefont{R.~C.} \bibnamefont{de~Paiva}},
  \bibnamefont{and} \bibinfo{author}{\bibfnamefont{R.}~\bibnamefont{da~Rocha}},
  \bibinfo{journal}{Eur. Phys. J. Plus} \textbf{\bibinfo{volume}{137}},
  \bibinfo{pages}{1185} (\bibinfo{year}{2022}), \eprint{2203.08740}.

\bibitem[{\citenamefont{Yang et~al.}(2023{\natexlab{a}})\citenamefont{Yang,
  Liu, \"Ovg\"un, Long, and Xu}}]{Yang:2022ifo}
\bibinfo{author}{\bibfnamefont{Y.}~\bibnamefont{Yang}},
  \bibinfo{author}{\bibfnamefont{D.}~\bibnamefont{Liu}},
  \bibinfo{author}{\bibfnamefont{A.}~\bibnamefont{\"Ovg\"un}},
  \bibinfo{author}{\bibfnamefont{Z.-W.} \bibnamefont{Long}}, \bibnamefont{and}
  \bibinfo{author}{\bibfnamefont{Z.}~\bibnamefont{Xu}}, \bibinfo{journal}{Phys.
  Rev. D} \textbf{\bibinfo{volume}{107}}, \bibinfo{pages}{064042}
  (\bibinfo{year}{2023}{\natexlab{a}}), \eprint{2203.11551}.

\bibitem[{\citenamefont{Li}(2023)}]{Li:2022hkq}
\bibinfo{author}{\bibfnamefont{Z.}~\bibnamefont{Li}}, \bibinfo{journal}{Phys.
  Lett. B} \textbf{\bibinfo{volume}{841}}, \bibinfo{pages}{137902}
  (\bibinfo{year}{2023}), \eprint{2212.08112}.

\bibitem[{\citenamefont{Islam and Ghosh}(2021)}]{Islam:2021dyk}
\bibinfo{author}{\bibfnamefont{S.~U.} \bibnamefont{Islam}} \bibnamefont{and}
  \bibinfo{author}{\bibfnamefont{S.~G.} \bibnamefont{Ghosh}},
  \bibinfo{journal}{Phys. Rev. D} \textbf{\bibinfo{volume}{103}},
  \bibinfo{pages}{124052} (\bibinfo{year}{2021}), \eprint{2102.08289}.

\bibitem[{\citenamefont{Wu et~al.}(2023{\natexlab{b}})\citenamefont{Wu, Guo,
  and Kuang}}]{Wu:2023wld}
\bibinfo{author}{\bibfnamefont{M.-H.} \bibnamefont{Wu}},
  \bibinfo{author}{\bibfnamefont{H.}~\bibnamefont{Guo}}, \bibnamefont{and}
  \bibinfo{author}{\bibfnamefont{X.-M.} \bibnamefont{Kuang}},
  \bibinfo{journal}{Phys. Rev. D} \textbf{\bibinfo{volume}{107}},
  \bibinfo{pages}{064033} (\bibinfo{year}{2023}{\natexlab{b}}),
  \eprint{2306.10467}.

\bibitem[{\citenamefont{Zi and Li}(2023)}]{Zi:2023omh}
\bibinfo{author}{\bibfnamefont{T.}~\bibnamefont{Zi}} \bibnamefont{and}
  \bibinfo{author}{\bibfnamefont{P.-C.} \bibnamefont{Li}},
  \bibinfo{journal}{Phys. Rev. D} \textbf{\bibinfo{volume}{108}},
  \bibinfo{pages}{084001} (\bibinfo{year}{2023}), \eprint{2306.02683}.

\bibitem[{\citenamefont{Ruffini and Wheeler}(1971)}]{Ruffini:1971bza}
\bibinfo{author}{\bibfnamefont{R.}~\bibnamefont{Ruffini}} \bibnamefont{and}
  \bibinfo{author}{\bibfnamefont{J.~A.} \bibnamefont{Wheeler}},
  \bibinfo{journal}{Phys. Today} \textbf{\bibinfo{volume}{24}},
  \bibinfo{pages}{30} (\bibinfo{year}{1971}).

\bibitem[{\citenamefont{Yuan and Narayan}(2014)}]{Yuan:2014gma}
\bibinfo{author}{\bibfnamefont{F.}~\bibnamefont{Yuan}} \bibnamefont{and}
  \bibinfo{author}{\bibfnamefont{R.}~\bibnamefont{Narayan}},
  \bibinfo{journal}{Ann. Rev. Astron. Astrophys.}
  \textbf{\bibinfo{volume}{52}}, \bibinfo{pages}{529} (\bibinfo{year}{2014}),
  \eprint{1401.0586}.

\bibitem[{\citenamefont{Bromley et~al.}(1997)\citenamefont{Bromley, Chen, and
  Miller}}]{Bromley:1996wb}
\bibinfo{author}{\bibfnamefont{B.~C.} \bibnamefont{Bromley}},
  \bibinfo{author}{\bibfnamefont{K.}~\bibnamefont{Chen}}, \bibnamefont{and}
  \bibinfo{author}{\bibfnamefont{W.~A.} \bibnamefont{Miller}},
  \bibinfo{journal}{Astrophys. J.} \textbf{\bibinfo{volume}{475}},
  \bibinfo{pages}{57} (\bibinfo{year}{1997}), \eprint{astro-ph/9601106}.

\bibitem[{\citenamefont{Wang et~al.}(2022)\citenamefont{Wang, Lin, and
  Wei}}]{Wang:2022yvi}
\bibinfo{author}{\bibfnamefont{H.-M.} \bibnamefont{Wang}},
  \bibinfo{author}{\bibfnamefont{Z.-C.} \bibnamefont{Lin}}, \bibnamefont{and}
  \bibinfo{author}{\bibfnamefont{S.-W.} \bibnamefont{Wei}},
  \bibinfo{journal}{Nucl. Phys. B} \textbf{\bibinfo{volume}{985}},
  \bibinfo{pages}{116026} (\bibinfo{year}{2022}), \eprint{2205.13174}.

\bibitem[{\citenamefont{Yang et~al.}(2023{\natexlab{b}})\citenamefont{Yang,
  Zhang, and Ma}}]{Yang:2022btw}
\bibinfo{author}{\bibfnamefont{J.}~\bibnamefont{Yang}},
  \bibinfo{author}{\bibfnamefont{C.}~\bibnamefont{Zhang}}, \bibnamefont{and}
  \bibinfo{author}{\bibfnamefont{Y.}~\bibnamefont{Ma}}, \bibinfo{journal}{Eur.
  Phys. J. C} \textbf{\bibinfo{volume}{83}}, \bibinfo{pages}{619}
  (\bibinfo{year}{2023}{\natexlab{b}}), \eprint{2211.04263}.

\bibitem[{\citenamefont{Bambi}(2013)}]{Bambi:2013nla}
\bibinfo{author}{\bibfnamefont{C.}~\bibnamefont{Bambi}},
  \bibinfo{journal}{Phys. Rev. D} \textbf{\bibinfo{volume}{87}},
  \bibinfo{pages}{107501} (\bibinfo{year}{2013}), \eprint{1304.5691}.

\bibitem[{\citenamefont{Ovalle}(2017)}]{Ovalle:2017fgl}
\bibinfo{author}{\bibfnamefont{J.}~\bibnamefont{Ovalle}},
  \bibinfo{journal}{Phys. Rev. D} \textbf{\bibinfo{volume}{95}},
  \bibinfo{pages}{104019} (\bibinfo{year}{2017}), \eprint{1704.05899}.

\end{thebibliography}
\bibliographystyle{apsrev}

\end{document}